\documentclass[12pt,prd,dvipdfm,nofootinbib,floatfix,tightenlines]{revtex4}
\usepackage[T1]{fontenc}
\usepackage[latin1]{inputenc}
\usepackage{amsmath}
\usepackage{graphicx}
\usepackage{color}

\makeatletter
\@ifundefined{textcolor}{}
{%
 \definecolor{BLACK}{gray}{0}
 \definecolor{WHITE}{gray}{1}
 \definecolor{RED}{rgb}{1,0,0}
 \definecolor{GREEN}{rgb}{0,1,0}
 \definecolor{BLUE}{rgb}{0,0,1}
 \definecolor{CYAN}{cmyk}{1,0,0,0}
 \definecolor{MAGENTA}{cmyk}{0,1,0,0}
 \definecolor{YELLOW}{cmyk}{0,0,1,0}
 }

\usepackage{hyperref}

\makeatother

\usepackage[figuresright]{rotating}

\renewcommand{\citet}[1]{\cite{#1}}

\begin{document}

\title{General-mass treatment for deep inelastic scattering \\ at two-loop
accuracy}

\author{Marco Guzzi,$^{1}$ Pavel M. Nadolsky,$^{1}$ 
Hung-Liang Lai,$^{2}$, C.-P. Yuan$^{3,4}$}

\affiliation{
$^{1}$Department of Physics, Southern Methodist University, 
Dallas, TX 75275, USA\\
$^{2}$Taipei Municipal University of Education, Taipei, Taiwan\\
$^{3}$Department of Physics \& Astronomy, Michigan State University, \\
 East Lansing, MI 48824, USA\\
$^{4}$Center for High Energy Physics, Peking University, Beijing 100871,
China}

\begin{abstract}
We present a next-to-next-to-leading order (NNLO) realization of a
general quark mass scheme (S-ACOT-$\chi$) in deep inelastic scattering
and explore the impact of NNLO terms on heavy-quark structure functions
$F_{2,L}^{c}(x,Q)$. An amended QCD factorization theorem for DIS
is discussed that validates the S-ACOT-$\chi$ scheme to all orders
in the QCD coupling strength. As a new feature, 
kinematical constraints on collinear production of heavy quarks 
that are crucial near the heavy-quark threshold are included in 
the amended factorization theorem. An algorithmic procedure is outlined
for implementing this scheme at NNLO by using mass-dependent and massless
results from literature. At two loops in QCD cut diagrams, the S-ACOT-$\chi$ scheme reduces
scale dependence of heavy-quark DIS cross sections as compared to
the fixed-flavor number scheme. 
\end{abstract}

\date{August 25, 2011}



\maketitle

\section{Introduction\label{sec:Introduction}}

In a modern global QCD analysis of parton distribution functions (PDFs),
several factors are comparable in magnitude to next-to-next-to-leading
order (NNLO) radiative contributions in the QCD coupling strength
$\alpha_{s}$. Among these factors, dependence of QCD cross sections
on masses of heavy quarks, $m_{c}$ and $m_{b}$, can be significant.
Global fits are sensitive to two types of mass effects, kinematical
suppression of production of $c$ and $b$ quarks near respective
mass thresholds in deep inelastic scattering (DIS), and large radiative
contributions to collinear production of $c\bar{c}$ or $b\bar{b}$
pairs at large collider energy. The first effect -- suppression of
DIS charm production near the threshold -- must be carefully estimated
when obtaining PDF parametrizations in order to accurately predict
key scattering rates at the Large Hadron Collider \citet{Tung:2006tb}.
The second effect is tied to an observation that $c$ and $b$ quarks
behave as practically massless and indistinguishable from other massless
flavors in typical Tevatron and LHC observables. It is therefore natural
to evaluate all fitted cross sections in a {}``general-mass'' (GM)
factorization scheme, which assumes that the number of (nearly) massless
quark flavors varies with energy, and at the same time includes dependence
on heavy-quark masses in relevant kinematical regions.

In this paper, we study NNLO quark mass terms in the default GM scheme
of CTEQ PDF analyses called {}``S-ACOT-$\chi$''. Here and in the following, 
the order of the calculation is defined by the number of QCD loops 
in Feynman cut diagrams, so that ``NNLO'' refers to the two-loop accuracy,
or ${\cal O}(\alpha_s^2)$, in the DIS coefficient functions.
Since its inception
in 1993 \citet{Aivazis:1993pi}, the ACOT scheme has undergone evolution
based on the work in \citet{Collins:1998rz,Kramer:2000hn,Tung:2001mv}.
The S-ACOT-$\chi$ version of the ACOT scheme is employed successfully
to compute heavy-quark cross sections in recent NLO CTEQ6.6, CT09,
and CT10 global fits \citet{Lai:2010vv,Nadolsky:2008zw,Pumplin:2009nk}. 

The S-ACOT-$\chi$ scheme is motivated by the QCD factorization theorem
for DIS with massive quarks \citet{Collins:1998rz}, which provides
the scheme's organizational backbone and key methods. In Sec.~\ref{sec:Theory},
we demonstrate how to amend the QCD factorization theorem
in order to validate the S-ACOT-$\chi$ scheme to all orders of $\alpha_{s}$.
We then apply this scheme at NNLO to neutral-current DIS production,
which provides the bulk of the DIS data, and for which all components
of the calculation are readily available.%
\footnote{For charged-current DIS, only massless \citet{Moch:2007gx,Moch:2008fj}
and some massive \citet{Buza:1997mg} NNLO coefficient functions have
been computed.%
}

Compared to other heavy-quark schemes available 
at NNLO \citet{Buza:1996wv, Chuvakin:1999nx, 
Thorne:1997uu,Thorne:1997ga,Thorne:2006qt,Alekhin:2009ni,Forte:2010ta},
our implementation aims to achieve more explicit analogy to the computation
of NNLO cross sections in the zero-mass (ZM) scheme \citet{SanchezGuillen:1990iq,vanNeerven:1991nn,Zijlstra:1991qc}. As another distinction, 
the S-ACOT-$\chi$ scheme quickly converges to the fixed-flavor number scheme 
near the heavy-quark threshold as a consequence of 
the amended factorization theorem, without requiring supplemental matching 
conditions  that are present in other general-mass schemes.

In Sec.~\ref{sec:Theory}, the S-ACOT-$\chi$ cross sections are
presented in the form that is reminiscent of counterpart ZM cross
sections, up to replacement of some massless components by their mass-dependent
expressions available in literature. This representation
is based on a few compact formulas that include 
the desirable features existing in other heavy-quark 
NNLO calculations. 
In Sec.~\ref{sec:Numerical-examples}, numerical predictions are
illustrated on the example of NNLO charm production cross sections.
They show that inclusion of the NNLO terms reduces theoretical uncertainties
compared to NLO.

Recent studies \citet{Tung:2006tb,Thorne:2010pa,Alekhin:2011sk,Placakyte:2010}
show that, at NLO, the LHC electroweak cross sections depend considerably
on the mass scheme and parametric input for the charm mass $m_{c}$
in the PDF analysis, even though the combined HERA-1 data 
set~\cite{:2009wt} itself
has a small total uncertainty. Yet, the upcoming combination of HERA-1
heavy-quark cross sections is expected to improve constraints on $m_{c}$.
The S-ACOT-$\chi$ implementation brings theoretical predictions up
to matching accuracy by including the NNLO terms.

\section{S-ACOT-$\chi$ scheme: theoretical framework \label{sec:Theory}}

Consider neutral-current DIS at energy that is sufficient to produce
$N_{l}$ light flavors (such as $l=u,\, d$ and $s$) and one heavy
flavor $h$ ({}``charm'') with mass $m_{h}$. The extension to production of several heavy flavors will be postponed until Sec.~\ref{sub:Several-heavy-flavors}.

The GM scheme is designed so as to enable quick convergence of perturbative
QCD series involving heavy quarks at any momentum transfer $Q$. Perturbative
QCD cross sections in the GM scheme must converge reliably near the
heavy-quark production threshold ($Q^{2}\approx m_{h}^{2}$), as well
as far above it ($Q^{2}\gg m_{h}^{2}$), and smoothly interpolate
between the limits. When $Q$ is of order $m_{h}$, it is most natural
to include all Feynman subgraphs with heavy-quark lines into the hard-scattering
function (Wilson coefficient function). Such approach is called a
{}``fixed-flavor number'' (FFN) factorization scheme. ${\cal O}(\alpha_s^2)$ 
coefficient functions for massive quark DIS production in this scheme have been
computed in \citet{Laenen:1992zk,Riemersma:1994hv,Harris:1995tu}.
At this $Q$, the NNLO coefficient functions in the GM scheme with
$N_{l}+1$ flavors are expected to reduce to the FFN massive cross
sections in the FFN scheme with $N_{l}$ flavors. On the other hand,
at high virtualities ($Q^{2}\gg m_{h}^{2}$), the NNLO GM cross sections
should be indistinguishable from the NNLO ZM cross sections \citet{SanchezGuillen:1990iq,vanNeerven:1991nn,Zijlstra:1991qc}. In this limit, 
the heavy-quark contributions are dominated 
by asymptotic collinear contributions 
that are also fully known to ${\cal O}(\alpha_s^2)$ \citet{Buza:1996wv,Buza:1997nv,Bierenbaum:2007qe,Bierenbaum:2009zt}. Mellin
moments for {\it some} structure functions and operator matrix elements \cite{Blumlein:2006mh,Bierenbaum:2009mv,Ablinger:2010ty,Blumlein:2012vq,Ablinger:2011pb,Ablinger:2012qj} 
and dominant logarithmic contributions \cite{Laenen:1998kp,Kawamura:2012cr} have been 
computed to ${\cal O}(\alpha_s^3)$.

A realization of such scheme called ``ACOT'' was developed in
Refs.~\citet{Aivazis:1993pi,Aivazis:1993kh} and proven for inclusive
DIS to all orders in Ref.~\citet{Collins:1998rz}. A non-zero PDF
is assigned in this scheme to each quark flavor that can be 
produced in the final state at the given $Q$ value. 
S-ACOT-$\chi$ is the most recent variant
of the ACOT scheme that adds two beneficial features. First, coefficient
functions derived from Feynman graphs with initial-state heavy quarks
are simplified by neglecting non-critical 
$m_h$ dependence \citet{Collins:1998rz,Kramer:2000hn}.
Second, threshold suppression is introduced by evaluating these coefficient
functions as a function of $\chi\equiv x\left(1+4m_h^{2}/Q^{2}\right)$
instead of Bjorken $x$ \citet{Tung:2001mv}. Both modifications follow
from the factorization theorem for inclusive DIS \citet{Collins:1998rz}
and produce predictions that are simpler, yet numerically accurate.
They are included as a part of the NNLO implementation that is now
presented.

\subsection{Overview of QCD factorization\label{sec:TheoryOverview}}

A DIS structure function $F(x,Q)$, such as $F_{2}$ or $F_{L}$,
is written in a factorized form as\begin{align}
F(x,Q) & =\sum_{i=1}^{N_{f}^{fs}}e_{i}^{2}\sum_{a=0}^{N_{f}}\int_{x}^{1}\frac{d\xi}{\xi}\, C_{i,a}\left(\frac{x}{\xi},\frac{Q}{\mu},\frac{m_{h}}{\mu},\alpha_{s}(\mu)\right)\,\Phi_{a/p}(\xi,\mu)\nonumber \\
 & \equiv\sum_{i=1}^{N_{f}^{fs}}e_{i}^{2}\sum_{a=0}^{N_{f}}\left[C_{i,a}\otimes\Phi_{a/p}\right](x,Q),\label{Ffactorization}\end{align}
where $\Phi_{a/p}(\xi,\mu)$ is a parton distribution function (PDF)
for a parton type $a$, light-cone momentum fraction $\xi,$ and factorization
scale $\mu.$ $C_{i,a}(\widehat{x},Q/\mu,m_{h}/\mu,\alpha_{s}(\mu))$
are Wilson coefficient functions evaluated at $\widehat{x}=x/\xi$.
Convolution integrals over $\xi$ are indicated by {}``$\otimes$''.
Two sums appear on the right-hand side of Eq.~(\ref{Ffactorization}),
over all quark flavors $i=1,...,N_{f}^{fs}$ that couple to the virtual
photon with fractional electric charges $e_{i}=2/3$ or $-1/3$, and
over parton flavors $a$ in the PDF $\Phi_{a/p}$. The index $a$
runs over quark flavors ($a=1,...N_{f}$ for $u,d,s,...$) and the
gluon ($a=0$). Perturbative coefficients of neutral-current DIS are
the same for quarks and antiquarks up to NNLO. For each combination
of flavors $i$ and $a$, summation of quark and antiquark contributions
of these flavors is always implied, but not shown for brevity. 

Eq.~(\ref{Ffactorization}) distinguishes between $N_{f}^{fs}$,
the number of quark flavors produced in the final state (\emph{fs}),
and $N_{f}$, the number of active quark flavors in $\alpha_{s}$
and PDFs. The distinction is important for the ensuing discussion,
as generally $N_{f}^{fs}$ is different from $N_{f}$ \citet{Tung:2006tb}.
$N_{f}^{fs}$ is equal to the number of final-state flavors that can
be produced at the given $\gamma^{*}p$ center-of-mass 
energy $W=Q\sqrt{1/x-1}$.
All produced quark states can couple to the photon, so that the outer
summation in Eq.~(\ref{Ffactorization}) runs up to $i=N_{f}^{fs}$. 

On the other hand, $N_{f}$ is a parameter of the renormalization
and factorization schemes. It is commonly set equal to the number
of quark flavors with the masses that are lighter than $Q$. Only flavors with
$a\leq N_{f}$ have non-zero PDFs in the inner summation, but their
actual number depends on the factorization scheme.

To determine $C_{i,a}$, we calculate auxiliary structure functions
for scattering on an initial-state parton $b,$ $F(e+b\rightarrow e+X)\equiv\sum_{i=1}^{N_{f}^{fs}}e_{i}^{2}F_{i,b}.$
The coefficient functions $C_{i,a}$ are infrared-safe parts of $F_{i,b}.$
They enter convolutions together with parton-level PDFs $\Phi_{a/b}(\xi,\mu)$
for splittings $b\rightarrow a$, as
\begin{equation}
F_{i,b}(\widehat{x},Q)=\sum_{a=0}^{N_{f}}\left[C_{i,a}\otimes\Phi_{a/b}\right](\widehat{x},Q).\label{Fib}\end{equation}

In the modified minimal subtraction ($\overline{\rm MS}$) scheme, the parton-level PDFs are given by
matrix elements of bilocal field operators that can be computed in
perturbation theory. For example, the PDF for finding a quark $q$
in a massless parton $b$, in the light-cone gauge, is

\begin{equation}
\Phi_{q/b}(\xi)=\int\frac{dy^{-}}{2\pi}e^{-\xi p^{+}y^{-}}\langle b(p)|\overline{\psi}(0,y^{-},\vec{0}_{T})\gamma^{+}\psi(0)|b(p)\rangle,\label{Phiab}\end{equation}
where the light-cone momentum components of the partons $b$ and $q$
are $p^{\mu}=\left\{ p^{+},0,\vec{0}_{T}\right\} $ and $k^{\mu}=\left\{ \xi p^{+},m_{q}^{2}/(2\xi p^{+}),\vec{0}_{T}\right\} $,
respectively, and $p^{\pm}\equiv(p^{0}\pm p^{3})/\sqrt{2}$.

The functions $F_{i,b}$, $C_{i,a}$, and $\Phi_{a/b}$ can be expanded
as a series of $a_{s}\equiv\alpha_{s}(\mu,N_{f})/(4\pi)$: \begin{eqnarray}
F_{i,b}(x) & = & F_{i,b}^{(0)}(x)+a_{s}\, F_{i,b}^{(1)}(x)+a_{s}^{2}\, F_{i,b}^{(2)}(x)+\dots,\label{FibPert}\\
C_{i,a}(\widehat{x}) & = & C_{i,a}^{(0)}(\widehat{x})+a_{s}\, C_{i,a}^{(1)}(\widehat{x})+a_{s}^{2}C_{i,a}^{(2)}(\widehat{x})+\dots,\label{CiaPert}\\
\Phi_{a/b}(\xi) & = & \delta_{ab}\delta(1-\xi)+a_{s}\, A_{ab}^{(1)}(\xi)+a_{s}^{2}A_{ab}^{(2)}(\xi)+\dots\,.\label{PhiabPert}\end{eqnarray}
In the last equation, $A_{ab}^{(k)}$ $(k=0,1,2,\dots)$ are perturbative
coefficients composed of Dokshitzer-Gribov-Lipatov-Altarelli-Parisi (DGLAP) splitting functions $P_{ab}^{(k)}$,
such as $A_{hg}^{(1)}(\xi)=2\: P_{hg}^{(1)}(\xi)\,\ln\left(\mu^{2}/m_{h}^{2}\right)$
for the $g\rightarrow h\bar{h}$ splitting. 

By equating coefficients on both sides of Eq.~(\ref{Fib}), order
by order in $a_{s}$, we obtain\begin{eqnarray}
C_{i,b}^{(0)}(\widehat{x}) & = & F_{i,b}^{(0)}(\widehat{x}),\nonumber \\
C_{i,b}^{(1)}(\widehat{x}) & = & F_{i,b}^{(1)}(\widehat{x})-\left[C_{i,a}^{(0)}\otimes A_{ab}^{(1)}\right](\widehat{x}),\nonumber \\
C_{i,b}^{(2)}(\widehat{x}) & = & F_{i,b}^{(2)}(\widehat{x})-\left[C_{i,a}^{(0)}\otimes A_{ab}^{(2)}\right](\widehat{x})-\left[C_{i,a}^{(1)}\otimes A_{ab}^{(1)}\right](\widehat{x}).\label{coef1}\end{eqnarray}
Perturbative terms $C_{i,a}^{(k)}$ in the coefficient functions
are thus derived 
from the perturbative expansions for $F_{i,b}$ and $\Phi_{a/b}$
upon implied summation over the repeating index $a$.

The coefficients $A_{ab}^{(k)}$ in $\Phi_{a/b}$ consist of large
or singular terms arising in $F_{i,b}^{(k)}$ when the momenta of
$a$ and $b$ are collinear. Subtraction of convolutions of the $A_{ab}^{(k)}$
terms from $F_{i,b}^{(k)}$ on the right-hand side of Eqs.~(\ref{coef1})
produces finite (infrared-safe) results for $C_{i,b}^{(k)}$.

Depending on the masses of $a$ and $b,$ two forms of $A_{ab}^{(k)}$
in these equations are possible. If both $a$ and $b$ are massless,
$A_{ab}^{(k)}$ contains a singular part, given in $n=4-\epsilon$
dimensions by $\sum_{p=0}^{k}(1/\epsilon)^{p}s_{p,ab}$, where the finite prefactor 
$s_{p,ab}$
contains a DGLAP splitting function; and a finite part (logs+finite
terms) of the form $\sum_{p=0}^{k}\ln^{p}(\mu^{2}/\mu_{IR}^{2})s_{p,ab}^{\prime}$,
where $\mu$ is the factorization scale, and $\mu_{IR}$ is the parameter
of the dimensional regularization in the infrared limit.

When these {}``mass singularities'' are subtracted as in Eqs.~(\ref{coef1}),
one obtains infrared-safe parts $\widehat{F}_{i,b}^{(k)}$ of $F_{i,b}^{(k)}$,
denoted by a caret:
\begin{eqnarray}
\widehat{F}_{i.b}^{(k)}\left(\widehat{x},\frac{Q^{2}}{\mu^{2}}\right)=F_{i,b}^{(k)}\left(\widehat{x},\frac{Q^{2}}{\mu_{IR}^{2}},\frac{1}{\epsilon}\right)-\sum_{p=0}^{k}\left[C_{i,a}^{(p)}\otimes A_{ab}^{(k-p)}\right]\left(\widehat{x},\frac{\mu^{2}}{\mu_{IR}^{2}},\frac{1}{\epsilon}\right)\,.\label{FibHat}\end{eqnarray}
The difference $\widehat{F}_{i,b}^{(k)}$ is finite, even though both
the {}``bare'' functions $F_{i,b}^{(k)}$ and the PDF coefficients
$A_{ab}^{(k)}$ contain the singular $1/\epsilon^{p}$ terms, where
$p$ is a positive integer. 

If a massive parton $a$ is produced from a massless parton $b$ (as
in $g\rightarrow c\bar{c}$ or $u\rightarrow g\rightarrow c\bar{c}$),
the coefficients $A_{ab}^{(k)}$ consist solely of logarithms and
finite terms involving mass $m_{a}$, 
\begin{equation}
A_{ab}^{(k)}\left(\xi,\frac{\mu^{2}}{m_{a}^{2}}\right)=\sum_{p=0}^{k}\ln^{p}(\mu^{2}/m_{a}^{2})s_{p,ab}^{'}(\xi).\end{equation}
The coefficient $A_{ab}^{(k)}$ is finite for $m_{a}\neq0,$ but $A_{ab}^{(k)}\rightarrow\infty\mbox{ when }\mu^{2}\gg m_{a}^{2}.$
For such massive quarks, the coefficients $A_{ab}^{(k)}$ appear as
subtractions from the massive $F_{ib}^{(k)}$ in 
the expressions for $C_{i,a}^{(k)}$.
In accordance with the S-ACOT scheme, the mass-dependent $A_{ab}^{(k)}$
only appear in explicit heavy-particle production, \emph{i.e.}, when
the transitions are of the type $b(m_{b}=0)\rightarrow a(m_{a}\neq0)$. All
other subprocesses use massless expressions, constructed from renormalized
parts $\widehat{F}_{i.b}^{(k)}\left(x,\frac{Q}{\mu}\right)$ given
by Eq.~(\ref{FibHat}).

\subsection{Heavy-quark component $F_{h}$ of inclusive $F(x,Q)$ \label{sec:Fheavy}}

To construct the Wilson coefficient functions explicitly,
we decompose $F(x,Q)$ according to the (anti-)quark couplings to
the photon \citep{Forte:2010ta}. Terms in which the photon couples
to the light ($l$) or heavy ($h$) quark are designated as $F_{l}$
and $F_{h}$, respectively:\begin{equation}
F=\sum_{l=1}^{N_{l}}F_{l}+F_{h},\end{equation}
 with\begin{equation}
F_{l}(x,Q)=e_{l}^{2}\sum_{a}\left[C_{l,a}\otimes\Phi_{a/p}\right](x,Q),\label{Flight}\end{equation}
 and\begin{equation}
F_{h}(x,Q)=e_{h}^{2}\sum_{a}\left[C_{h,a}\otimes\Phi_{a/p}\right](x,Q).\label{Fheavy}\end{equation}
Note that this separation is purely theoretical: $F_{l}$ and $F_{h}$
cannot be measured separately or distinguished in some other way. Furthermore,
the heavy-quark component $F_{h}$ is not the same as the semi-inclusive
heavy-quark structure function $F_{h,SI}$ measured in experiments.
The relation between $F_{h,SI}$ and $F_{h}$ will be 
clarified in Sec.~\ref{sub:FhSI},
with the explicit formula given by Eq.~(\ref{FhSI}).

Focusing first on the contribution $F_{h}$ with the photon coupled
to $h$, we obtain its Wilson coefficients $C_{h,a}(\widehat{x})$
from the parton-level functions $F_{h,b}$ via Eqs.~(\ref{coef1}):\begin{eqnarray}
 &  & C_{h,a}^{(0)}(\widehat{x})=\delta_{ha}\delta(1-\widehat{x});\nonumber \\
 &  & C_{h,g}^{(1)}=F_{h,g}^{(1)}-C_{h,h}^{(0)}\otimes A_{hg}^{(1)};\,\,\,\, C_{h,l}^{(1)}=C_{l,h}^{(1)}=0;\,\,\,\, C_{h,h}^{(1)}=F_{h,h}^{(1)}-C_{h,h}^{(0)}\otimes A_{hh}^{(1)};\nonumber \\
 &  & C_{h,g}^{(2)}=F_{h,g}^{(2)}-C_{h,h}^{(0)}\otimes A_{hg}^{(2)}-C_{h,h}^{(1)}\otimes A_{hg}^{(1)}-C_{h,g}^{(1)}\otimes A_{gg}^{(1)};\nonumber \\
 &  & C_{h,l}^{(2)}=F_{h,l}^{PS,(2)}-C_{h,h}^{(0)}\otimes A_{hl}^{PS,(2)}-C_{h,g}^{(1)}\otimes A_{gl}^{(1)};\nonumber \\
 &  & C_{h,h}^{(2)}=F_{h,h}^{(2)}-C_{h,h}^{(0)}\otimes A_{hh}^{(2)}-C_{h,h}^{(1)}\otimes A_{hh}^{(1)}-C_{h,g}^{(1)}\otimes A_{gh}^{(1)}.\label{coefFh1}\end{eqnarray}
 In these expressions, the coefficient $C_{h,l}^{(2)}$ with the initial-state
light quark depends on flavor-non-diagonal, or pure-singlet (PS),
components of $F_{i,j}^{(2)}$ and $A_{ij}^{(2)},$ defined by\begin{equation}
G_{i,j}\equiv G_{i,j}^{PS}+\delta_{ij}G_{i,j}^{NS},\mbox{ for }G_{i,j}=C_{i,j}^{(2)},\, F_{i,j}^{(2)},\mbox{ and }A_{i,j}^{(2)}.\label{PSNS}\end{equation}
 On the other hand, the coefficient $C_{h,h}^{(2)}$ with the initial-state
heavy quark depends both on the pure singlet (PS) and non-singlet
(NS) components, as will be shown below.

\begin{figure}

\includegraphics[width=1\columnwidth]{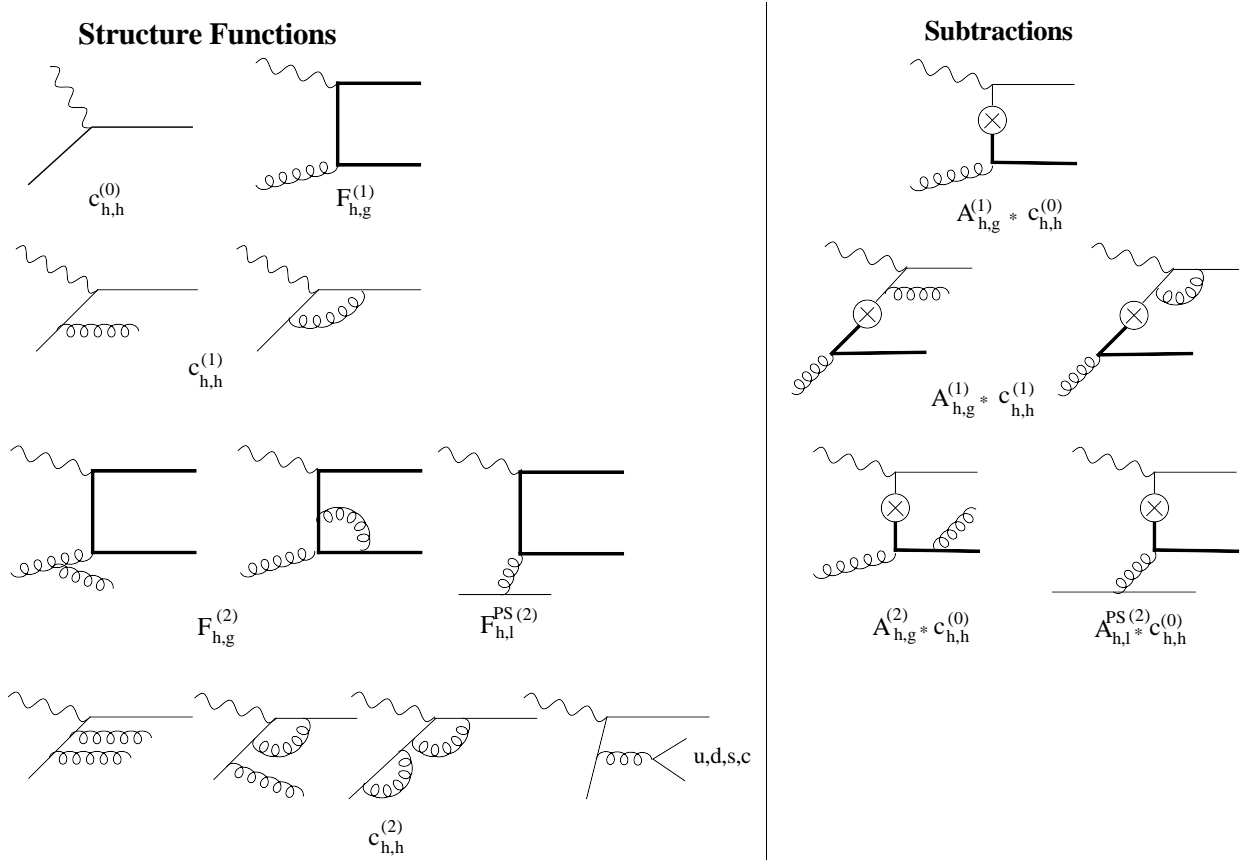}

\caption{Representative scattering contributions to $F_{h}(x,Q)$. \label{fig:FhDiagrams}}

\end{figure}

Representative diagrams for heavy-quark contributions in Eqs.~(\ref{coefFh1})
are shown in Fig.~\ref{fig:FhDiagrams}. The reader may consult this
figure frequently to identify various terms in the ensuing discussion.
Propagators and external legs for quarks that are indicated
by thick lines (thin lines) will eventually be computed with
full mass dependence (in the massless approximation) .

The heavy-quark diagrams fall into two categories, those that do not
involve a collinear approximation for scattering of heavy quarks (often
called {}``flavor-creation'', or FC, terms), and those that do (flavor-excitation,
or FE, terms). While FC contributions must be evaluated exactly, the
approximate nature of the FE terms allows some useful simplifications.

At ${\cal O}(\alpha_{s}^{2})$, the flavor-creation contributions
include the coefficients $F_{h,g}^{(1)}$, $F_{h,g}^{(2)}$, and $F_{h,l}^{PS,(2)}$.
The heavy quarks appear in these terms inside the $O(\alpha_{s})$
Feynman subgraph for $\gamma^{*}g\rightarrow h\bar{h}$, connected
by a gluon propagator to an initial-state gluon or a light quark.
These contributions are evaluated with the exact kinematical dependence
on $m_{h}$, and hence are defined unambiguously.

The FE cross sections 
are proportional to the heavy-quark PDF that approximates 
collinear production of heavy-quark pairs from light partons 
in the high-energy limit. Structure functions and coefficient functions
with an initial-state heavy quark, 
such as $F_{h,h}^{(k)}$ and $C_{h,h}^{(k)}$,
fall into this class. The FE coefficient functions 
reduce to unique $\overline{\rm MS}$
expressions when $m_{h}$ is negligible \citet{Collins:1998rz}, but,
near the threshold, they may differ by non-unique powerlike contributions
$\left(m_{h}^{2}/Q^{2}\right)^{p}$ with $p>0$. Within the ACOT scheme,
several conventions have been proposed to include the powerlike contributions
in a way compatible with the QCD factorization theorem.%
\footnote{The differences between these conventions are formally of a higher
order in $\alpha_{s}$, but some conventions lead to faster perturbative
convergence.}

Among these conventions, the {}``full ACOT scheme'' \citet{Aivazis:1993pi}
evaluates the FE coefficient functions ($C_{h,h}^{(k)},$ etc.) 
with their complete mass dependence. 
The simplified ACOT (S-ACOT) scheme \citet{Collins:1998rz,Kramer:2000hn}
neglects \emph{all} mass terms in $C_{h,h}^{(k)}$ and thereupon reduces
tedious computations typical for the full ACOT scheme. The S-ACOT-$\chi$
scheme \citet{Tung:2001mv} adopted in our computation includes \emph{the
most important} mass dependence in $C_{h,h}^{(k)}$ and uses simpler
zero-mass expressions everywhere else. It generalizes the slow-rescaling 
prescription for single heavy quark production in neutrino DIS 
at leading order \cite{Barnett:1976ak} 
to other heavy final states and higher QCD orders. 

If we use uppercase and lowercase letters to denote 
mass-dependent and massless quantities, and 
a caret to indicate renormalized ZM functions, the
S-ACOT-$\chi$ convention for functions with initial-state heavy quarks
can be summarized as\begin{equation}
C_{h,h}^{(k)}\left(\frac{x}{\xi},\frac{Q}{\mu},\frac{m_{h}}{Q}\right)=c_{h,h}^{(k)}\left(\frac{\chi}{\xi},\frac{Q}{\mu},m_{h}=0\right)\,\theta(\chi\leq\xi\leq1),\label{ChhChi}\end{equation}
 and\begin{equation}
F_{h,h}^{(k)}\left(\frac{x}{\xi},\frac{Q}{\mu},\frac{m_{h}}{Q}\right)=\widehat{f}_{h,h}^{(k)}\left(\frac{\chi}{\xi},\frac{Q}{\mu},m_{h}=0\right)\,\theta(\chi\leq\xi\leq1),\label{FhhChi}
\end{equation}
 where\begin{equation}
\chi=x\,\left(1+\frac{(\sum_{fs} m_h)^{2}}{Q^{2}}\right),\label{chi}
\end{equation}
and $\sum_{fs} m_h$ is the net mass of all heavy particles produced in the final state. 
With the exception of a few very rare subprocesses identified below,  
at most one heavy-quark pair is produced in all cases that we consider. Without losing accuracy,
we therefore assume $\chi= x \left(1+\frac{4m_h^2}{Q^{2}}\right)$ throughout the computation.

This form is motivated by an observation that the largest powerlike terms
are associated with the constraint imposed on the convolution 
by energy conservation in production of heavy final states \citet{Nadolsky:2009ge}.
The ZM functions here depend on the variable $\chi$
instead of Bjorken $x$ as an input parameter, and 
the momentum fraction in their convolutions 
is integrated over the range $\chi\leq\xi\leq1$.
The $\chi$ convention is justified in the context 
of the QCD factorization theorem in Sec.~\ref{sec:ChiConvention}.
Its numerical impact is discussed in Sec.~\ref{sec:Threshold-effects}.

When the S-ACOT-$\chi$ scheme is adopted, Eqs.~(\ref{coefFh1})
become
\begin{eqnarray}
 &  & c_{h,a}^{(0)}=\delta_{ha}\delta(1-\widehat{\chi});\,\,\,\, c_{\, h,h}^{(1)}=\widehat{f}_{\, h,h}^{(1)};\label{coefFh21}\\
 &  & C_{h,g}^{(1)}=F_{h,g}^{(1)}-A_{\, hg}^{(1)};\label{coefFh22}\\
 &  & C_{h,g}^{(2)}=\widehat{F}_{h,g}^{(2)}-A_{\, hg}^{(2)}-c_{\, h,h}^{(1)}\otimes A_{\, hg}^{(1)};\label{coefFh23}\\
 &  & c_{\, h,h}^{(2)}=\widehat{f}_{\, h,h}^{(2)};\,\,\,\, C_{h,l}^{(2)}=\widehat{F}_{h,l}^{PS,(2)}-A_{\, hl}^{PS,(2)}.\label{coefFh24}
\end{eqnarray}
Lowercase functions in these equations are
given by ZM expressions. Among all terms, only the structure functions
$F_{h,g}^{(1)},$ $\widehat{F}_{h,g}^{(2)},$ and $\widehat{F}_{h,l}^{PS,(2)}$
are calculated with the exact mass dependence. The carets 
above $\widehat{F}_{h,g}^{(2)}$
and $\widehat{F}_{h,l}^{PS,(2)}$ indicate that the massless $1/\epsilon$ pole
terms, $C_{h,g}^{(1)}\otimes A_{gg}^{(1)}$ and $C_{h,g}^{(1)}\otimes A_{gl}^{(1)}$,
have been subtracted from them. 

In the rest of the terms, the input longitudinal variable is set to
be $\widehat{\chi}=\chi/\xi$. The convolution of such term $f(\widehat{\chi})$
with the PDF $\Phi(\xi)$ is \begin{equation}
\left[f\otimes\Phi\right](\zeta)\equiv\int_{\zeta}^{1}\frac{d\xi}{\xi}\, f\left(\frac{\zeta}{\xi}\right)\,\Phi(\xi),\label{ConvChi}\end{equation}
 where $\zeta=\chi$. {[}The naive massless approximation is $\zeta=x$.{]}

The one-loop expressions $\widehat{f}_{h,h}^{(k)},$ $F_{h,g}^{(1)}$,
and $A_{h,g}^{(1)}$ can be found in \citet{Witten:1975bh,Aivazis:1993pi,Buza:1996wv}.
$\widehat{F}_{h,g}^{(2)}$ and $\widehat{F}_{h,l}^{PS,(2)}$ coincide
with the massive structure functions with initial-state gluons and
pure-singlet light quarks in \citet{Laenen:1992zk,Riemersma:1994hv}. They
are independent of $N_{l}$. The expressions for $A_{hg}^{(2)}$ and
$A_{hl}^{(2)}$ are computed as $A_{Hg}^{(2)}$ and $A_{Hq}^{(2)}$
in \citet{Buza:1996wv}.

The ${\cal O}(\alpha_{s}^{2})$ contribution $c_{h,h}^{(2)}=\widehat{f}_{h,h}^{(2)}$
corresponds to radiation of up to $N_{l}+1$ flavors of $q\bar{q}$
pairs off an incoming quark $h$. It can be found as a sum of the
pure-singlet and non-singlet ZM coefficient functions 
from Refs.~\citet{SanchezGuillen:1990iq,vanNeerven:1991nn,Zijlstra:1991qc,Moch:2004xu,Vermaseren:2005qc}:
\begin{equation}
c_{h,h}^{(2)}=c_{h,h}^{PS,(2)}+c_{h,h}^{NS,(2)}=\widehat{f}_{h,h}^{(2)}=\widehat{f}_{h,h}^{PS,(2)}+\widehat{f}_{h,h}^{NS,(2)}.\label{c2hh}
\end{equation}
In this equation, both pure-singlet and non-singlet parts of $\widehat{f}_{h,h}^{(2)}$
are taken to be massless, which is one of the choices possible within
the S-ACOT scheme. Note that the $c_{h,h}^{(2)}$ functions with three indicated 
final-state $h$ (anti-)quarks and an unobserved fourth $h$ (anti-)quark among the target remnants
could justifiably use a more restrictive rescaling variable, $\chi= x \left(1+\frac{16m_h^2}{Q^{2}}\right)$, as their parameter.
Since their respective contributions are vanishingly small, 
it suffices to evaluate them with the same variable $\chi= x \left(1+\frac{4m_h^2}{Q^{2}}\right)$
as in the rest of the terms to simplify the implementation.

It is equally acceptable to evaluate the pure-singlet
$F_{h,h}^{PS,(2)}$ with a massive $\gamma^* g\rightarrow h\bar h$ subgraph, 
so that the corresponding coefficient function is given by the massive $C_{h,l}^{(2)}$ in Eq.~(\ref{coefFh24}).
In this case, $F_{h,h}^{PS,(2)}$ can be combined with the pure-singlet
contribution with initial-state light quarks, also given by $C_{h,l}^{(2)}$.
The complete ${\cal O}(\alpha_{s}^{2})$ part of $F_{h}(x,Q)$ then
takes a simple form
\begin{equation}
F_{h}^{(2)}=e_{h}^{2}\left\{ c_{h,h}^{NS,(2)}\otimes(\Phi_{h/p}+\Phi_{\bar{h}/p})+C_{h,l}^{(2)}\otimes\Sigma+C_{h,g}^{(2)}\otimes\Phi_{g/p}\right\} ,\label{FheavyFinal}
\end{equation}
 where $\Sigma(x,\mu)$ is the singlet quark PDF summed over $N_{f}=N_{l}+1$
flavors:\begin{equation}
\Sigma(x,\mu)=\sum_{i=1}^{N_{f}}\left[\Phi_{i/p}(x,\mu)+\Phi_{\bar{i}/p}(x,\mu)\right].\label{SingletPDF}\end{equation}
We use Eq.~(\ref{FheavyFinal}) for practical implementation of
$F_{h}$, with the coefficient functions computed as in Eqs.~(\ref{coefFh23})
and (\ref{coefFh24}). 

\begin{figure}
\centering{}\includegraphics[width=0.8\textwidth,height=7cm,keepaspectratio]{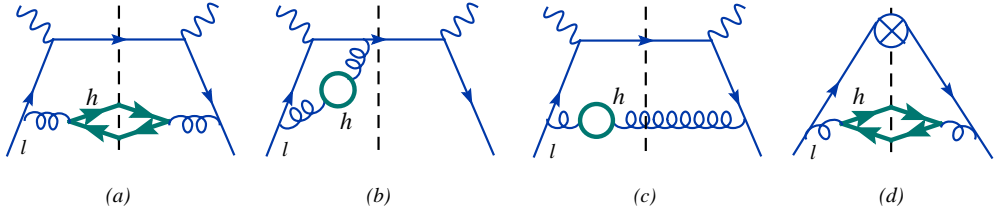}
\caption{Disconnected heavy-quark contributions to $F_{l,l}^{NS,(2)}$ (a,b,c)
and $A_{l,l}^{NS,(2)}$ (d).\label{fig:FNSll} }

\end{figure}

\subsection{Light-quark component of $F(x,Q)$\label{sec:Flight}}

By a similar argument, Eq.~(\ref{coef1}) serves as a starting point
for finding coefficient functions for the light-quark component $F_{l}(x,Q)$.
The corresponding Wilson coefficients are\begin{eqnarray}
 &  & c_{l,a}^{(0)}=\delta_{la}\delta(1-x);\nonumber \\
 &  & C_{l,l}^{(1)}=F_{l,l}^{(1)}-C_{l,l}^{(0)}\otimes A_{ll}^{(1)};\,\,\, C_{l,g}^{(1)}=F_{l,g}^{(1)}-C_{l,l}^{(0)}\otimes A_{lg}^{(1)};\,\,\,\, C_{l,h}^{(1)}=C_{l,l'}^{(1)}=0;\nonumber \\
 &  & C_{l,l}^{(2)}=F_{l,l}^{PS,(2)}+F_{l,l}^{NS,(2)}-C_{l,l}^{(0)}\otimes\left[A_{ll}^{PS,(2)}+A_{ll}^{NS,(2)}\right]-C_{l,l}^{(1)}\otimes A_{ll}^{(1)}-C_{l,g}^{(1)}\otimes A_{gl}^{(1)};\nonumber \\
 &  & C_{l,l'}^{(2)}=F_{l,l'}^{PS,(2)}-C_{l,l}^{(0)}\otimes A_{ll'}^{PS,(2)}-C_{l,g}^{(1)}\otimes A_{gl'}^{(1)},\mbox{ for }l^\prime \neq l;\nonumber \\
 &  & C_{l,h}^{(2)}=F_{l,h}^{PS,(2)}-C_{l,l}^{(0)}\otimes A_{lh}^{PS,(2)}-C_{l,g}^{(1)}\otimes A_{gh}^{(1)};\nonumber \\
 &  & C_{l,g}^{(2)}=F_{l,g}^{(2)}-C_{l,l}^{(0)}\otimes A_{lg}^{(2)}-C_{l,l}^{(1)}\otimes A_{lg}^{(1)}-C_{l,g}^{(1)}\otimes A_{gg}^{(1)}.\label{coefFl1}\end{eqnarray}
The quark-to-quark Wilson coefficients $C_{l,l}^{(2)}$, $C_{l,h},$
and $C_{l,l'}^{(2)}$ are decomposed into their
PS and NS components as in Eq.~(\ref{PSNS}). Non-singlet contributions
$F_{l,l}^{NS,(2)}$ and $A_{ll}^{NS,(2)}$ contain squared matrix elements 
with heavy-quark
lines that are disconnected from the initial-state proton, as in Fig.~\ref{fig:FNSll}.
These diagrams must be evaluated with full dependence on $m_{h}$.
The rest of the coefficients in Eqs.~(\ref{coefFl1}) do not contain
disconnected heavy-quark lines. They are computed according to ZM formulas.

Explicitly, the non-singlet functions with mass dependence consist
of two parts, arising either from Feynman diagrams with light partons
only (designated as $g_{light}$), or with a heavy quark in the final-state
emission or virtual loop (denoted by $G_{heavy}(m_{h})$): \[
G=g_{light}+G_{heavy}(m_{h}),\]
where $G=F_{l,l}^{NS,(2)}$ or $A_{l,l}^{NS,(2)}$. The function
$G_{heavy}(m_{h})$, provided by the graphs of the type 
shown in Fig.~\ref{fig:FNSll}, retains complete $m_h$ dependence.
The function $g_{light}$, obtained from the same graphs 
as in Fig.~\ref{fig:FNSll}, 
but with the heavy quark $h$
replaced by one of the light quarks ($u, d, ...$),
is evaluated in the ZM approximation. 

Masses can be neglected in the rest
of Eqs.~(\ref{coefFl1}), so we get\begin{eqnarray}
 &  & c_{l,a}^{(0)}=\delta_{la}\delta(1-x);\\
 &  & c_{l,l}^{(1)}=\widehat{f}_{l,l}^{(1)};\,\,\, c_{l,g}^{(1)}=\widehat{f}_{l,g}^{(1)};\,\,\, c_{l,h}^{(1)}=c_{l,l'}^{(1)}=0;\\
 &  & C_{l,l}^{(2)}=C_{l,l}^{NS,(2)}+c^{PS,(2)},\mbox{ where }\label{C2ll}\\
 &  & C_{l,l}^{(2),NS}=\widehat{f}_{l,l,light}^{NS,(2)}+F_{l,l,heavy}^{NS,(2)}-A_{ll,heavy}^{NS,(2)};\label{C2NS}\\
 &  & c_{l,h}^{(2)}=c_{l,l'}^{(2)}=c^{PS,(2)};\,\,\,\, c_{l,g}^{(2)}=\widehat{f}_{l,g}^{(2)}.\label{coefFl2}\end{eqnarray}
 The one-loop coefficients $c_{l,a}^{(1)}$ have been known for a long time \citet{Bardeen:1978yd,Altarelli:1978id,Humpert:1980uv}.
The two-loop massless contributions in Eqs.~(\ref{C2NS}) and (\ref{coefFl2})
can be derived from the published ZM results according to the following
procedure. Using the decomposition Eq.~(\ref{PSNS}) for $c_{i,j}$
in the ZM scheme,\begin{equation}
c_{i,j}\equiv c^{PS}+\delta_{ij}c^{NS,(2)},\label{PSNS2}\end{equation}
 where $c^{PS}$ and $c^{NS}$ are independent of the quark flavors
$i$ or $j$ given that the masses are neglected, we write
\begin{eqnarray}
 & F(x,Q)=\sum_{i,a}e_{i}^{2}\left[c_{i,a}\otimes\Phi_{a/p}\right]=\sum_{i}e_{i}^{2}\left\{ \sum_{j}\left(c^{PS}+\delta_{ij}c^{NS}\right)\otimes\Phi_{j/p}+c_{g}\otimes\Phi_{g/p}\right\} \nonumber \\
 & =\left[c^{NS}\otimes\Sigma^{+,NS}\right]+\frac{\left(\sum_{i}e_{i}^{2}\right)}{N_{f}}\left\{ \left[c^{S}\otimes\Sigma\right]+N_{f}\,\left[c_{g}\otimes\Phi_{g/p}\right]\right\},\label{FPSNS}\end{eqnarray}
with
\[
c^{S}\equiv c^{NS}+N_{f}c^{PS}.\]
The singlet PDF $\Sigma(x,\mu)$ is given by Eq.~(\ref{SingletPDF}),
and the non-singlet sum of (anti-)quark PDFs is\[
\Sigma^{+,NS}(x,\mu)=\sum_{i=1}^{N_{f}}e_{i}^{2}\left(\Phi_{i/p}(x,\mu)+\Phi_{\bar{i}/p}(x,\mu)-\frac{1}{N_{f}}\Sigma(x,\mu)\right).\]

Eq.~(\ref{FPSNS}) expresses $F(x,Q)$ in the same representation
as Eq.~(4.1) in the N$^{3}$LO calculation of DIS cross sections
\citet{Vermaseren:2005qc}. Comparing Eqs.~(\ref{C2NS}) and (\ref{coefFl2})
with ZM coefficient functions in Section 4 of that reference (which
are indicated here by an asterisk $"*"$), we find that\begin{equation}
c^{PS,(2)}=c_{I,ps}^{(2,*)}/N_{f},\label{cPS2}\end{equation}
\begin{equation}
\widehat{f}_{i,g}^{(2)}=c_{I,g}^{(2,*)}/N_{f},\end{equation}
 and \begin{equation}
\widehat{f}_{l,l,light}^{NS,(2)}=c_{I,ns}^{(2,*)}(n_{f}=N_{l}),\end{equation}
 with $I=2$ or $L$ for $F_{2}(x,Q)$ and $F_{L}(x,Q),$ respectively. 

The non-singlet heavy-quark coefficient function,\begin{equation}
F_{l,l,heavy}^{NS,(2)}(x,Q^{2}/m_{h}^{2})=
\left(L_{I,q}^{NS,(2)}(x,Q^{2}/m_{h}^{2})\right)_{+}+\frac{2}{3}\ln\left(\frac{Q^{2}}{m_{h}^{2}}\right)\, c_{l,l}^{(1)}(x),\label{FNS2llheavy}\end{equation}
is composed of contributions of several classes shown in Figs.~\ref{fig:FNSll}(a)-(c).
Diagrams with real emission of a heavy-quark pair (as in Fig.~\ref{fig:FNSll}(a))
in $F_{I}(x,Q)$ contribute a function $L_{I,q}^{NS,(2)}(x,Q^{2}/m_{h}^{2})$
in Eqs.~(A.1) and (A.2) of Ref.~\citet{Buza:1995ie}. This contribution
is combined with the virtual two-loop diagrams, cf. Fig.~\ref{fig:FNSll}(b),
to produce the first term on the right-hand side of Eq.~(\ref{FNS2llheavy}),
in which $L_{I,q}^{NS,(2)}(x,Q^{2}/m_{h}^{2})$ is regularized by
the plus prescription at $x\rightarrow1$. Contributions with a heavy-quark
polarization graph inserted into a one-loop $\gamma^{*}q$ scattering
diagram, of the kind shown in Fig.~\ref{fig:FNSll}(c), produce the
second term in Eq.~(\ref{FNS2llheavy}), where $c_{l,l}^{(1)}$ is
available from Refs.~\citet{Bardeen:1978yd,Altarelli:1978id,Humpert:1980uv}.

In this derivation, we do not explicitly compute the virtual loop
contribution in Fig.~\ref{fig:FNSll}(b), but deduce it from the
Adler sum rule \citet{Adler:1965ty,Altarelli:1981ax,Dokshitzer:1995qm,Brock:1993sz}.
The sum rule states that the sum of the real and virtual 
contributions to $F_{l,l,heavy}^{NS,(2)}(x,Q^{2}/m_{h}^{2})$
satisfies\begin{equation}
\int_{0}^{1}F_{l,l,heavy}^{NS,(2)}(x,Q^{2}/m_{h}^{2})\, dx=0.\label{Adler}\end{equation}
 With this rule, it can be demonstrated that the virtual contribution
amounts to imposing the plus prescription on $L_{I,q}^{NS,(2)}(x,Q^{2}/m_{h}^{2})$
as in Eq.~(\ref{FNS2llheavy}).

In the asymptotic limit $Q^{2}\gg m_{h}^{2},$ $F_{l,l,heavy}^{NS,(2)}$
for the inclusive $F_{2}$ contains large terms proportional to $\ln(Q^{2}/m_{h}^{2})$.
Those coincide with the ${\cal O}(\alpha_{s}^{2})$ non-singlet part
$A_{l,l,heavy}^{NS,(2)}$ of the light-quark PDF that arises from radiation
of a heavy-quark pair as shown in Fig.~\ref{fig:FNSll}(d). 
$A_{l,l,heavy}^{NS,(2)}$
is computed as $A_{qq,H}^{NS,(2)}(z,m_{h}^{2}/\mu^{2})$ in Eq.~(B.4)
of Ref.~\citet{Buza:1996wv}, which we evaluate as a function 
of $z=\chi/\xi$ in accord with the S-ACOT-$\chi$ scheme. 

When $A_{l,l,heavy}^{NS,(2)}$ is subtracted from $F_{l,l,heavy}^{NS,(2)}$
as in Eq.~(\ref{C2NS}), the difference is free of the collinear
logs. After the difference is combined with the light-quark-only contributions
$\widehat{f}_{l,l,light}^{NS,(2)}$, we obtain the full non-singlet
coefficient function $C_{l,l}^{(2),NS},$ which coincides in the limit
$m_{h}^{2}/Q^{2}\rightarrow0$ with its zero-mass $\overline{\rm MS}$
expression in Eq.~(8) of Ref.~\citet{vanNeerven:1991nn}. For the
longitudinal function $F_{L}$, the heavy-quark subtraction $A_{l,l,heavy}^{NS,(2)}$
is zero. Putting everything together, we obtain the final expression
for the NNLO light-quark component,\begin{equation}
F_{l}^{(2)}=e_{l}^{2}\left\{ C_{l,l}^{NS,(2)}\otimes(\Phi_{l/p}+\Phi_{\bar{l}/p})+c^{PS,(2)}\otimes\Sigma+c_{l,g}^{(2)}\otimes\Phi_{g/p}\right\} ,\label{FlightFinal}\end{equation}
where the Wilson coefficients are listed in Eqs.~(\ref{C2ll})-(\ref{coefFl2})
and (\ref{cPS2})-(\ref{FNS2llheavy}).

\subsection{Several heavy flavors \label{sub:Several-heavy-flavors}}
\begin{figure}
\centering{}\includegraphics[width=0.8\textwidth,height=7cm,keepaspectratio]{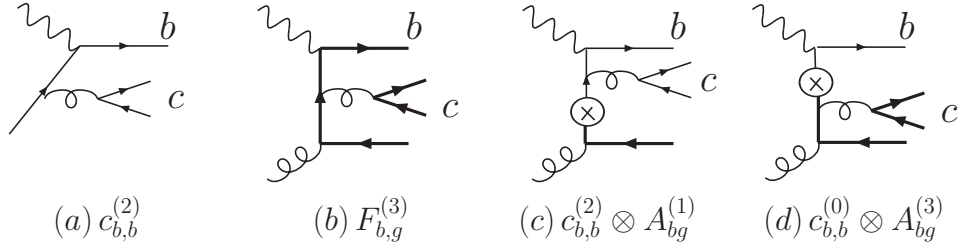}
\caption{Examples of $\alpha_s^2$ and $\alpha_s^3$ contributions 
with heavy-quark lines of different flavors.\label{fig:c2bb} }

\end{figure}

The structure functions $F_{l}$ and $F_{h}$ in Eqs.~(\ref{FlightFinal})
and (\ref{FheavyFinal}) are all that is needed to compute inclusive
$F(x,Q).$ Our expressions can be readily extended to include two or
more heavy-quark flavors:\begin{equation}
F=\sum_{l=1}^{N_{l}}F_{l}+\sum_{h=N_{l+1}}^{N_{f}^{fs}}F_{h},\label{FSeveralHeavyFlavors}\end{equation}
 where the sum runs over all quark flavors satisfying $4m_{h}^{2}\leq W^{2}$
(\emph{i.e.}, up to the number $N_{f}^{fs}$ of the kinematically allowed
final-state flavors). 

Beginning at ${\cal O}(\alpha_s^2)$, some contributions 
to $F_h$ include heavy quarks of two different flavors, say, 
$c$ and $b$. For instance, 
Fig.~\ref{fig:c2bb}(a) shows a two-loop diagram in $F_b$ 
in which an incoming bottom quark 
radiates a $c\bar c$ pair before or after the scattering on the photon.
This flavor-excitation contribution, relevant at $Q$ large enough, 
is evaluated 
by a massless expression, $c^{(2)}_{b,b}(m_c=m_b=0)\otimes \Phi_{b/p}(\chi,\mu)$, 
in accord with the main rule of 
the S-ACOT scheme. We take the rescaling variable to be 
$\chi=x\left(1+4m_b^2/Q^2\right)$, given the numerical smallness of the cross section, even though 
a more restrictive choice $\chi=x\left(1+(2\,m_b+2\,m_c)^2/Q^2)\right)$ conforms better with the exact
momentum conservation in production of a $b\bar b$ + $c\bar c$ pair.

The above contribution resums the large-$Q$ logarithmic behavior of a three-loop function 
$F^{(3)}_{b,g}$ for the process $\gamma^* g \rightarrow b\bar b c\bar c$. A representative Feynman
diagram is shown in Fig.~\ref{fig:c2bb}(b). The rest of the diagrams in the class are
related to the shown diagram by re-attaching the $g\rightarrow c\bar c$ branch
to one of the external legs ($g$, $b$, or $\bar b$). 
 
In the $Q^2\rightarrow \infty$ limit,
$F^{(3)}_{b,g}$ {\it simultaneously} contains logarithms $\ln(Q^2/m_c^2)$ and $\ln(Q^2/m_b^2)$ 
that must be subtracted in order to obtain an infrared-safe coefficient function  $C^{(3)}_{b,g}$.
The subtraction is realized by applying the perturbative expansion 
procedure discussed in Section~\ref{sec:Fheavy} to the three-loop level. We get
\begin{equation}
C_{b,g}^{(3)}=F_{b,g}^{(3)}-c_{b,b}^{(2)}\otimes A_{bg}^{(1)}-c_{b,b}^{(0)}\otimes A_{bg}^{(3)},\label{C3bg} 
\end{equation} 
where the last two terms on the right-hand side are the subtractions associated with 
the diagrams of the type shown in Figs.~\ref{fig:c2bb}(c) and (d). 
The functions $F_{b,g}^{(3)}$, $A_{bg}^{(1)}$, and $A_{bg}^{(3)}$ 
are evaluated with full dependence on $m_c$ and $m_b$. 
[The coefficient $A_{bg}^{(3)}$ has been recently computed in Ref.~\cite{Ablinger:2012qj}.] 
The coefficient functions $c_{b,b}^{(0)}$ and $c_{b,b}^{(2)}$ are massless. 

Based on the general structure of the S-ACOT scheme, we expect 
the first and third term in Fig.~\ref{fig:c2bb} to cancel when 
$Q^2 \approx m_b^2$, and the fourth term to cancel the ${\cal O}(\alpha_s^3)$ contribution 
to the $c^{(0)}_{b,b}\otimes \Phi_{b/p}$ term in the same limit. 
The third and fourth terms 
cancel the $\ln(Q^2/m_c^2)$ and $\ln(Q^2/m_b^2)$ contributions to the second term, 
$F^{(3)}_{b,g}$, in the limit $Q^2\gg m_b^2$. The numerical realization of these cancellations 
at three loops is yet to be demonstrated in the future, 
pending on the calculation 
of the unknown massive three-loop coefficients. The factorization theorem 
is indicative of the structure of the S-ACOT-$\chi$ subtraction terms 
that will arise at that order.

\subsection{Semi-inclusive heavy quark production \label{sub:FhSI}}

A clarification is needed that the heavy-quark component $F_{h}$
of inclusive $F(x,Q)$ (defined as the part proportional to the heavy-quark
electric charge $e_{h}^{2}$) is not directly measurable. Rather,
experiments publish the semi-inclusive (SI) heavy-quark structure
function $F_{h,SI}(x,Q)$ that is determined from the cross section
with at least one registered heavy meson. In the case of $F_{2}$
at NNLO, $F_{h,SI}(x,Q)$ with $h=c$ essentially coincides with the
charm structure function $F_{2}^{(c)}$ that is commonly measured
by HERA experiments.

While $F_{h,SI}$ must be defined with care to obtain infrared-safe
results at all $Q$ \citet{Chuvakin:1999nx}, for a global fit it
is sufficient to approximate $F_{h,SI}$ 
in the following way \citet{Forte:2010ta}.
At moderate $Q$ values accessible at HERA, we define it as\begin{equation}
F_{h,SI}(x,Q)=F_{h}(x,Q)+\sum_{l=1}^{N_{l}}e_{l}^{2}L_{I,q}^{NS,(2)}\otimes(\Phi_{l/p}+\Phi_{\bar{l}/p}).\label{FhSI}\end{equation}
Here $F_{h}(x,Q)$ is the component with the heavy quark struck by
the photon, cf. Eqs.~(\ref{Fheavy}) and (\ref{FheavyFinal}). $L_{I,q}^{NS,(2)}$,
given by Eqs.~(A.1) and (A.2) in Ref.~\citet{Buza:1995ie},
is the non-singlet part of the light-quark component $F_{l}(x,Q)$
that contains radiation of a $h\bar{h}$ pair in the final state,
as shown in Fig.~\ref{fig:FNSll}(a). This is the same function that
was discussed below Eq.~(\ref{FNS2llheavy}). 
However, since the virtual diagram 
in Fig.~\ref{fig:FNSll}(b) does not contribute to $F_{h,SI}$,
the plus prescription is not imposed on $L_{I,q}^{NS,(2)}$ in this case.

The $F_{h,SI}(x,Q)$ function that is thus defined is numerically
stable in comparisons to the existing data \citet{Forte:2010ta}. 
Our numerical analysis shows that the contribution associated with 
$L_{I,q}^{NS,(2)}$ provides between 0 and 3\% of the semi-inclusive 
charm cross sections at $Q<10$ GeV, 
which is insignificant compared to typical experimental errors.

\subsection{Factorization and $\chi$ convention \label{sec:ChiConvention}}

In the remainder of this section, we show that the S-ACOT-$\chi$
scheme is fully compatible with the QCD factorization theorem for
DIS. 

To see why the $\chi$ convention is needed, consider again the heavy-quark
contribution to $F(x,Q)$ of the proton from Eq.~(\ref{Fheavy}),
\begin{equation}
F_{h}(x,Q)=e_{h}^{2}\sum_{l=1}^{N_{l}}\int_{\chi}^{1}\frac{d\xi}{\xi}C_{h,l}\left(\xi p,m_{h},\frac{Q}{\mu_{0}}\right)\Phi_{l/p}(\xi,\mu_{0}).\label{FheavyMu0}\end{equation}
This expression is for the same $Q$ value as in Eq.~(\ref{Fheavy}), 
but the factorization scale $\mu_{0}\approx1$ GeV 
is taken to be below the switching-point scale for $N_l+1$ flavors, 
so that only PDFs
for light parton flavors $(l=0,...N_{l}$) are present. For this scale
choice, the PDFs $\Phi_{l/p}(\xi,\mu_{0})$ do not include subgraphs
with the heavy-quark lines: those are contained solely in the Wilson
coefficient functions $C_{h,l}\left(\xi p,m_{h},\frac{Q}{\mu_{0}}\right)$. 
The right-hand side is non-zero when the light parton $l$ 
carries enough energy to produce at least 
one $h\bar{h}$ pair in the final state. 
This condition is reflected in the 
integration limits $\chi \leq \xi \leq 1$ that are imposed on the
convolution by the energy conservation constraints inside the
coefficient functions $C_{h,l}$. 

If $\mu$ is gradually increased above $\mu_0$, 
a coefficient function $C_{h,h}$ with an initial-state 
heavy quark is introduced when $\mu$ crosses 
the switching point from $N_l$ to $N_l+1$
active flavors. This function does not automatically
vanish outside of the physical range $\chi\leq\xi\leq 1$. 
If $C_{h,h}$ is defined so as to contribute in a wider range 
$\xi_{min} \leq \xi \leq 1$ at the switching point, 
with $x\leq \xi_{min} < \chi$, then the DGLAP evolution preserves 
the same wider range at all $\mu$ above the switching point.

If $\widehat{W}$ is the center-of-mass energy of the photon scattering
on a \emph{light} parton $a$,\begin{equation}
\widehat{W}^{2}\equiv(p_{a}+q)^{2}=Q^{2}\left(\xi/x-1\right),\label{What2}\end{equation}
a final state with several heavy particles of the net mass $\sum m_h$ is produced when 
\begin{equation}
\left(\sum m_{h}\right)^2\leq\widehat{W}^{2}\leq W^{2}=Q^{2}(1/x-1).\label{MomConservationWhat}
\end{equation}
According to this condition, the momentum fraction $\xi$ must be in the range
\begin{equation}
\chi\leq\xi\leq1,\label{MomConservationXi}
\end{equation}
for production to occur, where $\chi=x\, \bigl(1+(\sum m_{h})^{2}/Q^{2}\bigr)\geq x$.

If collinear approximations for flavor-excitation (FE) and subtraction
terms in $C_{h,a}$ violate this fundamental requirement, 
large spurious contributions from the unphysical kinematical
region cancel to each order of $\alpha_{s}$, but survive as higher-order
logarithmic terms. They can be eliminated by a supplemental condition
that the correct integration limits are always to be preserved, as in
Eq.~(\ref{MomConservationXi}).%
\footnote{Even in the $Q^{2}\gg m_{h}^{2}$ limit, convolutions with FE terms
and subtractions could be in principle extended to include contributions
from $0\leq\xi\leq x$. This would not violate QCD factorization order
by order, but will destabilize higher-order terms. This is avoided by an
implicit assumption that the FE convolutions in the ZM limit 
are restricted to the physical range $x\leq\xi\leq1$. %
} 

We will now show how to apply this condition at any order by including
it into the QCD factorization theorem. For this purpose, we examine the
projection operator $Z$ that encapsulates the main rules of each factorization
scheme \citet{Collins:1998rz}. It applies a set of approximations
to the Feynman graphs with leading momentum regions in order to enable
all-order factorization. 

We will closely follow the derivation and notations in 
Ref.~\citet{Collins:1998rz}. In this approach, 
the Feynman graphs containing the leading DIS contributions are composed
of two-particle irreducible subgraphs $H$ and $T$, joined by one
parton line on each side of the unitarity cut. Each leading graph
$H\cdot T$ involves integration over the momentum $k^{\mu}$ of the
intermediate parton and summation over its spin components, \begin{equation}
H\cdot T\equiv\sum_{a=g,u,\bar{u},d,\bar{d},...}\int\frac{d^{4}k}{(2\pi)^{4}}\sum_{spins}H_{a}(q,k)T_{a}(k,p).\end{equation}
Virtualities of all momenta are of order $Q$ in the hard subgraph
$H_{a}(q,k)$, and they are much smaller than $Q$ in the target subgraph
$T_{a}(k,p)$. $H$ eventually contributes to the Wilson coefficient functions, 
and $T$ to the PDFs. $q^{\mu}$ and $p^{\mu}$ are the photon's and proton's
4-momenta. The nearly massless proton moves in the $+z$ direction
in the Breit reference frame. 

The purpose of the $Z$ operator is to approximate the leading-power
(logarithmically divergent) part of $H\cdot T$ by a simpler expression,
denoted by $H\cdot Z\cdot T$, and to recast $H\cdot T$ as \begin{align}
H\cdot T & =\sum_{a}\int\frac{d^{4}k}{(2\pi)^{4}}\int\frac{d^{4}l}{(2\pi)^{4}}\sum_{spins}H_{a}(q,l)\, Z_{a}(l,k;\widehat{l})\, T_{a}(k,p)+\mbox{non-leading power term. }\label{HZT1}\end{align}
The leading-power approximation $H\cdot Z\cdot T$ provides the bulk
of $H\cdot T$. The non-leading power part is suppressed by terms
of order \[
\left(\frac{\mbox{highest virtuality in }T}{\mbox{lowest virtuality in }H}\right)^{r},\mbox{ with }r>0.\]
When it is recursively applied to all leading subgraphs, the $Z$
projection generates the factorized expression for the structure function,
$F=\sum_{a}\left[C_{a}\otimes f_{a}\right]+{\cal O}(\Lambda_{QCD}/Q)$.
\citet{Collins:1998rz}

The $Z$ operation simplifies integration over the momentum and summation
over the spin components of the intermediate parton, and it also simplifies the
hard graph $H$. The momentum $l^{\mu}$ of the parton $a$ that enters
$H$ is replaced by a simpler momentum $\widehat{l}^{\mu}$, \emph{e.g.},
$\widehat{l}^{\mu}=\xi p^{\mu}$ if $a$ is massless (where
$0\leq\xi\leq 1$). The $Z$ operation discards power-suppressed terms 
in $H$, such as the masses of the light quarks that are always
negligible compared to $Q$. It also specifies when the heavy-quark
mass terms are to be retained in $H$, depending on the type of
the factorization scheme 
and the partonic scattering subprocess.

In all variants of the ACOT scheme, 
the $Z$ operator is the same in all partonic channels 
except for the $H$ subgraphs with an incoming heavy-quark
line. The target parts $T_{a}$, operators $Z_{l}$ and
$Z_{g}$ for the $H$ subgraphs with initial light-quark and gluon
lines are identical in all variants, while the operator $Z_h$ for the  
$H$ subgraphs with incoming heavy quarks is not.
The PDFs in $T_{a}$ are defined
by the operator matrix elements as in Eq.~(\ref{Phiab}) and retain dependence
on quark masses of all contributing flavors. 

The $Z_{h}$ operator is of the form \[
Z_{h}(l,k;\widehat{l})=\frac{1}{4}\,(2\pi)^{4}\, S_{H}(\widehat{l})\, S_{T}\,\delta(l^{+}-\widehat{l}^{+})\delta(l^{-}-\widehat{l}^{-})\delta^{2}(\vec{l}_{T}),\]
where $S_{H}(\widehat{l})$ and $S_{T}=\gamma^{+}$ are projectors
on the leading spin components in $H$ and $T$, respectively. The
incoming heavy quark in $H_{h}$ has an approximate momentum $\widehat{l}^{\mu},$
where the light-cone components of $\widehat{l}^{\mu}$ in the Breit
frame are $\widehat{l}^{\pm}\equiv\left(\widehat{l}^{0}\pm\widehat{l}^{3}\right)/\sqrt{2}$
and $\vec{l}_{T}=0$. With this representation, the $H_{h}\cdot Z_{h}\cdot T_{h}$
integral assumes the form of a convolution over $\xi$, 
\begin{align}
H_{h}\cdot Z_{h}\cdot T_{h} & =\int\frac{d\xi}{\xi}\mbox{tr}\left[H_{h}(q,\widehat{l})\frac{S_{H}(\widehat{l})}{2}\right]\,\int\frac{dk^{-}d^{2}\vec{k_{T}}}{(2\pi)^{4}}\mbox{tr}\left[\frac{\gamma^{+}}{2}T_{h}(k,p)\right],\label{HZT2}\end{align}
where $\xi=\widehat{l}^{+}/p^{+}$ is the ratio of the large {}``+''
momentum components. %
\begin{table}
\begin{centering}
\begin{tabular}{|c|c|c|c|c|}
\hline 
\noalign{\vskip3pt}
Scheme & $\widehat{l}^{\mu}$ in $Z_{h}$ & $S_{H}$ & $m_{h}$ in $H_{h}(q,\widehat{l})$ & $\xi$ range in $H_{h}\cdot T_{h}$\tabularnewline[3pt]
\hline
\hline 
ACOT & $\left(\xi p^{+},\frac{m_{h}^{2}}{2\xi p^{+}},\vec{0}_{T}\right)$ & $\frac{\widehat{l}\cdot\gamma+m_{h}}{\xi p^{+}}$ & $m_{h}\neq0$ & $\frac{x}{2}\,\left(1+\sqrt{1+\frac{4m_{h}^{2}}{Q^{2}}}\right)\leq\xi\leq1$\tabularnewline[3pt]
\hline 
\noalign{\vskip3pt}
S-ACOT & $\left(\xi p^{+},0,\vec{0}_{T}\right)$ & $\gamma^{-}$ & $m_{h}=0$ & $x\leq\xi\leq1$\tabularnewline[3pt]
\hline 
\noalign{\vskip3pt}
S-ACOT-$\chi$ & $\left(\xi\frac{p^{+}}{1+4m^{2}/Q^{2}},0,\vec{0}_{T}\right)$ & $\gamma^{-}$ & $m_{h}=0$ & $x\,\left(1+4m_{h}^{2}/Q^{2}\right)\leq\xi\leq1$\tabularnewline[3pt]
\hline
\end{tabular}
\par\end{centering}

\caption{Components of the projection operator $Z_{h}(l,k;\widehat{l})$ in
three versions of the ACOT scheme. \label{tab:Zh}}

\end{table}

Table~\ref{tab:Zh} collects expressions for $\widehat{l}^{\mu}$
and $S_{H}(\widehat{l})$ in the ACOT, S-ACOT, and S-ACOT-$\chi$
schemes. It also lists the integration ranges in the $H_{h}\cdot T_{h}$
convolutions and indicates if $m_{h}$ is set to zero in the $H_{h}$
subgraphs. We re-emphasize that the three schemes listed in the table 
are distinguished only by the hard subgraphs, or Wilson coefficients, 
with initial-state heavy quarks, 
\emph{i.e.}, in the flavor-excitation channel. The differences arise 
solely in the terms proportional 
to powers of $m_{h}^{2}/Q^{2}$  in $H_{h}(q,l)$.
One could simplify the FE hard-scattering contributions by
setting $m_{h}=0$ as in the S-ACOT scheme. In the full ACOT scheme,
the lower limit of integration in $H_{h}\cdot Z_{h}\cdot T_{h}$ is
set by the kinematics of scattering of a massive quark into a massive
quark, which violates the momentum conservation condition of Eq.~(\ref{FheavyMu0})
for pair production of massive quarks from light-quark scattering.
In the S-ACOT scheme, one is tempted to set the integration range
to $x\le\xi\leq1,$ which is also incompatible with momentum conservation.
One cannot just restrict the integration range to $\chi\leq\xi\leq1,$
as this disallows the lowest-order FE contribution $c_{h,h}^{(0)}\otimes h$
that contributes at $\xi=x$.%
\begin{figure}[p]
\begin{centering}
\includegraphics[width=1\textwidth]{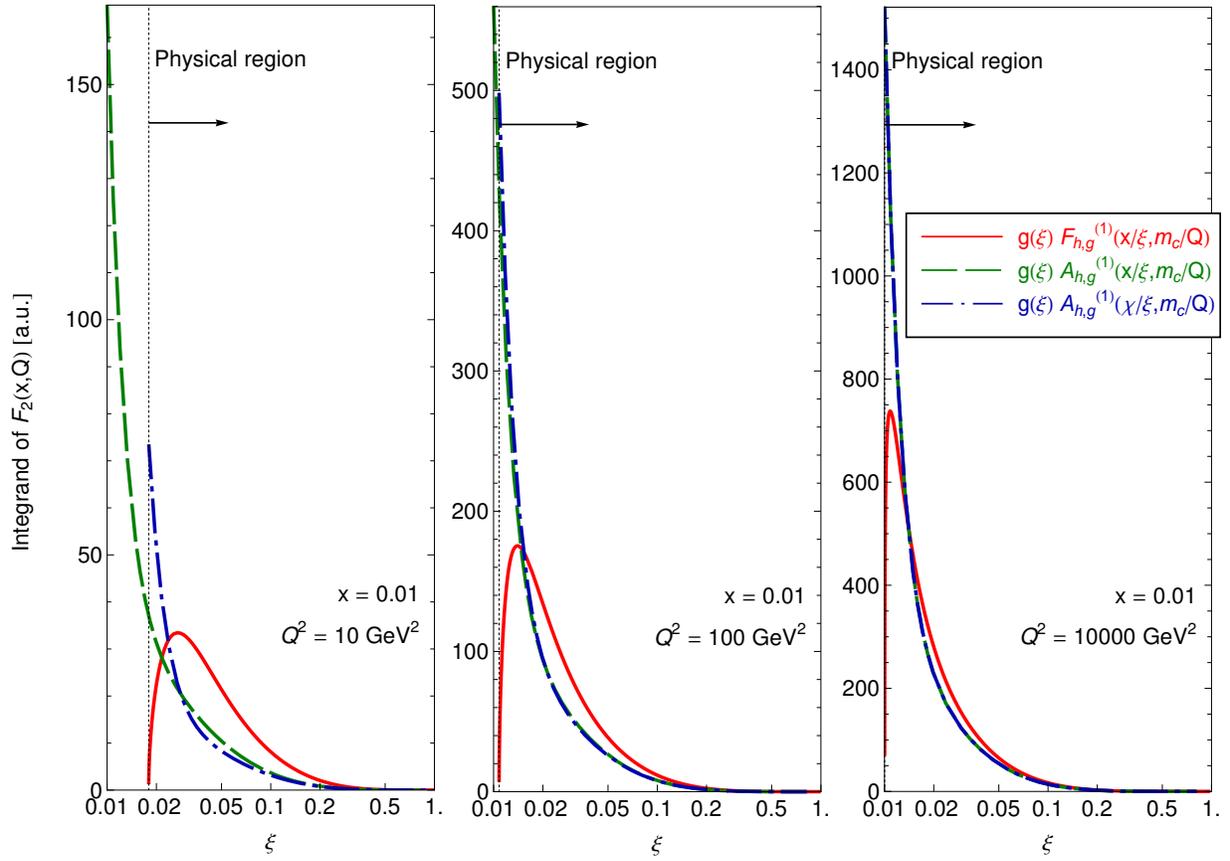}
\par\end{centering}

\caption{Integrands of convolution integrals $F_{h,g}^{(1)}\otimes g$ and
$A_{h,g}^{(1)}\otimes g$ with and without the $\chi$ prescription,
plotted as a function of the momentum fraction $\xi$. \label{fig:Integrands}}

\end{figure}

A better way is provided by the rescaling $\xi\rightarrow\widetilde{\xi}=\kappa\xi,$
$p^{+}\rightarrow\widetilde{p}^{+}=p^{+}/\kappa$, which leaves $H(q,\widehat{l})$
invariant, as it does not change $\widehat{l}^{\mu}$, $\widehat{x}=Q^{2}/(2\widehat{l}\cdot q)$,
or other kinematical variables in $H(q,\widehat{l})$. At the same
time, rescaling changes the integration range for $\xi$ in the convolution.

Choosing $\kappa=1+(\sum m_{h})^{2}/Q^{2},$ we obtain the S-ACOT-$\chi$
scheme that has all desirable features:
\begin{enumerate}
\item The proof of QCD factorization for the S-ACOT scheme in \citet{Collins:1998rz}
also applies to the S-ACOT-$\chi$ scheme, since the S-ACOT and S-ACOT-$\chi$
schemes have the same $H(q,\widehat{l})$. The $Z$ operation of the 
S-ACOT-$\chi$ scheme upholds all expected properties that are listed 
in Sec.~9C of Ref.~\citet{Collins:1998rz}.
\item The integration over $\xi$ proceeds over the physical range $\chi\leq\xi\leq1$
in all channels. It includes all physically possible scattering channels for $\xi \geq \chi$,
but excludes kinematically prohibited values for $\xi < \chi$. Since the form of $\chi$ is
associated with a specific coefficient function, the same form is to be used in convolutions of
this coefficient function in subtraction terms at higher orders.
\item The S-ACOT-$\chi$ coefficient functions $H_h$ in the flavor-excitation
channels are given by ZM expressions evaluated at $\widehat{x}=\chi/\xi$.
Kinematical prefactors outside of the coefficient functions are independent of $\xi$ and 
not affected by the rescaling.
\item The target subgraphs $T_a$, 
corresponding to the PDFs, are given by universal
operator matrix elements that are the same in all ACOT-like schemes.  
\item The same value $N_f$ is used in the evolution of $\alpha_s(\mu)$, PDFs, and hard graphs
in each $Q$ range.   
\item When $Q$ is much larger than $m_{h},$ the coefficient functions of the S-ACOT-$\chi$ scheme
reduce to those of the zero-mass $\overline{\rm MS}$ scheme, without additional
finite renormalizations.
\item When $Q$ is of order $m_{h}$, the S-ACOT-$\chi$ scheme is generally closer
to the FFN scheme than the S-ACOT scheme as a consequence 
of its $Z$ operation that satisfies energy
conservation. In this scheme, matching onto the FFN scheme 
does not rely on propositions beyond the factorization
theorem with energy conservation, 
such as conditions for derivatives of $F(x,Q)$ \citet{Thorne:1997ga}
or damping factors \citet{Forte:2010ta}. 
\end{enumerate}

\subsection*{An illustration for the $\chi$ convention}

The advantages of $\chi$ rescaling can be demonstrated on the example
of the ${\cal O}(\alpha_{s})$ $\gamma^{*}g$ contribution, consisting
of the gluon-initiated box graph and corresponding subtraction, and
shown by the second and third graphs on the upper row of Fig.~\ref{fig:FhDiagrams}:
\begin{equation}
[C_{h,g}^{(1)}\otimes g](x,Q)=\int_{\chi}^{1}\frac{d\xi}{\xi}g(\xi,Q)F_{h,g}^{(1)}\left(\frac{\chi}{\xi}\right)-\int_{\zeta}^{1}\frac{d\xi}{\xi}g(\xi,Q)A_{h,g}^{(1)}\left(\frac{\zeta}{\xi}\right).\label{C1hg2}\end{equation}
The integrands of the convolution integrals on the right-hand side,
$g(\xi,Q)\: F_{h,g}^{(1)}(\chi/\xi)$ and $g(\xi,Q)\: A_{h,g}^{(1)}(\zeta/\xi)$,
where $\zeta=x$ or $\chi,$ are plotted in Fig.~\ref{fig:Integrands}(a-c)
as a function of $\xi$. The computation follows the numerical setup
described in the next section. The scale on the $\xi$ axis is logarithmic:
the convolution integrals are proportional to the areas under the respective
integrand curves. For definiteness, we choose $x=0.01$ and $Q^{2}=$10,
100, and 10000 $\mbox{GeV}^{2}$, but the same features are observed
for other $x$ and $Q$ values.

In the charm creation contribution, $g\otimes F_{hg}^{(1)}$, the
integrand (red solid line) vanishes outside of the physical range
$\chi\leq\xi\leq1$, where $\chi\approx0.018,$ $0.0108,$ and $0.010008$
for $Q^{2}=$10, 100, and 10000 $\mbox{GeV}^{2}$. On the other hand,
the naive choice $\zeta=x$ of the S-ACOT scheme allows the 
integrand $g(\xi,Q)\: A_{h,g}^{(1)}(\zeta/\xi)$ (green dashed curve) 
in the second term on the right-hand side of Eq.~(\ref{C1hg2})
to contribute in the unphysical region
$x\leq\zeta\leq\chi$. Its spurious contribution is comparatively
large at the smallest $Q$. It is not fully canceled by the counterpart
FE term $\left[c_{h,h}^{(0)}\otimes c\right](x,Q)$ in the first upper
graph of Fig.~\ref{fig:FhDiagrams}, leading to a bloated higher-order
uncertainty.

The S-ACOT-$\chi$ integrand $g(\xi,Q)\: A_{h,g}^{(1)}(\chi/\xi)$
vanishes below $\xi=\chi$ (cf.~the blue dash-dotted line). It is
numerically moderate at physical $\xi$ values, $\xi>\chi$, and its
integral cancels well with $\left[c_{h,h}^{(0)}\otimes c\right](\chi,Q)$,
as will be further demonstrated in Sec.~\ref{sec:CancellationsLowQ}.
Note also that the difference between the two definitions for the
$g(\xi,Q)\: A_{h,g}^{(1)}(\zeta/\xi)$ integrand is small in most
of the physical range $\chi\leq\xi\leq1.$

As the virtuality $Q$ increases, the difference between $\chi$ and
$x$ progressively reduces, and $\xi$ varies in a wider interval.
Finally in (c), for very large $Q$, the S-ACOT and S-ACOT-$\chi$
subtractions become identical. $A_{h,g}^{(1)}\otimes g$ approximates
well the collinear splitting contribution that drives much of the
shape of $F_{h,g}^{(1)}(\chi/\xi)g(\xi)$. When $A_{h,g}^{(1)}\otimes g$
is subtracted from $F_{h,g}^{(1)}\otimes g$ as in Eq.~(\ref{C1hg2}),
it produces a moderate \emph{negative} ${\cal O}(\alpha_{s})$ contribution,
which is further reduced at NNLO. These cancellations are further
examined in Sec.~\ref{sec:CancellationsHighQ}.

\section{Numerical examples \label{sec:Numerical-examples}}

In this section, we show representative plots from our validation
tests for the NNLO inclusive structure functions $F_{2}(x,Q)$ and $F_{L}(x,Q)$
computed according to the S-ACOT-$\chi$ scheme. We focus on the partial
contributions in which the photon strikes a charm quark, given by
$F_{h}(x,Q)$ in Eq.~(\ref{Fheavy}) for $h=c$, and for structure
functions $F=F_{2}$ or $F_{L}$. These contributions are referred
to as $F_{2c}(x,Q)$ and $F_{Lc}(x,Q)$ in the figures. The same comparisons
have been repeated for the bottom-quark functions $F_{2b}$ and $F_{Lb},$
as well as for the full inclusive functions $F=\sum_{l=1}^{N_{l}}F_{l}+\sum_{h=N_{l}+1}^{N_{f}^{fs}}F_{h}$
and alternative values of Bjorken $x.$ The results of other tests
show similar patterns and can be viewed at \citet{SACOTNNLO2011}.

NNLO coefficient functions $F_h^{(2)}(x,Q)$ for massive quarks 
are computed using a program available
from \citet{Riemersma:1994hv}. This program tabulates two-loop heavy-quark
coefficient functions in a form that allows fast evaluation of convolution
integrals in the $Q$ range covered by the experimental data.

The PDFs in all comparisons are obtained by using the Les Houches
Accord toy parametrization \citet{Giele:2002hx,Whalley:2005nh} at
the starting scale $Q_{0}=m_{c}=\sqrt{2}$ GeV. Other input parameters
are $\alpha_{s}(Q_{0})=0.36$ and the pole mass $m_{c}$.\footnote{Our program 
can alternatively read  $\overline{\rm MS}$ masses as the
input. In this case, the $\overline{\rm MS}$ masses are later converted
into the respective pole masses, because the operator matrix elements $A_{ab}^{(k)}$
are published as functions of the pole masses.%
} The switching between 3 and 4 flavors happens at $Q=m_{c}$. The
$\alpha_{s}$ and PDFs are evolved to higher $Q$ values by the HOPPET
computer 
code \citet{Salam:2008qg}.\footnote{Bottom-quark contributions 
are omitted in this comparison. The charm
PDF is zero at $Q<Q_{0}=m_{c}$, but acquires a non-negligible value
immediately above $Q_{0}$ through an ${\cal O}(\alpha_{s}^{2})$
discontinuity existing at the switching point.%
}

\subsection{$Q$ dependence}

\begin{figure}[ht]
\begin{centering}
\includegraphics[width=0.48\columnwidth]{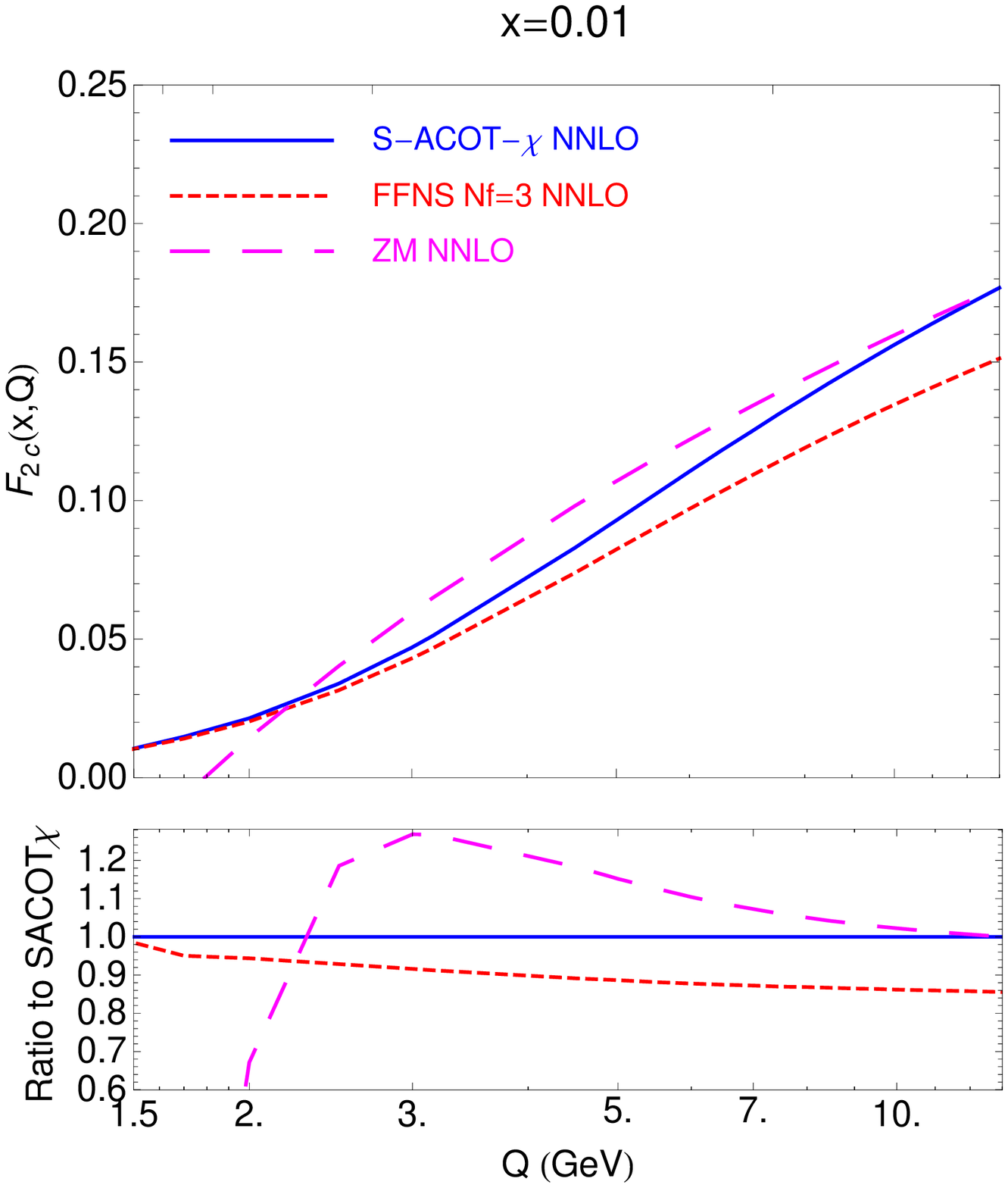}
~~\includegraphics[width=0.48\columnwidth]{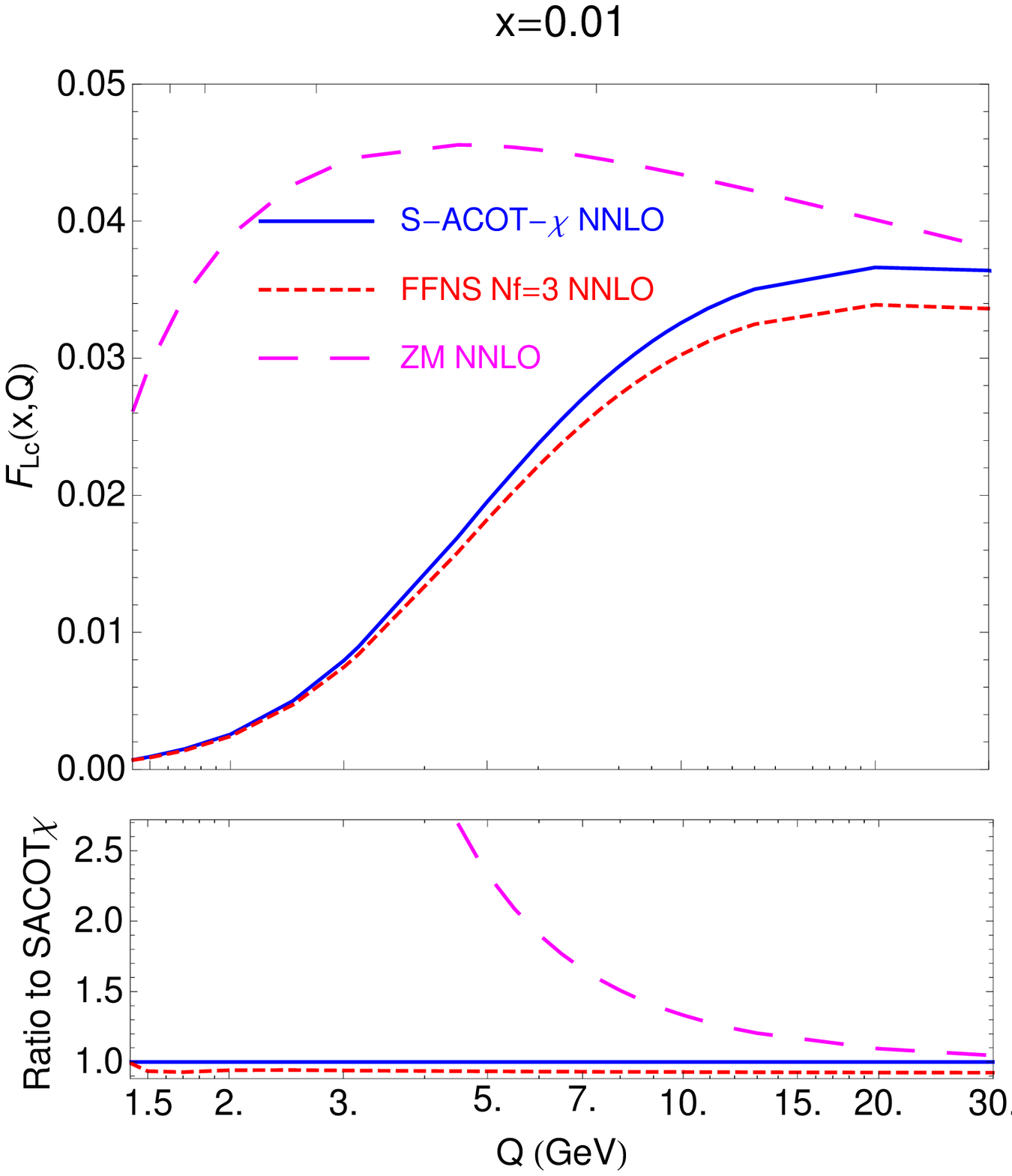} 
\par\end{centering}

\caption{{\small Comparison of $F_{2c}(x,Q)$ (left) and $F_{Lc}(x,Q)$ (right), computed
at ${\cal {\cal O}}(\alpha_{s}^{2})$ in the S-ACOT-$\chi$ (solid),
FFN $N_{f}=3$ (short dashed), and ZM $N_{f}=4$ (long dashed) schemes,
shown as a function of $Q$ at $x=10^{-2}$. }}
{\small \label{fig:QDependence}} 
\end{figure}

Fig.~\ref{fig:QDependence} examines $Q$ dependence of charm structure
functions $F_{2c}$ (left panel) and $F_{Lc}$ (right panel). They
are computed to order $\alpha_{s}^{2}$ in all schemes, referred 
to as ``NNLO'' by the counting convention for 
the {\it inclusive} structure functions considered here. 
[In predictions for {\it semi-inclusive} charm production, 
the ${\cal O}(\alpha_s^2)$ cross section in the FFN scheme is often
counted as NLO, since the ${\cal O}(\alpha_s^0)$ flavor-excitation cross
section is absent in this scheme.]
The upper insets in both panels show predictions at $x=10^{-2}$ in
the S-ACOT-$\chi$ scheme, FFN scheme with $N_{f}=3$, and ZM scheme
with $N_{f}=4$. The lower insets show ratios of the FFN and ZM predictions
to the S-ACOT-$\chi$ prediction.

The left panel shows that the S-ACOT-$\chi$ theory prediction for
$F_{2c}(x,Q)$ (blue solid line) is numerically close to the FFN prediction
(red short-dashed line) at $Q\approx m_{c}$ and to the ZM prediction
(magenta long-dashed line) at $Q>10$ GeV.

Similarly, in the right
panel, the S-ACOT-$\chi$ prediction for the longitudinal function
$F_{Lc}(x,Q)$ coincides with the corresponding FFN prediction at
$Q\approx m_{c}$ and approaches the ZM prediction at $Q>30$ GeV.
{[}$F_{Lc}(x,Q)$ is sensitive to mass-dependent corrections to scattering
off longitudinally polarized photons. Its matching on the ZM prediction
happens at higher $Q$ values than in $F_{2c}$.{]} The S-ACOT-$\chi$
prediction interpolates between the FFN and ZM predictions at intermediate
$Q$ values, precisely as expected. %

\subsection{Dependence on the factorization scale}

\begin{figure}[p]
\begin{centering}
\includegraphics[width=0.48\columnwidth]{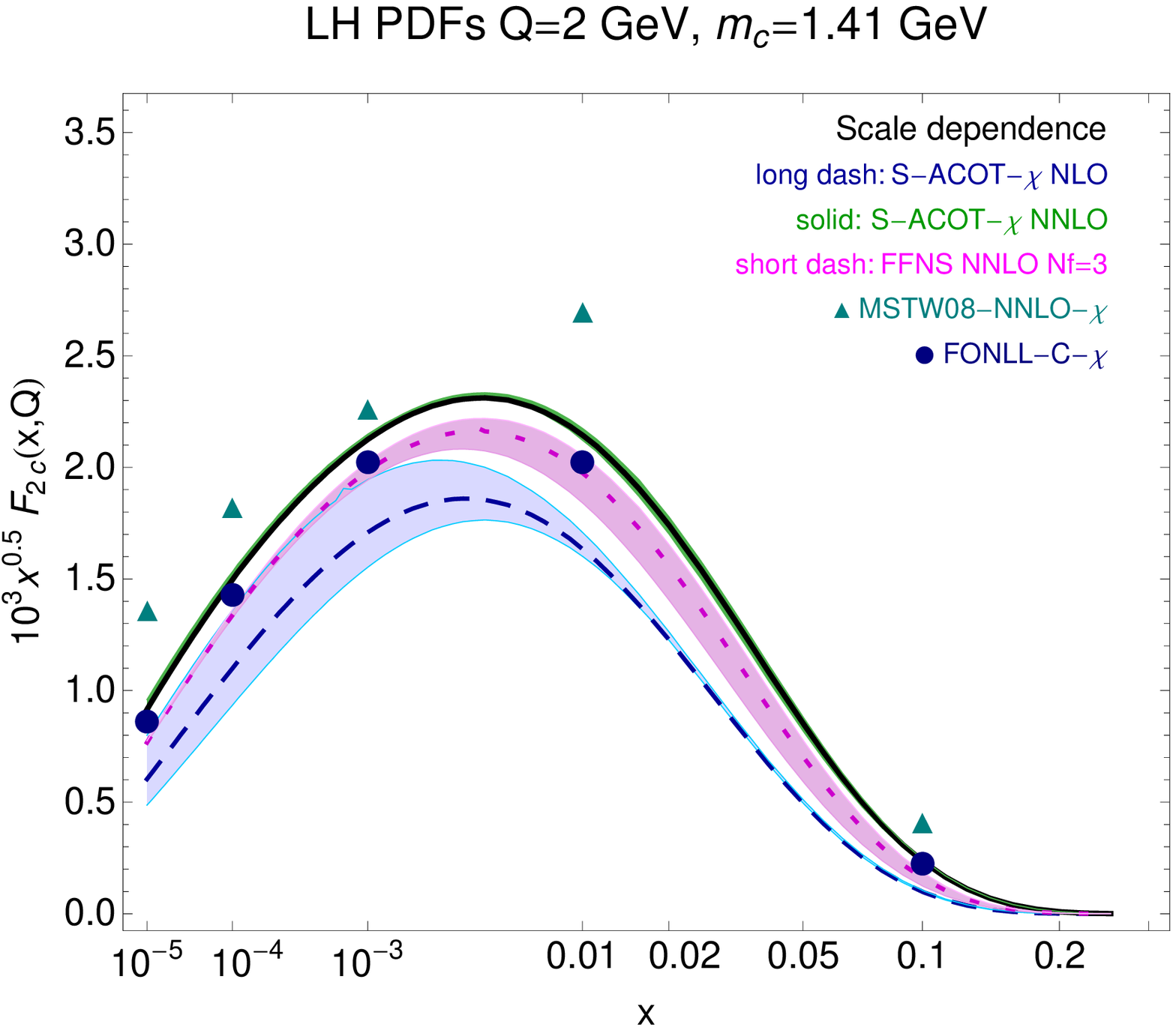}~~\includegraphics[width=0.48\columnwidth]{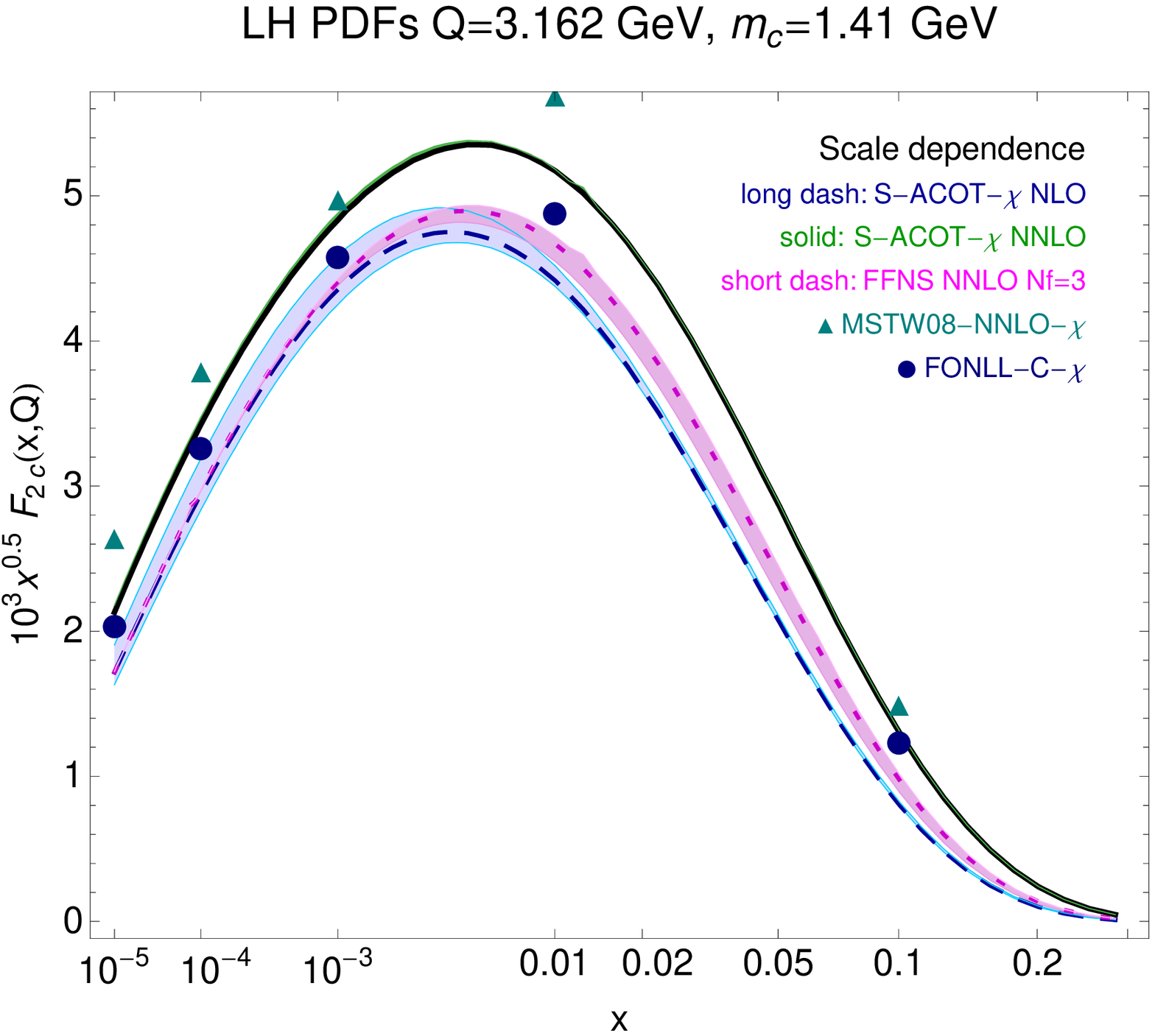}\\
 (a)\hspace{5cm}(b) 
\par\end{centering}

\begin{centering}
\includegraphics[width=0.48\columnwidth]{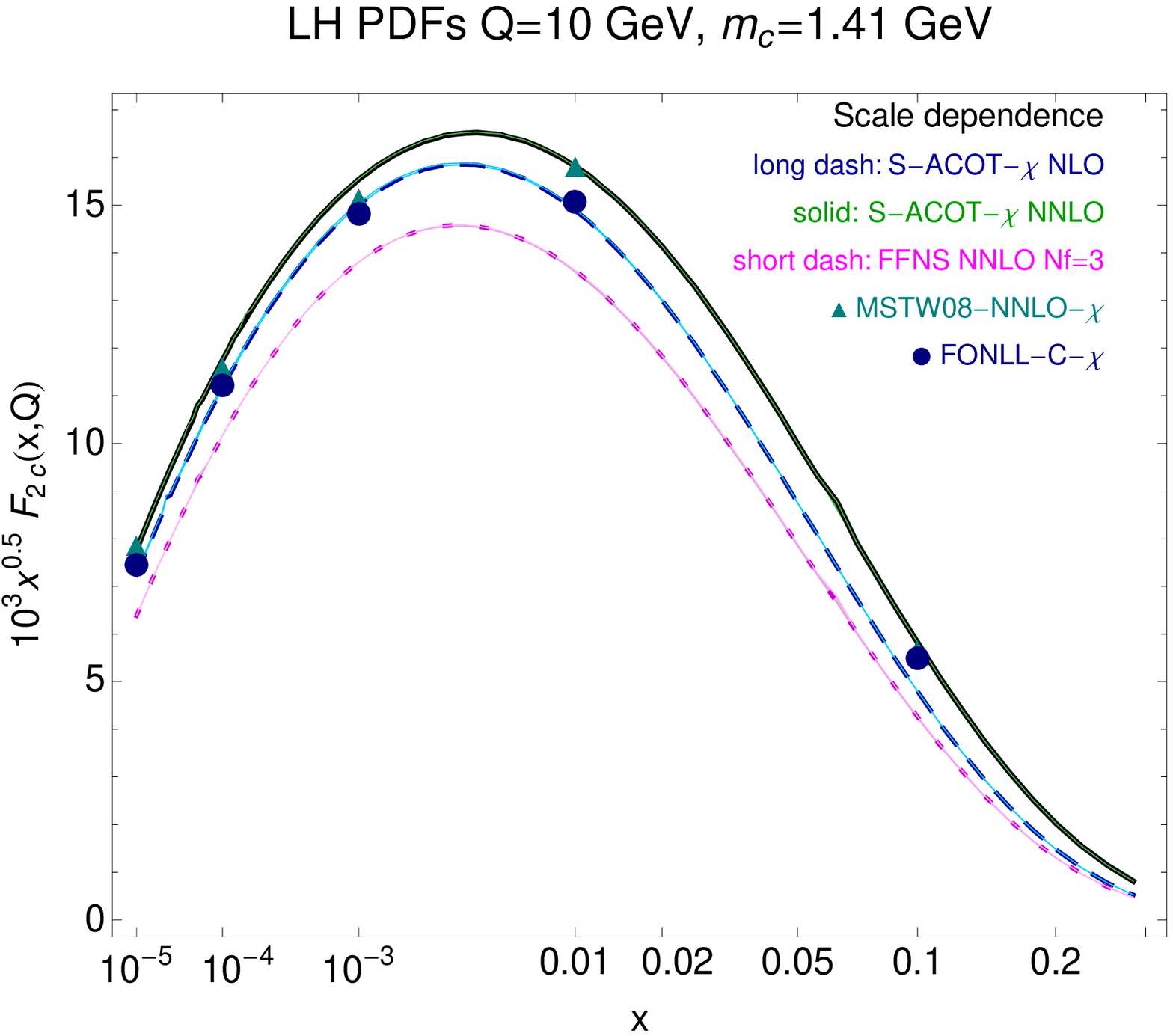}~~\includegraphics[width=0.48\columnwidth]{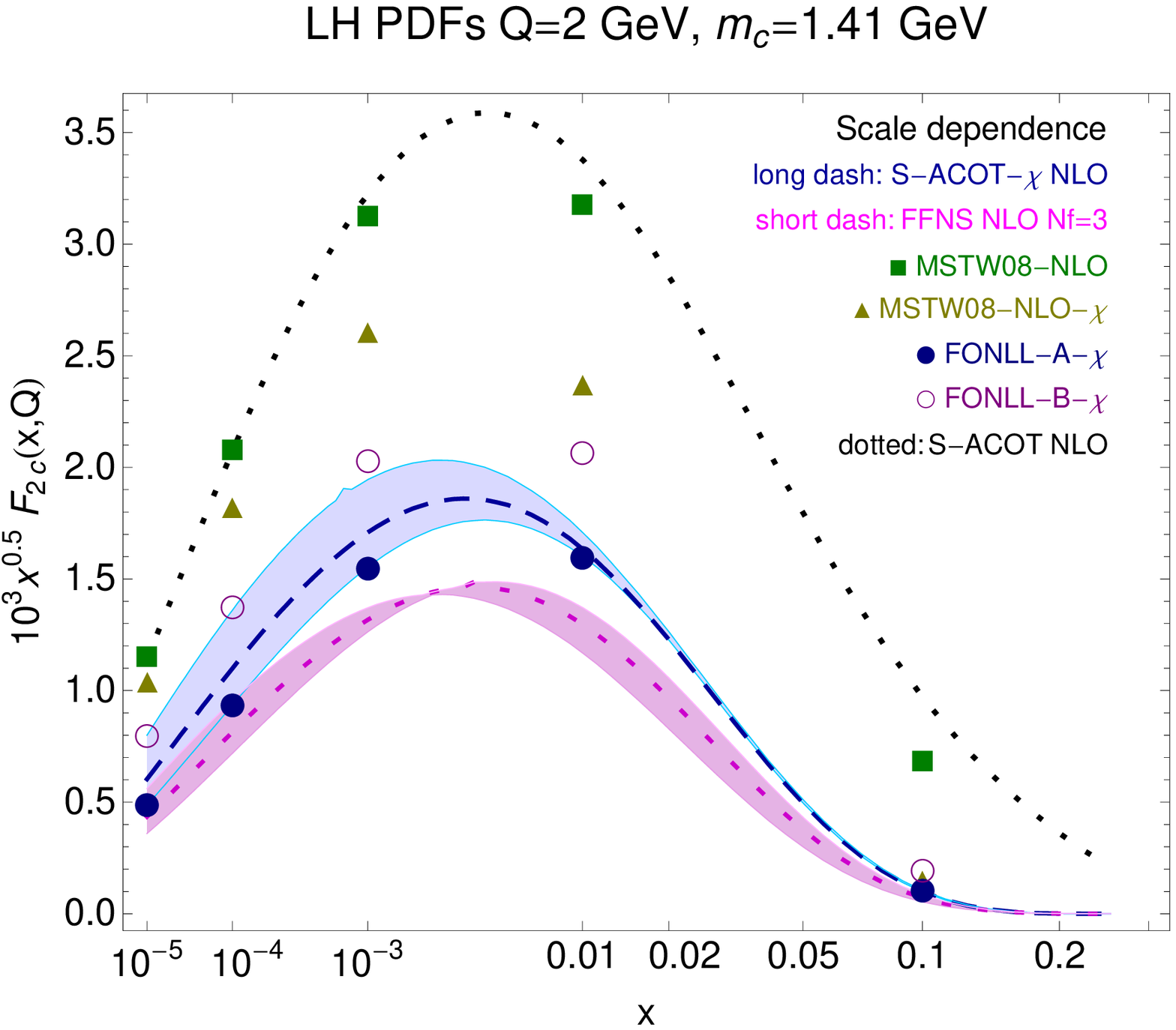}\\
 (c)\hspace{5cm}(d) 
\par\end{centering}

\caption{{\small Comparison of predictions for $F_{2c}(x,Q)$ in the S-ACOT-$\chi$
scheme and alternative theoretical approaches at NLO ($O(\alpha_s)$) and NNLO ($O(\alpha_s^2)$). Central
predictions are for $\mu=\sqrt{Q^{2}+m_{c}^{2}}$, and the error bands
are for }$Q\leq\mu\leq\sqrt{Q^{2}+4m_{c}^{2}}$.}

{\small \label{fig:ScaleDependence}} 
\end{figure}

NLO computations leave substantial uncertainty in the DIS charm-quark
contributions due to the choice of the renormalization/factorization
scale and differences in the FE terms in the threshold region. NNLO terms
drastically reduce these uncertainties. Factorization scale dependence,
and its reduction from NLO to NNLO, is illustrated by Fig.~\ref{fig:ScaleDependence}.
Reduction of uncertainties in the modeling of 
kinematical threshold effects is discussed 
in Sec.~\ref{sec:Threshold-effects}. 

In Fig.~\ref{fig:ScaleDependence}(a)-(c), predictions for $F_{2c}(x,Q)$
in the S-ACOT-$\chi$ and FFN ($N_{f}=3$) schemes are plotted versus
Bjorken $x$ at representative $Q^{2}$ values 
of 4, 10, and 100 $\mbox{GeV}^{2}$.
The $F_{2c}(x,Q)$ values on the $y$ axis are multiplied by $10^{3}\sqrt{x}$
to better visualize the accessible $x$ region. Central predictions
are computed for $\mu=\sqrt{Q^{2}+m_{c}^{2}}$, the default scale
in heavy-quark DIS cross sections 
in the CT10 global analysis \citet{Lai:2010vv}.
The error bands are obtained by varying the scale in the range $Q\leq\mu\leq\sqrt{Q^{2}+4m_{c}^{2}}$.

At $Q=2$ GeV in Fig.~\ref{fig:ScaleDependence}(a), the NNLO S-ACOT-$\chi$
central prediction (black solid line inside a green band) is slightly
above the NNLO FFN prediction (short-dashed line inside a magenta
band) and has a smaller scale uncertainty than FFN. At $Q$ below 2
GeV (not shown), the NNLO S-ACOT-$\chi$ and FFN predictions get even
closer. In contrast, the NLO S-ACOT-$\chi$ prediction (a long-dashed
line inside a blue band) underestimates the NNLO FFN result and has
wider scale dependence.

As $Q$ increases to 10 GeV (Fig.~\ref{fig:ScaleDependence}(c)),
S-ACOT-$\chi$ predicts more event rate than the FFN scheme both at
NLO and NNLO. Altogether, the $Q$ dependence in these figures is
fully compatible with the matching of the S-ACOT-$\chi$ results on
the FFN and ZM results in the limits $Q^{2}\approx m_{c}^{2}$ and
$Q^{2}\gg m_{c}^{2}$, respectively.

\subsection{NNLO vs. NLO predictions \label{sub:NNLOvsNLO}}

Improved stability of the NNLO prediction in Fig.~\ref{fig:ScaleDependence}(a)
can be appreciated by comparing it to the counterpart NLO result shown
in Fig.~\ref{fig:ScaleDependence}(d). Here, we collect 
NLO $F_{2c}(x,Q)$ values at $Q=2$ GeV, obtained in the 
FFN and S-ACOT schemes. We also show NLO predictions 
in the modified Thorne-Roberts (TR') scheme \citet{Thorne:1997ga,Thorne:1997uu,Thorne:2006qt}),
as implemented by the MSTW'08 PDF analysis \citet{Martin:2009iq},
and in the 
FONLL schemes A and B used by the NNPDF collaboration \citet{Forte:2010ta}.
The S-ACOT and MSTW predictions
are shown with the $\chi$ scaling as well as without it.

The spread in NLO values of $F_{2c}(x,Q)$ observed in the figure
is extensive, nominally suggesting a large uncertainty in the resulting
NLO PDF sets. However, when included in the PDF fits, the most extreme
predictions for $F_{2c}(x,Q)$ in this figure are excluded by the
fitted DIS data, which prefer the values that are about the same as
the (relatively unambiguous) NNLO result. In the CT10 NLO fit, the scale $\mu$
is set equal to $\sqrt{Q^{2}+m_{c}^{2}}$, which brings the NLO S-ACOT-$\chi$
prediction in agreement with the measured cross sections. Thus, according
to the past global fits, the NLO cross sections can be reconciled
with the heavy-quark data, but at the expense of tuning of the scale
parameter, for each value of $m_{c}$ and rescaling variable. The
key benefit of the NNLO calculation for $F_{2c}(x,Q)$ is to automatically
achieve such a good agreement, nearly independently of the factorization
scale.

\subsection{Threshold effects \label{sec:Threshold-effects}}

\begin{figure}[ht]
\begin{centering}
\includegraphics[width=0.48\columnwidth]{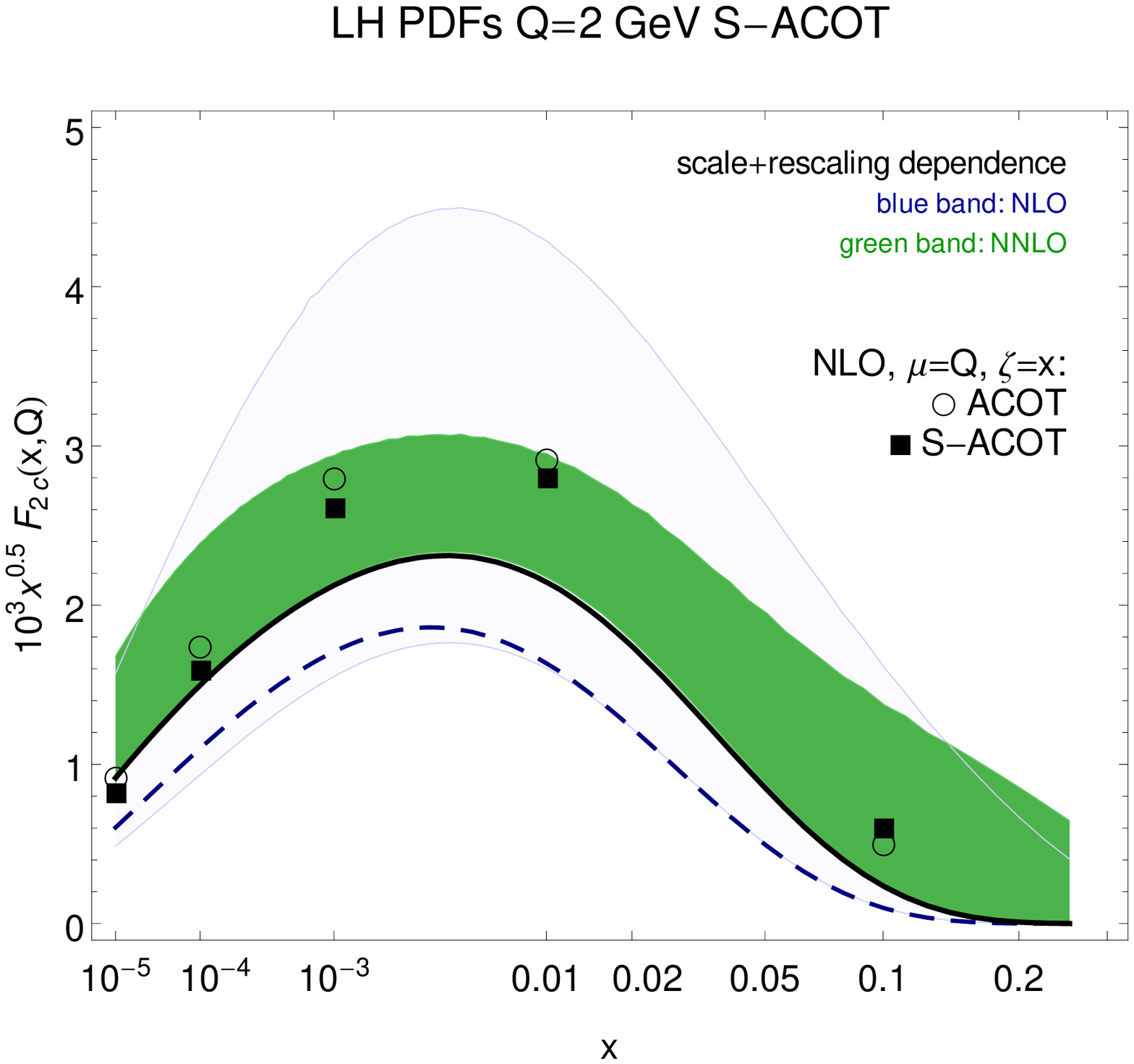}~~~~\includegraphics[width=0.48\columnwidth]{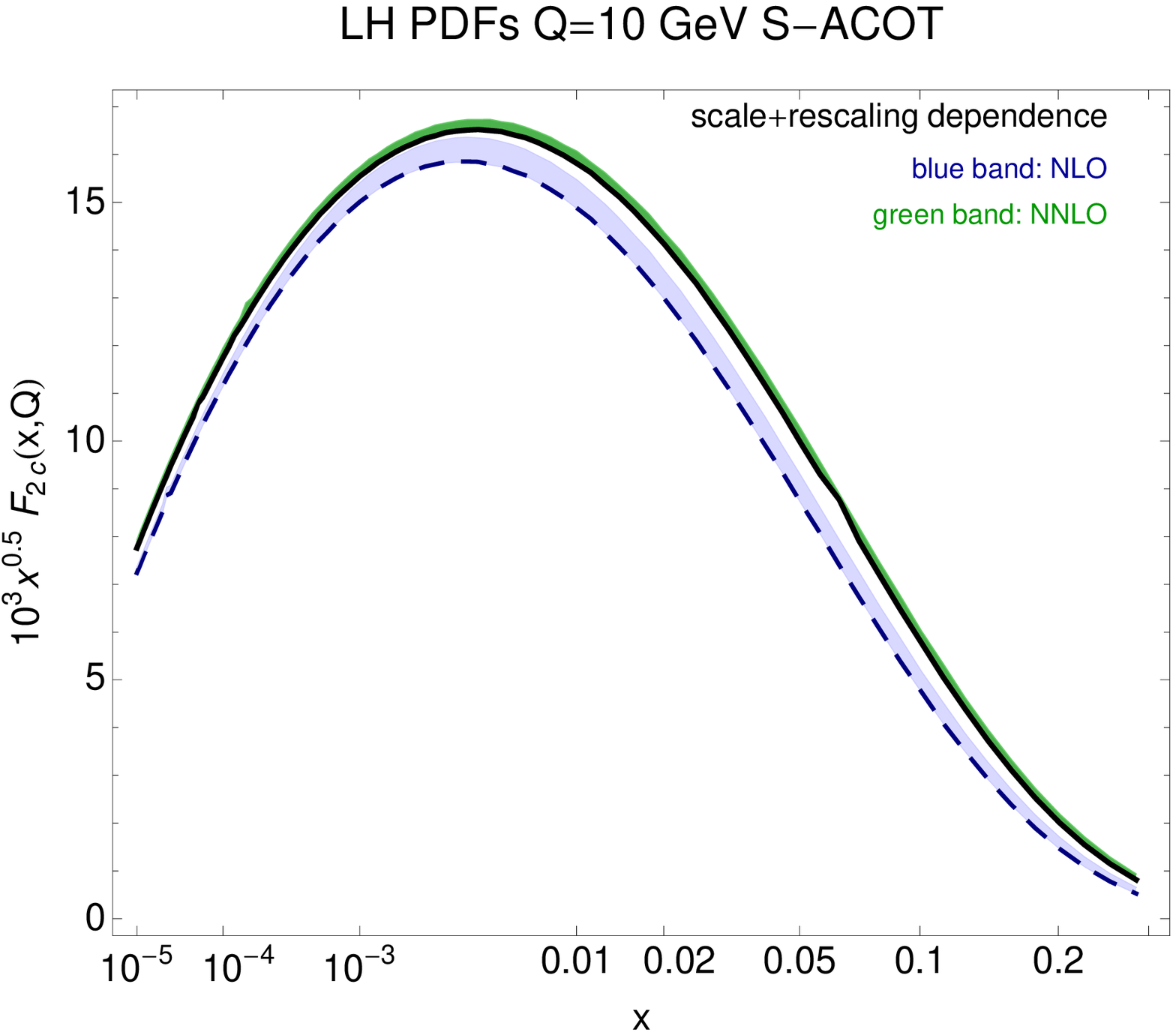}\\
 (a)\hspace{5cm}(b) 
\par\end{centering}

\caption{{\small Dependence on the factorization scale $\mu$ and the rescaling
variable $\zeta(\lambda)$. Left: $Q=2$ GeV; right: $Q=10$ GeV.}}
{\small \label{fig:RescalingDependence}} 
\end{figure}

In the above discussion, our NNLO structure functions are computed
using the optimal rescaling variable $\zeta=\chi$ in the FE heavy-quark
contributions. The rescaling variable significantly improves convergence
near the threshold by excluding contributions to the FE convolution
integrals that are kinematically disallowed.

Dependence on the rescaling prescription can be explored with the help
of the variable $\zeta$ that generalizes the $\chi$ variable as 
proposed in Ref.~\citet{Nadolsky:2009ge}. 
The generalized rescaling variable $\zeta$
is implicitly defined by\begin{eqnarray}
x=\frac{\zeta}{1+\zeta^{\lambda}\cdot(4m_{c}^{2})/Q^{2}},\end{eqnarray}
 where $\lambda$ is a real number. Various choices of positive $\lambda$
produce a family of GM schemes in which $\zeta$ takes continuous
values between $x$ (no rescaling) and $\chi$ (full rescaling). Specifically,
$\lambda=0$ produces $\zeta=\chi$ of the S-ACOT-$\chi$ scheme,
and $\lambda\gg1$ produces $\zeta\approx x$ that corresponds to
the plain S-ACOT scheme without rescaling. 
Negative $\lambda$ values
(not shown) strongly suppress the FE contributions by setting $\zeta>\chi$.

Fig.~\ref{fig:RescalingDependence} shows the bands of variations
in the NLO and NNLO S-ACOT $F_{2c}$ values at $Q=2$ GeV (left panel)
and 10 GeV (right panel) when $\mu$ and $\lambda$ are varied in
the ranges $Q\leq\mu\leq\sqrt{Q^{2}+4m_{c}^{2}}$ and $0\leq\lambda\leq100$,
respectively. If $\lambda$ is naively varied in the full range, at
smallest $Q$ values one obtains a large excursion in the NLO predictions
(light blue band), which is considerably reduced when going to NNLO
(green band). 

The ACOT and S-ACOT schemes without rescaling (corresponding to  $\lambda \gg 1$) 
are less motivated  than the S-ACOT-$\chi$ scheme, since they include kinematically
disallowed small-$x$ contributions that destabilize perturbation 
theory. To give an idea about these less favored schemes, 
the upper boundary of the green
band in Fig. 7 corresponds to the NNLO S-ACOT calculation (without
rescaling) for the scale $\mu = \sqrt{Q^2 + 4 m_c^2}$. Lower values of
the $\mu$ scale, such as $\mu=Q$, reduce the S-ACOT cross section and
bring it closer to the S-ACOT-$\chi$ cross section for $\mu=Q$ (the
black solid line). [The scale dependence of the S-ACOT cross section
is larger than that of the S-ACOT-$\chi$ cross section.]

The distinction between the ACOT and S-ACOT schemes arises from 
the additional ``dynamic'' mass terms in the ACOT flavor-excitation 
Wilson coefficients, which have little numerical effect \cite{Kramer:2000hn}.
An NNLO calculation in the full ACOT scheme is difficult 
and was not completed, but already at NLO the ACOT and S-ACOT schemes
become very close. The left panel compares the NLO $F_{2c}$ values 
in the ACOT and S-ACOT schemes for $\mu=Q$ without rescaling from Ref.~\cite{Binoth:2010ra}, 
as indicated by the circles and squares, respectively. 
The discrepancy between the ACOT and S-ACOT values is small already at NLO. 
It is likely to further decrease when going to the NNLO  as a term of order $\alpha_s^3$.

As $Q$ increases above a few GeV, dependence on $\lambda$
diminishes practically to nil, as in the right panel of the figure
for $Q=10$ GeV. The ACOT and S-ACOT scheme produce essentially coinciding predictions 
at such a large $Q$ value \cite{Binoth:2010ra}.  
Together with Fig.~\ref{fig:ScaleDependence}, Fig.~\ref{fig:RescalingDependence}
indicates that, at NNLO, the physically motivated rescaling variable is more
important at low $Q$ than the factorization scale choice or the difference
between the ACOT and S-ACOT schemes.

\subsection{Alternative mass schemes}

\subsubsection{TR' and FONLL schemes}
Figs.~\ref{fig:ScaleDependence}(a)-(c) also show NNLO predictions
in the alternative GM schemes, indicated by scattered symbols: the modified
Thorne-Roberts (TR') scheme and FONLL scheme C.
Their values are computed in the 2009 Les Houches benchmark study of
GM schemes \citet{Binoth:2010ra} by assuming the
same $\chi$ rescaling as the S-ACOT-$\chi$ scheme.

The three schemes are seen to be in good overall agreement, apart
from minor differences traced to subtle variations in the NNLO implementations
that the schemes provide.

At $Q=2$ GeV, the NNLO S-ACOT-$\chi$ prediction lies slightly above
the FONLL-C prediction and below the MSTW prediction. At $Q=10$ GeV,
the NNLO S-ACOT-$\chi$ prediction becomes closer to the MSTW prediction
and is still above the FONLL-C result. These differences can be understood
by noticing that the compared schemes may differ in subleading perturbative
terms. For example, the FONLL-C scheme includes a threshold damping
factor to match on the 3-flavor result 
near the threshold \citet{Forte:2010ta}. The S-ACOT-$\chi$ scheme
is not using the damping factor and is expectedly close to the FONLL-C
scheme at $Q\rightarrow m_{c}$, but not strictly identical.
The TR'/MSTW prediction includes a constant higher-order term (of
order ${\cal O}(\alpha_{s}^{3})$) to improve smoothness of switching
from 3 to 4 active flavors at $Q=m_{c}$. 
Neither S-ACOT-$\chi$ nor FONLL-C include this artificial constant term, 
which is why they may predict smaller $F_{2c}$ values at low $Q$.

\subsubsection{The BMSN scheme}

Structural similarities between the S-ACOT-$\chi$, TR', and FONLL
schemes reflect their conceptual origin from the Collins-Wilczek-Zee
(CWZ) renormalization method \cite{Collins:1978wz}. The CWZ procedure
is applied frequently to renormalize QCD quantities dependent on several
mass scales. It introduces a sequence of renormalization schemes
and associated differential equations that operate
with the number $N_{f}$ of active flavors that changes across the particle
mass thresholds. The CWZ procedure is invoked, for example, 
in the common definition of the QCD running coupling
$\alpha_{s}$ and by the zero-mass VFN scheme.

The family trait of the CWZ renormalization -- a hierarchy of fixed-flavor
number subschemes with sequentially incrementing $N_{f}$ values --
is also present in the ACOT-like general mass schemes. In practical
realizations of these schemes, scale dependence of \emph{both} $\alpha_{s}$
and PDFs is found by solving renormalization group equations with
a shared $N_{f}$ value in each mass range. The $N_{f}$ and $N_{f}+1$
expressions for $\alpha_{s}$ and PDFs are related at the switching
momentum scales through the known matching conditions.

A different path is taken in the approach of Buza, Matiuonine, Smith,
and van Neerven (BMSN \cite{Buza:1996wv}), which is adopted in the
fits by the ABM group \cite{Alekhin:2009ni,Alekhin:2012ig}. In the
BMSN and CSN \cite{Chuvakin:1999nx} 
frameworks, only $\alpha_{s}(\mu)$ is found from a renormalization
group equation according to the CWZ procedure. However the 4-flavor PDFs are
\emph{constructed} from the $3$-flavor PDFs at $\mu\geq m_{c}$ by
solving the matching equations for each $\mu$ value. Only the $3$-flavor
PDFs are evolved by the DGLAP equations in this case. Here we see
the key difference with the ACOT approach, which resums higher-order
corrections to the heavy-flavor PDFs at $\mu\geq m_{c}$ with the
help of DGLAP equations. The BMSN method does not provide this resummation,
crucial for implementing the collider data from $Q^{2}\gg m_{c}^{2},m_{b}^{2}$
into the global fit. To resum the heavy-quark collinear logs in their
published 5-flavor PDFs, ABM evolve them from 
the initial scale $m_{b}$ to higher energies \emph{after} 
the fit, starting from the best-fit parametrization
found in the BMSN approach.

In the BMSN framework, the number $N_{f}$ of active quark flavors in
$\alpha_{s}$ is incremented from 3 to 5 according to the usual convention
as the energy increases. $4$-flavor PDFs $f_{a}(4,x,\mu^{2})$ are
derived from the $3$-flavor PDFs $f_{a}(3,x,\mu^{2})$
as
\begin{equation}
f_{a}(4,x,\mu^{2})=\sum_{b}\left[A_{ab}\otimes f_{b}\right](3,x,\mu^{2}).\label{4from3}\end{equation}
The functions $A_{ab}(\widehat{x})=\delta_{ab}\delta(1-\widehat{x})+a_{s}A_{ab}^{(1)}(\widehat{x})+a_{s}^{2}A_{ab}^{(2)}(\widehat{x})+...$
are comprised of the coefficients
$A_{ab}^{(k)}$ in the perturbative expansion of the massive parton-level PDFs that were discussed 
in Sec.~\ref{sec:TheoryOverview}.
The BMSN 4-flavor structure function is given by \begin{equation}
F^{BMSN}(4,x,Q)=F(3,x,Q,m_{c}\neq0)+F(4,x,Q,m_{c}=0)-F^{asymp}(3,x,Q,m_{c}\neq0),\label{FBMSN}
\end{equation}
where $F(3,x,Q,m_{c}\neq0)$ is obtained for three massless quarks ($u,$
$d,$ $s$) and one massive quark ($c$).
$F^{asymp}(3,x,Q,m_{c}\neq0)$ is the dominant part of $F(3,x,Q,m_{c}\neq0)$
in the asymptotic limit $Q^{2}\gg m_{c}^{2}$, and $F(4,x,Q,m_{c}=0)$
is computed with 4 massless quarks in the Wilson coefficients, with
the 4-flavor PDFs defined by Eq.~(\ref{4from3}). 

This arrangement provides a nearly ideal matching of the $4$-flavor
$F^{BMSN}(4,x,Q)$ onto the 3-flavor $F(3,x,Q,m_{c}\neq0)$ as $\mu\rightarrow m_{c}$,
possible only in the absence of resummation 
of collinear logs $\ln(\mu^{2}/m_{c}^{2})$. 
At ${{\cal O}}(\alpha_{s}^{2})$,
the last two terms in Eq.~(\ref{FBMSN}) are related by the replacement of the 3-flavor QCD coupling by the 4-flavor coupling,
\begin{equation}
F(4,x,Q,m_{c}=0)=
\Biggl(F^{asymp}(3,x,Q,m_{c}\neq0)\Biggr)_{\alpha_{s}(3,\mu)\, \rightarrow\, \alpha_{s}(4,\mu)-\frac{1}{6\pi}\alpha_{s}^{2}\,\ln\left(\frac{\mu^{2}}{m_{h}^{2}}\right)} .
\label{F4asymp}
\end{equation}
Since $\alpha_{s}(N_f,\mu)$ is nearly continuous at the
switching point between 3 and 4 flavors (apart from a mild discontinuity
that first enters at ${{\cal O}}(\alpha_{s}^{2})$), it follows from
Eqs.~(\ref{FBMSN}) and (\ref{F4asymp}) that $F^{BMSN}(4,x,Q)\approx F(3,x,Q,m_{c}\neq0)$
at $Q\rightarrow m_{c}$. The matching onto the FFN scheme at $Q^2\approx m_c^2$
is achieved by dropping numerical DGLAP evolution of 4-flavor PDFs.

In deriving these relations, BMSN allow only \emph{one} parton flavor
to be massive in each $\mu$ range. This assumption is untrue at some
level for $\mu$ comparable to $m_{b}$, since $m_{c}$ is not negligible
compared to $m_{b}.$ It creates conceptual difficulties in extending
the VFN scheme proposed by BMSN to three loops \cite{Ablinger:2011pb,Ablinger:2012qj},
since both the parton-level structure functions $F_{a,b}^{(3)}$ and operator
matrix elements $A_{ab}^{(k)}$ may depend on $m_{c}$ and $m_{b}$
at the same time.

In fact, this assumption is not necessary for proving QCD factorization
with heavy quarks and is not made in the S-ACOT approach. The 
proof of the general-mass factorization scheme 
does not assume that all quarks but one are massless.
Masses of heavy quarks that may be comparable to $Q$ are never neglected 
in the target subgraphs associated with 
the PDFs and $A_{ab}^{(k)}$, cf.~Ref.~\cite{Collins:1998rz}
and Sec.\,\ref{sec:ChiConvention}. 
As an illustration of their role, 
consider again the 
coefficient function $C^{(3)}_{b,g}$ with $c$ and $b$ quark lines 
in Eq.~(\ref{C3bg}) of 
Sec.~\ref{sub:Several-heavy-flavors}. This function is
found by subtracting convolutions of massive 
operator matrix elements $A^{(k)}_{bg}$ from a massive parton-level function
$F^{(3)}_{b,g}$. This is expected to produce $C^{(3)}_{b,g}$ that is
free of the $\ln(\mu^2/m_c^2)$ and $\ln(\mu^2/m_b^2)$ terms and 
coincides with $C^{(3)}_{b,g}$ in the effective $\overline{MS}$ scheme
with 5 massless flavors when $Q^2$ is unequivocally larger than
$m_c^2$  and $m_b^2$. [Verification of this prediction still awaits an
explicit calculation of the massive function $F^{(3)}_{b,g}$].
The operator matrix element
$A_{bg}^{(3)}$ that depends both on $m_{c}$ and $m_{b}$, and 
which caused concern in Refs.~\cite{Ablinger:2011pb,Ablinger:2012qj}, 
therefore naturally appears in the factorization formula (\ref{C3bg}) 
when deriving the infrared-safe $C_{b,g}^{(3)}$ with two quark species.
It is not anticipated to pose a problem from the S-ACOT-$\chi$ viewpoint.

Numerically, the S-ACOT-$\chi$ and BMSN approaches provide close
predictions for charm production at $Q\approx m_{c}$ \cite{Binoth:2010ra}.
While both approaches are in good agreement with the current HERA
data, they may lead to numerical differences in future precise DIS
analyses, both at scales of order $m_{b},$ where 
the resummed $\ln(Q^{2}/m_{c}^{2})$ terms
may already play some role, and at electroweak scales, where the expected 
differences of a few percent may be comparable to the PDF uncertainties. 
More generally, S-ACOT-$\chi$ points out a way to include heavy-quark
mass dependence and resummation of heavy-quark collinear logs in one step, 
and to implement three-loop DIS amplitudes with two massive quark
flavors along the guidelines of the CWZ renormalization.

\section{Cancellations between Feynman graphs\label{sec:Cancellations}}

\subsection{Cancellations at low $Q$ \label{sec:CancellationsLowQ}}

In order to match on the FFN and ZM predictions, certain classes of
Feynman diagrams inside the S-ACOT-$\chi$ NNLO coefficient functions
must cancel in the respective low-$Q$ and large-$Q$ regions. We
will show how these cancellations come about in the case of the charm-quark
function $F_{2c}(x,Q)$, but the pattern holds for the bottom quark
and other structure functions with suitable modifications.

The cancellations are revealed by plotting differences between various
matrix elements and collinear subtractions discussed in Section \ref{sec:TheoryOverview},
which are established by applying the factorization formula at the
parton level.

In the $Q\approx m_{c}$ region, all FE contributions in Eqs.~(\ref{coefFh21})-(\ref{coefFh24})
must cancel to a high degree in order for $F_{2c}(x,Q)$ to reduce
to the FFN matrix elements $F_{h,g}^{(1,2)}$ and $F_{h,l}^{(2)}$.
In the threshold region, the evolved charm PDF is effectively of order
${\cal O}(a_{s})$, \begin{equation}
\lim_{Q^{2}\rightarrow m_{c}^{2}}c(x,Q)\approx a_{s}(Q)\left[A_{hg}^{(1)}\otimes g\right](x,Q);\end{equation}
a FE contribution to $F_{2c}(x,Q)$ containing a coefficient $c_{h,h}^{(k)}$
is effectively of order ${\cal {O}}(a_{s}^{k+1}$) . Keeping this
in mind, at order ${\cal {\cal O}}(a_{s})$ the virtual-photon-charm
scattering diagram with $c_{h,h}^{(0)}$ in Eq.~(\ref{coefFh21})
cancels the gluon-initiated subtraction term with $A_{hg}^{(1)}$
in Eq.~(\ref{coefFh22}), and only the $\gamma^{*}g$ fusion diagram
$F_{hg}^{(1)}$ in Eq.~(\ref{coefFh22}) survives in the total $F_{2c}(x,Q).$
In this case, the difference \begin{equation}
D_{C^{(0)}}^{(1)}(x,Q)=\left[c_{h,h}^{(0)}\otimes c\right](x,Q)-a_{s}\,\left[c_{h,h}^{(0)}\otimes A_{hg}^{(1)}\otimes g\right](x,Q),\label{D1C0}\end{equation}
 where $c(x,Q)$ and $g(x,Q)$ represent the charm and gluon PDFs,
must be close to zero.

As the next order is included, the cancellation present in $D_{C^{(0)}}^{(1)}$
must further improve. Two differences quantify the cancellations to
this order: \begin{equation}
D_{C^{(0)}}^{(2)}(x,Q)=D_{C^{(0)}}^{(1)}(x,Q)-a_{s}^{2}\,\left[c_{h,h}^{(0)}\otimes A_{hg}^{(2)}\otimes g\right](x,Q)-a_{s}^{2}\,\left[c_{h,h}^{(0)}\otimes A_{hl}^{PS,(2)}\otimes\Sigma\right](x,Q),\label{D2C0}\end{equation}
 in which the convolutions of $c_{h,h}^{(0)}$ with $O(\alpha_{s}^{2})$
operator matrix elements $A_{hg}^{(2)}$ and $A_{hl}^{PS,(2)}$ are
subtracted from $D_{C^{(0)}}^{(1)}(x,Q)$; and \begin{equation}
D_{C^{(1)}}^{(2)}(x,Q)=a_{s}\,\left[c_{h,h}^{(1)}\otimes c\right](x,Q)-a_{s}^{2}\,\left[c_{h,h}^{(1)}\otimes A_{hg}^{(1)}\otimes g\right](x,Q),\label{D2C1}\end{equation}
which probes the cancellation between convolutions involving the coefficient
$c_{h,h}^{(1)}$. By comparing $D_{C^{(0)}}^{(2)}$ with $D_{C^{(0)}}^{(1)}$,
we quantify how the ${\cal O}(a_{s})$ cancellation in $D_{C^{(0)}}^{(1)}$,
proportional to $c_{h,h}^{(0)},$ improves upon the inclusion of the NNLO
corrections. The difference $D_{C^{(1)}}^{(2)}$ quantifies yet another 
${\cal O}(a_{s}^{2})$ cancellation that is independent of $D_{C^{(0)}}^{(1)}$. It has
the same structure as $D_{C^{(0)}}^{(1)}$, 
but includes the convolutions with $c_{h,h}^{(1)}$ instead of
$c_{h,h}^{(0)}$.

\begin{figure}[ht]
\begin{centering}
\includegraphics[width=0.49\columnwidth]{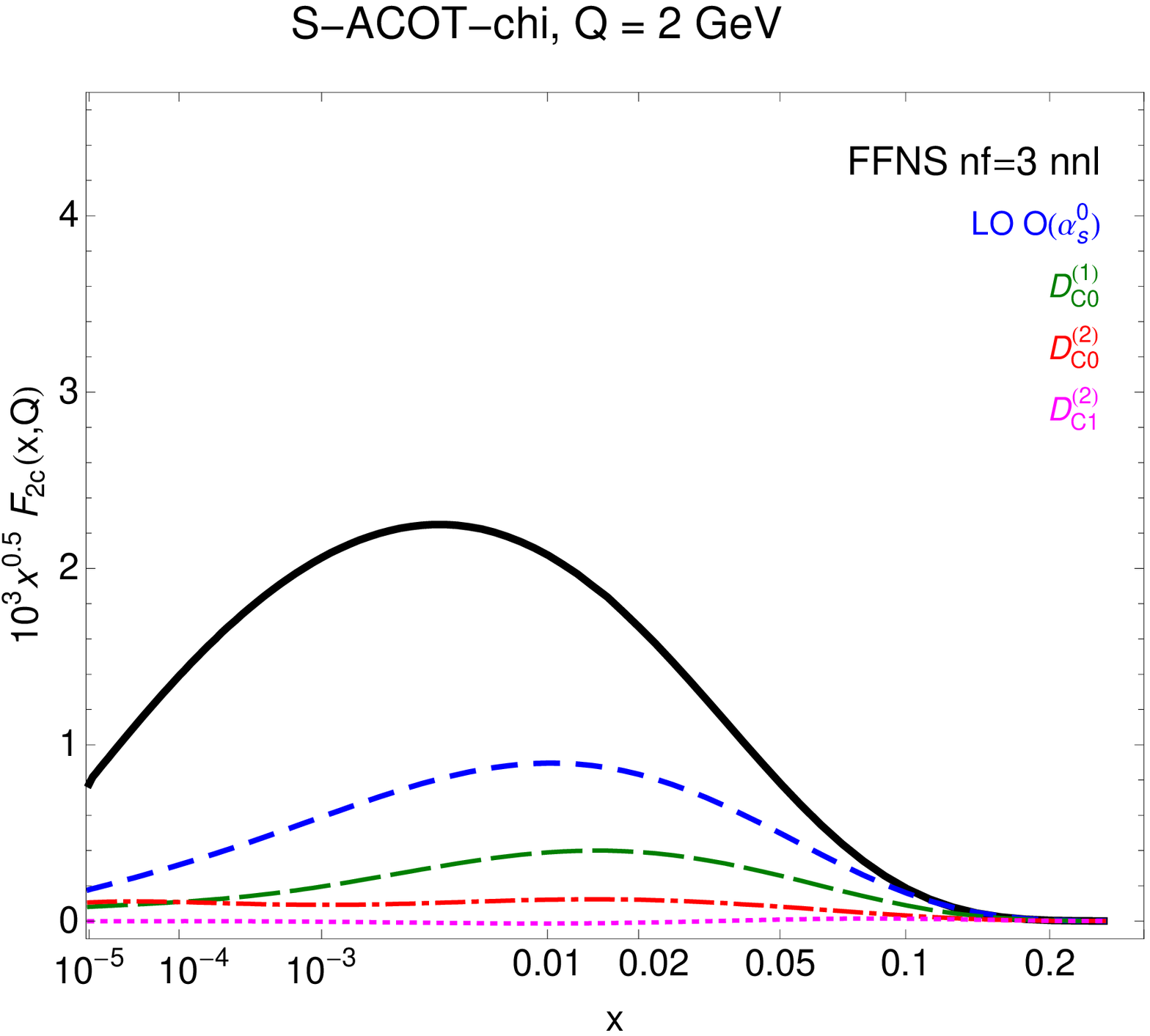}~~~~\includegraphics[width=0.49\columnwidth]{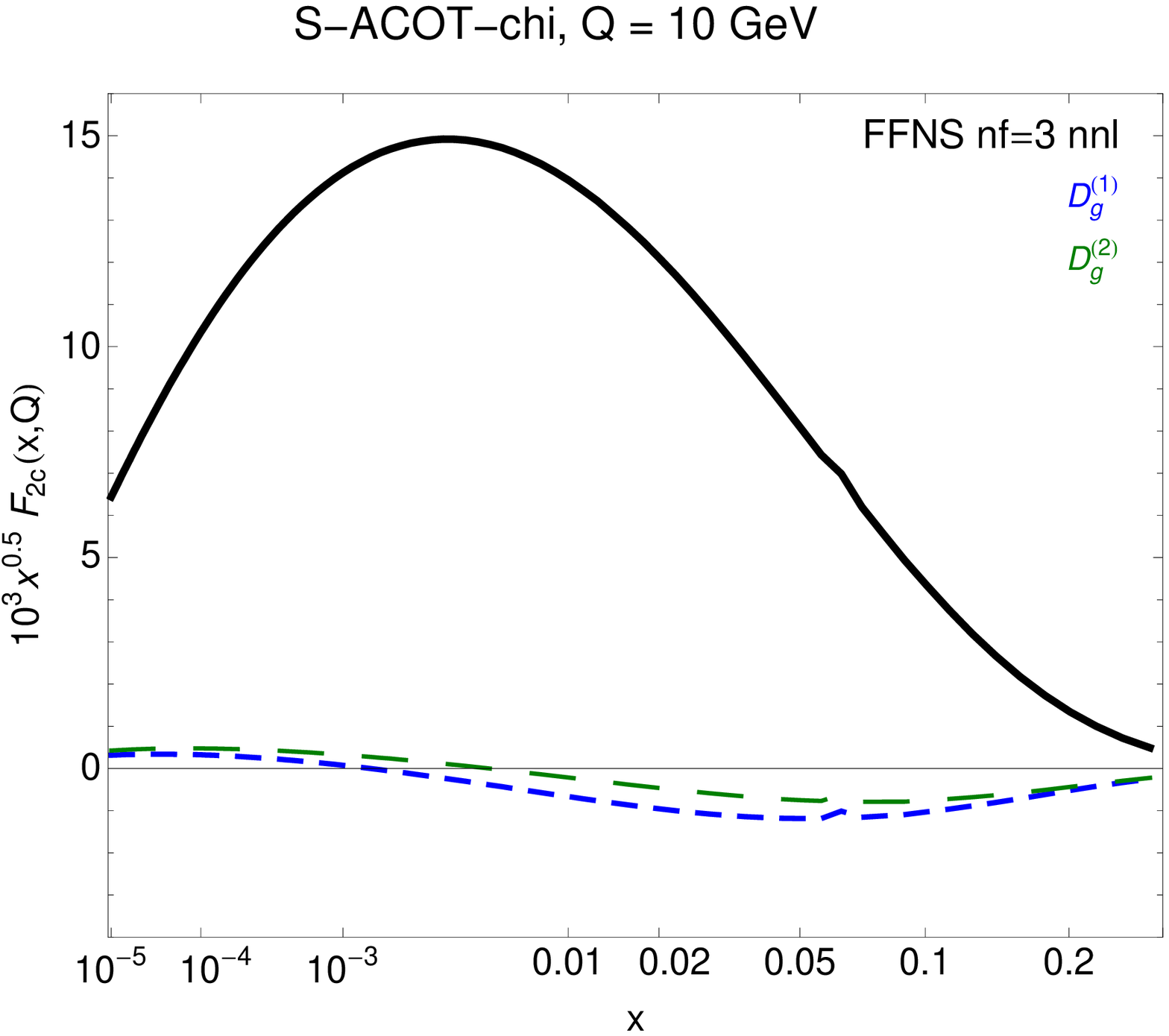} 
\par\end{centering}

\caption{{\small Cancellations in the S-ACOT-$\chi$ scheme at $Q^{2}\approx m_{c}^{2}$
(left) and $Q^{2}\gg m_{c}^{2}$ (right). \label{fig:Cancellations1}}}

\end{figure}

The left panel of Fig.~\ref{fig:Cancellations1} shows the $x$ dependence
of $D_{C^{(0)}}^{(1)}$, $D_{C^{(0)}}^{(2)}$, and $D_{C^{(1)}}^{(2)}$
at $Q=2$ GeV. To provide visual guidance, these differences are compared
to the FFN $N_{f}=3$ prediction at ${\cal O}(a_{s}^{2})$
(solid black line), which is roughly equal to the total rate at this
$Q$ (cf. the previous subsection). We also plot the S-ACOT-$\chi$
contribution of ${\cal O}(a_{s}^{0})$ provided by $c_{h,h}^{(0)}$
(dashed blue line), nominally counted as the lowest-order contribution.
While the LO contribution on its own is substantial comparatively
to the FFN result, it is mostly canceled by the subtraction in Eq.~(\ref{D1C0}),
so that the resulting difference $D_{C^{(0)}}^{(1)}$ (long-dashed
green line) is small.

The cancellation in $D_{C^{(0)}}^{(1)}$ is further improved by including
the next-order terms in $D_{C^{(0)}}^{(2)}$ as in Eq.~(\ref{D2C0}).
The difference $D_{C^{(0)}}^{(2)}$ (dot-dashed red line) and especially
the counterpart difference $D_{C^{(1)}}^{(2)}$ (dotted purple line)
give decreasingly small contributions. They satisfy \begin{eqnarray}
\left|D_{C^{(1)}}^{(2)}\right|\ll\left|D_{C^{(0)}}^{(2)}\right|\ll\left|D_{C^{(0)}}^{(1)}\right|\leq F_{2,c}(x,Q).\end{eqnarray}
Therefore, as $Q\rightarrow m_{c}$, the S-ACOT-$\chi$ scheme exhibits
an almost perfect match on the FFN computation by the virtue of perturbative
cancellations that improve with each order of $\alpha_{s}$. %

\subsection{Cancellations at large $Q$ \label{sec:CancellationsHighQ}}

A different cancellation pattern is observed when $m_{c}$ is negligible
compared to $Q$, when the large logarithms collected in $A_{hg}^{(1)}$,
etc. must be subtracted from the massive contributions $F^{(k)}$ to obtain
the infrared-safe coefficient functions $C^{(k)}$. These cancellations are illustrated in
the right panel of Fig.~\ref{fig:Cancellations1} by $D_{g}^{(1)}$
and $D_{g}^{(2)}$. They quantify the collinear subtractions in the contributions
containing the $\gamma^{*}g$ box subgraph. The lowest-order difference
$D_{g}^{(1)}$ is equal to the convolution of the coefficient $C_{h,g}^{(1)}$
as defined by Eq.(\ref{coefFh22}): \begin{eqnarray}
D_{g}^{(1)}(x,Q)\equiv\left[C_{h,g}^{(1)}\otimes g\right](x,Q)=a_{s}\left\{ \left[F_{h,g}^{(1)}\otimes g\right](x,Q)-\left[c_{h,h}^{(0)}\otimes A_{hg}^{(1)}\otimes g\right](x,Q)\right\} .\end{eqnarray}
 In this expression the subtraction term matches on the ${\cal O}(a_{s})$
photon-gluon contribution represented by $F_{h,g}^{(1)}$. The $x$
dependence of this matching is shown in the right panel of Fig.~\ref{fig:Cancellations1}
for $Q=10$ GeV. It can be seen that $D_{g}^{(1)}$ (blue short-dashed
line) is quite small compared to the ${\cal O}(a_{s}^{2})$
FFN result.

The $\alpha_{s}^{2}$-order difference can be constructed as \[
D_{g}^{(2)}(x,Q)=D_{g}^{(1)}(x,Q)+a_{s}^{2}\left\{ \left[C_{h,g}^{(2)}\otimes g\right](x,Q)+\left[C_{h,l}^{(2)}\otimes\Sigma\right](x,Q)\right\} ,\]
 which can be cast into the form

\begin{eqnarray}
 &  & D_{g}^{(2)}=D_{g}^{(1)}+a_{s}^{2}\left\{ \widehat{F}_{h,g}^{(2)}\otimes g+\widehat{F}_{h,l}^{PS,(2)}\otimes\Sigma-c_{h,h}^{(1)}\otimes A_{hg}^{(1)}\otimes g\right.\nonumber \\
 &  & \left.-c_{h,h}^{(0)}\otimes A_{hg}^{(2)}\otimes g-c_{h,h}^{(0)}\otimes A_{hl}^{PS,(2)}\otimes\Sigma\right\} \end{eqnarray}
by virtue of Eqs.~(\ref{coefFh23}) and (\ref{coefFh24}). At this
order, the collinear logarithms arising in $\widehat{F}_{h,g}^{(2)}$
are canceled by $c_{h,h}^{(0)}\otimes A_{hg}^{(2)}$ and $c_{h,h}^{(1)}\otimes A_{hg}^{(1)}$,
and, similarly, the collinear term in $\widehat{F}_{h,l}^{PS,(2)}$
is removed by $c_{h,h}^{(0)}\otimes A_{hl}^{PS,(2)}$ . The net effect
of the subtractions is that $D_{g}^{(2)}$ (the green long-dashed
line) provides a small correction to $D_{g}^{(1)}$. The perturbative
series converge well for $D_{g}^{(k)}$: \begin{eqnarray}
\left|D_{g}^{(2)}-D_{g}^{(1)}\right|\ll\left|D_{g}^{(1)}\right|\ll F_{2}^{c}(x,Q)\,.\end{eqnarray}

\subsection{Cancellations without kinematic rescaling\label{sec:Cancelnorescaling}}

\begin{figure}[ht]

\begin{centering}
\includegraphics[width=0.49\columnwidth]{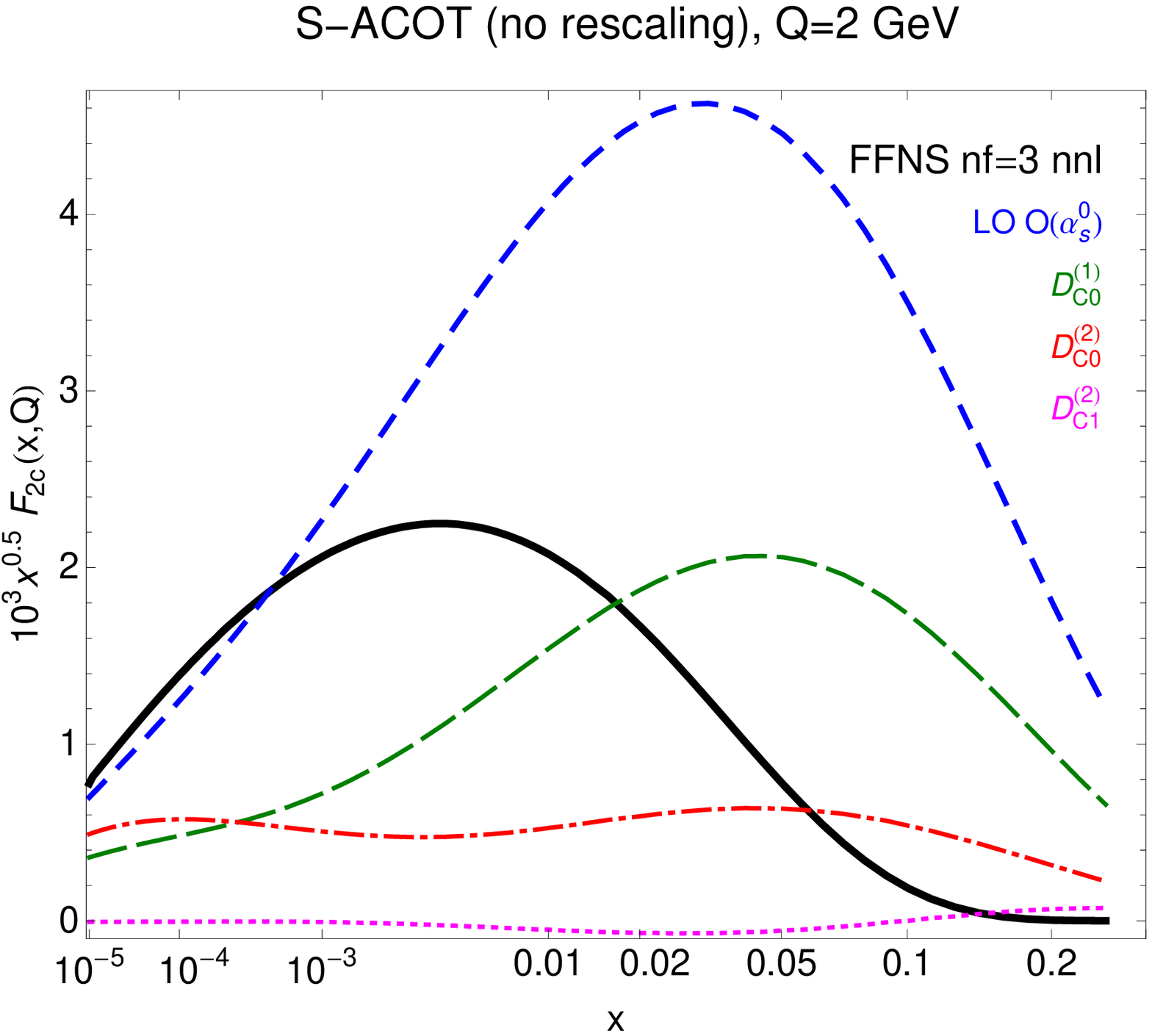} \includegraphics[width=0.49\columnwidth]{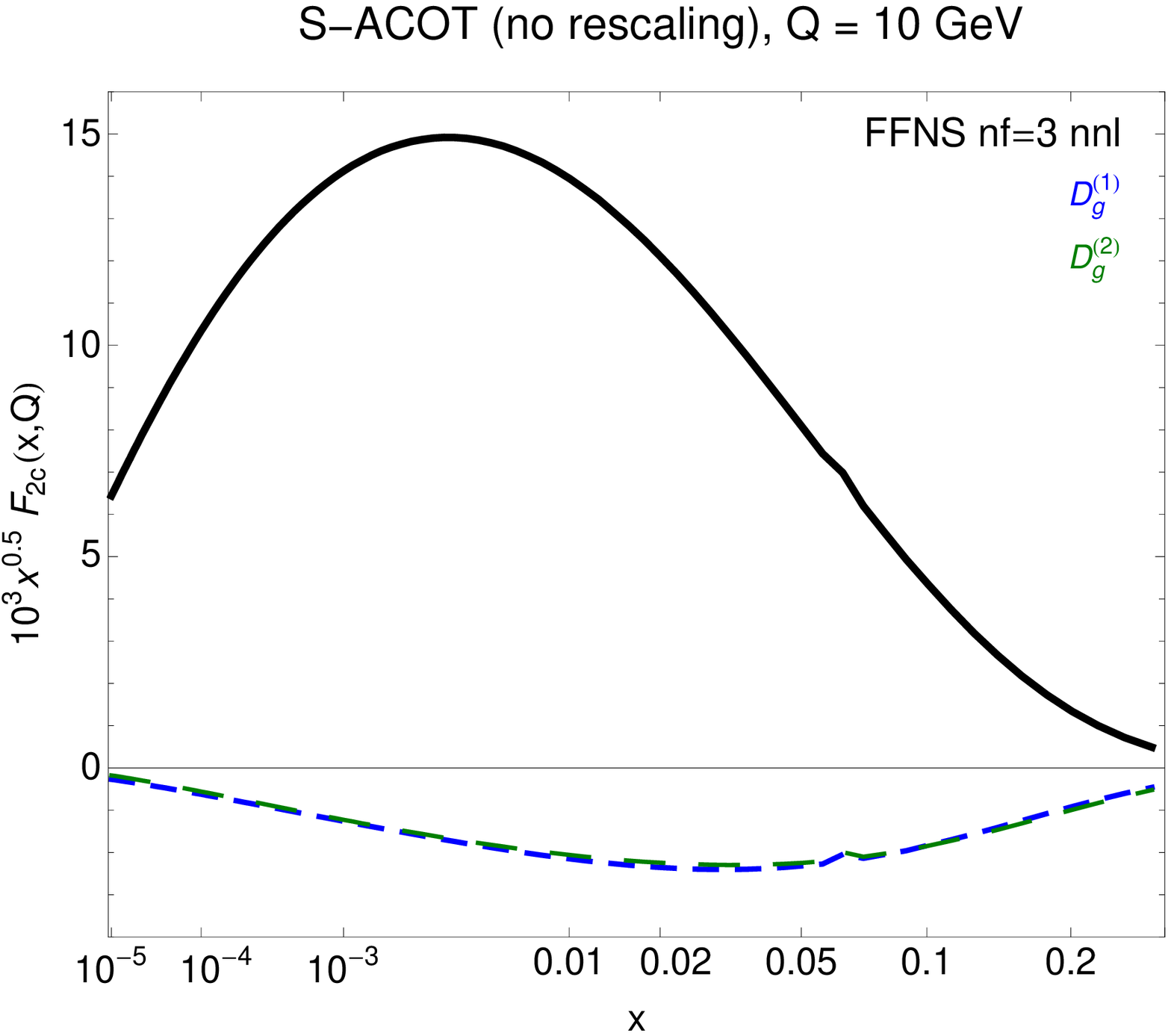} 
\par\end{centering}

\caption{Same as Fig.~\ref{fig:Cancellations1}, in the S-ACOT scheme without
rescaling.}

{\small \label{fig:Cancellations2}} 
\end{figure}

Although the cancellations happen for any rescaling variable $\zeta$,
their perturbative convergence is slower for a non-optimal choice,
such as $\zeta=x$. The differences $D_{C^{(0)}}^{(1)},$ etc. for
$\zeta=x$ are shown in Fig.~\ref{fig:Cancellations2} and, as one
can see, they are generally larger than in the case of $\zeta=\chi.$
Nonetheless, the differences are reduced by going to NNLO, although
not as fast as for the optimal rescaling choice. 

\section{Conclusions\label{sec:Conclusions}}

We examined connections between multi-loop calculations for massive
quark production and fundamental concepts behind QCD factorization.
An NNLO calculation for neutral-current DIS with massive quarks is
documented in a form that bears structural similarity to the NNLO
computation in the zero-mass scheme \citet{SanchezGuillen:1990iq,vanNeerven:1991nn,Zijlstra:1991qc}.
This calculation is algorithmic and utilizes readily available NNLO
expressions. The main formulas are presented by Eqs.~(\ref{FSeveralHeavyFlavors}),
(\ref{FheavyFinal}), and (\ref{FlightFinal}). The theoretical 
derivation presented in Sec.~\ref{sec:Theory} can be readily 
extended to two loops in charged-current DIS, 
after all needed heavy-quark matrix elements are computed. 

The conceptual foundation for the presented results is provided 
by the S-ACOT-$\chi$ factorization scheme. 
The discussion emphasized several strong
features of this scheme: its direct origin from the proof of QCD factorization
for DIS \citet{Collins:1998rz}, relative simplicity, and compliance
with phase space constraints on heavy-quark production at all energies. 

Throughout this study, we highlighted phenomenological importance
of energy conservation in massive particle production. We have shown
how the constraints from energy conservation can be satisfied in all
channels as a part of the QCD factorization theorem. These constraints
are included in the definition of the $Z$ operation in the Collins'
proof of QCD factorization by rescaling the partonic
momentum fraction in flavor-excitation Wilson coefficients. 
The rescaling variable depends on the mass $\sum m_h$ of heavy particles in 
the final state as $\chi=x\,\Bigl(1+(\sum m_h)^{2}/Q^{2}\Bigr)$, where
  $\sum m_h$=$2\, m_h$ and $m_h$ at the lowest order in neutral-current DIS 
and charged-current DIS, respectively.  
The S-ACOT-$\chi$
scheme thus realizes correct kinematical dependence solely by the
means of the QCD factorization theorem and momentum conservation. 

Schemes of the ACOT family differ only in mass-dependent terms in
heavy-quark Wilson coefficient functions. PDFs are given by the same
operator matrix elements in all schemes, such as Eq.~(\ref{Phiab}).
Estimates of these PDFs from global fits converge to unique universal
functions as order of the QCD coupling increases. Convergence 
is the fastest in the S-ACOT-$\chi$ scheme.

At NNLO, dependence of S-ACOT-$\chi$ predictions on the factorization
scale and other tunable parameters is reduced compared to NLO. Cancellations
between classes of Feynman diagrams are stabilized once NNLO terms 
are included. 

After the first version of this paper has been submitted,
an independent S-ACOT-$\chi$ calculation for NC DIS
has been realized in Ref.~\cite{Stavreva:2012bs}. In that approach, 
full mass dependence is included at ${\cal O}(\alpha_s)$, 
while {\it approximate} matrix elements are used in all heavy-quark channels 
at ${\cal O}(\alpha_s^2)$ and ${\cal O}(\alpha_s^3)$. They are obtained
from ZM matrix elements evaluated with a rescaling variable that mimic 
the dominant kinematic contributions, in an approach that resembles
the ``intermediate-mass'' scheme proposed in Ref.~\cite{Nadolsky:2009ge}.
In our study, 
the ${\cal O}(\alpha_s^2)$ contributions to flavor-creation channels 
and threshold matching 
coefficients are computed exactly, so that it reduces to the FFN scheme 
at $Q\approx m_c$. Since the kinematical mass terms 
dominate over the dynamical terms in most practical 
situations \cite{Nadolsky:2009ge}, the 
calculation in Ref.~\cite{Stavreva:2012bs} is beneficial for obtaining estimates of 
yet unknown heavy-quark coefficient functions, notably for heavy-quark
contributions to
neutral-current DIS at three loops and charged-current DIS at two loops.

The derivation
of S-ACOT-$\chi$ predictions is simpler than in some other GM schemes
\citet{Buza:1996wv,Buza:1997nv}, as it is carried out by assuming 
a unique number of active flavors ($N_{f}$) and one set of universal PDFs 
in every $Q$ range. 
It is minimal, in the sense that it does not impose conditions on
the $Q$ derivatives of structure functions \citet{Thorne:1997uu}
or introduce a damping factor \citet{Forte:2010ta}. Yet, after the
NNLO terms are included, the S-ACOT-$\chi$ predictions result in
good agreement with the other GM schemes. As the default heavy-quark
scheme of CTEQ PDF analyses, the S-ACOT-$\chi$ scheme is going to
play a crucial role in global fits at NNLO.

\subsubsection*{Acknowledgments}

Many ideas in this paper were inspired by Wu-Ki Tung. We thank J.
Smith for providing the computer code for computing massive NNLO DIS
cross sections and for helpful comments on this work. We benefited from
discussions with J. Huston, F. Olness, J. Pumplin, D. Stump, 
and other CTEQ members.
M.G. and P.M.N. appreciate stimulating communications with S. Alekhin, 
J. Bl\"umlein, A. Cooper-Sarkar, S. Forte, A. Mitov, S. Moch, J. Rojo, 
and R. Thorne.
This work was supported in part by the U.S. DOE Early Career Research Award
DE-SC0003870; by the U.S. National Science Foundation under grant
PHY-0855561; by the National Science Council of Taiwan under grants
NSC-98-2112-M-133-002-MY3 and NSC-99-2918-I-133-001; and by Lightner-Sams
Foundation. CPY appreciates hospitality of the National Center for
Theoretical Sciences in Taiwan, where a part of this work was done. 
MG thanks the hospitality of DESY during the work on the updated version.


\begin{thebibliography}{62}
\expandafter\ifx\csname natexlab\endcsname\relax\def\natexlab#1{#1}\fi
\expandafter\ifx\csname bibnamefont\endcsname\relax
  \def\bibnamefont#1{#1}\fi
\expandafter\ifx\csname bibfnamefont\endcsname\relax
  \def\bibfnamefont#1{#1}\fi
\expandafter\ifx\csname citenamefont\endcsname\relax
  \def\citenamefont#1{#1}\fi
\expandafter\ifx\csname url\endcsname\relax
  \def\url#1{\texttt{#1}}\fi
\expandafter\ifx\csname urlprefix\endcsname\relax\def\urlprefix{URL }\fi
\providecommand{\bibinfo}[2]{#2}
\providecommand{\eprint}[2][]{\url{#2}}

\bibitem[{\citenamefont{Tung et~al.}(2007)}]{Tung:2006tb}
\bibinfo{author}{\bibfnamefont{W.~K.} \bibnamefont{Tung}} \bibnamefont{et~al.},
  \bibinfo{journal}{JHEP} \textbf{\bibinfo{volume}{02}}, \bibinfo{pages}{053}
  (\bibinfo{year}{2007}).

\bibitem[{\citenamefont{Aivazis
  et~al.}(1994{\natexlab{a}})\citenamefont{Aivazis, Collins, Olness, and
  Tung}}]{Aivazis:1993pi}
\bibinfo{author}{\bibfnamefont{M.~A.~G.} \bibnamefont{Aivazis}},
  \bibinfo{author}{\bibfnamefont{J.~C.} \bibnamefont{Collins}},
  \bibinfo{author}{\bibfnamefont{F.~I.} \bibnamefont{Olness}},
  \bibnamefont{and} \bibinfo{author}{\bibfnamefont{W.-K.} \bibnamefont{Tung}},
  \bibinfo{journal}{Phys. Rev.} \textbf{\bibinfo{volume}{D50}},
  \bibinfo{pages}{3102} (\bibinfo{year}{1994}{\natexlab{a}}).

\bibitem[{\citenamefont{Collins}(1998)}]{Collins:1998rz}
\bibinfo{author}{\bibfnamefont{J.~C.} \bibnamefont{Collins}},
  \bibinfo{journal}{Phys. Rev.} \textbf{\bibinfo{volume}{D58}},
  \bibinfo{pages}{094002} (\bibinfo{year}{1998}).

\bibitem[{\citenamefont{Kramer et~al.}(2000)\citenamefont{Kramer, Olness, and
  Soper}}]{Kramer:2000hn}
\bibinfo{author}{\bibfnamefont{M.}~\bibnamefont{Kramer}, \bibfnamefont{1}},
  \bibinfo{author}{\bibfnamefont{F.~I.} \bibnamefont{Olness}},
  \bibnamefont{and} \bibinfo{author}{\bibfnamefont{D.~E.} \bibnamefont{Soper}},
  \bibinfo{journal}{Phys. Rev.} \textbf{\bibinfo{volume}{D62}},
  \bibinfo{pages}{096007} (\bibinfo{year}{2000}).

\bibitem[{\citenamefont{Tung et~al.}(2002)\citenamefont{Tung, Kretzer, and
  Schmidt}}]{Tung:2001mv}
\bibinfo{author}{\bibfnamefont{W.-K.} \bibnamefont{Tung}},
  \bibinfo{author}{\bibfnamefont{S.}~\bibnamefont{Kretzer}}, \bibnamefont{and}
  \bibinfo{author}{\bibfnamefont{C.}~\bibnamefont{Schmidt}},
  \bibinfo{journal}{J. Phys.} \textbf{\bibinfo{volume}{G28}},
  \bibinfo{pages}{983} (\bibinfo{year}{2002}).

\bibitem[{\citenamefont{Lai et~al.}(2010)}]{Lai:2010vv}
\bibinfo{author}{\bibfnamefont{H.-L.} \bibnamefont{Lai}} \bibnamefont{et~al.},
  \bibinfo{journal}{Phys. Rev.} \textbf{\bibinfo{volume}{D82}},
  \bibinfo{pages}{074024} (\bibinfo{year}{2010}).

\bibitem[{\citenamefont{Nadolsky et~al.}(2008)}]{Nadolsky:2008zw}
\bibinfo{author}{\bibfnamefont{P.~M.} \bibnamefont{Nadolsky}}
  \bibnamefont{et~al.}, \bibinfo{journal}{Phys. Rev.}
  \textbf{\bibinfo{volume}{D78}}, \bibinfo{pages}{013004}
  (\bibinfo{year}{2008}).

\bibitem[{\citenamefont{Pumplin et~al.}(2009)}]{Pumplin:2009nk}
\bibinfo{author}{\bibfnamefont{J.}~\bibnamefont{Pumplin}} \bibnamefont{et~al.},
  \bibinfo{journal}{Phys. Rev.} \textbf{\bibinfo{volume}{D80}},
  \bibinfo{pages}{014019} (\bibinfo{year}{2009}).

\bibitem[{\citenamefont{Moch and Rogal}(2007)}]{Moch:2007gx}
\bibinfo{author}{\bibfnamefont{S.}~\bibnamefont{Moch}} \bibnamefont{and}
  \bibinfo{author}{\bibfnamefont{M.}~\bibnamefont{Rogal}},
  \bibinfo{journal}{Nucl.Phys.} \textbf{\bibinfo{volume}{B782}},
  \bibinfo{pages}{51} (\bibinfo{year}{2007}).

\bibitem[{\citenamefont{Moch et~al.}(2009)\citenamefont{Moch, Vermaseren, and
  Vogt}}]{Moch:2008fj}
\bibinfo{author}{\bibfnamefont{S.}~\bibnamefont{Moch}},
  \bibinfo{author}{\bibfnamefont{J.}~\bibnamefont{Vermaseren}},
  \bibnamefont{and} \bibinfo{author}{\bibfnamefont{A.}~\bibnamefont{Vogt}},
  \bibinfo{journal}{Nucl.Phys.} \textbf{\bibinfo{volume}{B813}},
  \bibinfo{pages}{220} (\bibinfo{year}{2009}).

\bibitem[{\citenamefont{Buza and van Neerven}(1997)}]{Buza:1997mg}
\bibinfo{author}{\bibfnamefont{M.}~\bibnamefont{Buza}} \bibnamefont{and}
  \bibinfo{author}{\bibfnamefont{W.}~\bibnamefont{van Neerven}},
  \bibinfo{journal}{Nucl.Phys.} \textbf{\bibinfo{volume}{B500}},
  \bibinfo{pages}{301} (\bibinfo{year}{1997}).

\bibitem[{\citenamefont{Buza et~al.}(1998)\citenamefont{Buza, Matiounine,
  Smith, and van Neerven}}]{Buza:1996wv}
\bibinfo{author}{\bibfnamefont{M.}~\bibnamefont{Buza}},
  \bibinfo{author}{\bibfnamefont{Y.}~\bibnamefont{Matiounine}},
  \bibinfo{author}{\bibfnamefont{J.}~\bibnamefont{Smith}}, \bibnamefont{and}
  \bibinfo{author}{\bibfnamefont{W.~L.} \bibnamefont{van Neerven}},
  \bibinfo{journal}{Eur. Phys. J.} \textbf{\bibinfo{volume}{C1}},
  \bibinfo{pages}{301} (\bibinfo{year}{1998}).

\bibitem[{\citenamefont{Chuvakin et~al.}(2000)\citenamefont{Chuvakin, Smith,
  and van Neerven}}]{Chuvakin:1999nx}
\bibinfo{author}{\bibfnamefont{A.}~\bibnamefont{Chuvakin}},
  \bibinfo{author}{\bibfnamefont{J.}~\bibnamefont{Smith}}, \bibnamefont{and}
  \bibinfo{author}{\bibfnamefont{W.~L.} \bibnamefont{van Neerven}},
  \bibinfo{journal}{Phys. Rev.} \textbf{\bibinfo{volume}{D61}},
  \bibinfo{pages}{096004} (\bibinfo{year}{2000}).

\bibitem[{\citenamefont{Thorne and
  Roberts}(1998{\natexlab{a}})}]{Thorne:1997uu}
\bibinfo{author}{\bibfnamefont{R.~S.} \bibnamefont{Thorne}} \bibnamefont{and}
  \bibinfo{author}{\bibfnamefont{R.~G.} \bibnamefont{Roberts}},
  \bibinfo{journal}{Phys. Lett.} \textbf{\bibinfo{volume}{B421}},
  \bibinfo{pages}{303} (\bibinfo{year}{1998}{\natexlab{a}}).

\bibitem[{\citenamefont{Thorne and
  Roberts}(1998{\natexlab{b}})}]{Thorne:1997ga}
\bibinfo{author}{\bibfnamefont{R.~S.} \bibnamefont{Thorne}} \bibnamefont{and}
  \bibinfo{author}{\bibfnamefont{R.~G.} \bibnamefont{Roberts}},
  \bibinfo{journal}{Phys. Rev.} \textbf{\bibinfo{volume}{D57}},
  \bibinfo{pages}{6871} (\bibinfo{year}{1998}{\natexlab{b}}).

\bibitem[{\citenamefont{Thorne}(2006)}]{Thorne:2006qt}
\bibinfo{author}{\bibfnamefont{R.~S.} \bibnamefont{Thorne}},
  \bibinfo{journal}{Phys. Rev.} \textbf{\bibinfo{volume}{D73}},
  \bibinfo{pages}{054019} (\bibinfo{year}{2006}).

\bibitem[{\citenamefont{Alekhin et~al.}(2010)\citenamefont{Alekhin,
  \protect{Bl\"umlein}, Klein, and Moch}}]{Alekhin:2009ni}
\bibinfo{author}{\bibfnamefont{S.}~\bibnamefont{Alekhin}},
  \bibinfo{author}{\bibfnamefont{J.}~\bibnamefont{\protect{Bl\"umlein}}},
  \bibinfo{author}{\bibfnamefont{S.}~\bibnamefont{Klein}}, \bibnamefont{and}
  \bibinfo{author}{\bibfnamefont{S.}~\bibnamefont{Moch}},
  \bibinfo{journal}{Phys. Rev.} \textbf{\bibinfo{volume}{D81}},
  \bibinfo{pages}{014032} (\bibinfo{year}{2010}).

\bibitem[{\citenamefont{Forte et~al.}(2010)\citenamefont{Forte, Laenen, Nason,
  and Rojo}}]{Forte:2010ta}
\bibinfo{author}{\bibfnamefont{S.}~\bibnamefont{Forte}},
  \bibinfo{author}{\bibfnamefont{E.}~\bibnamefont{Laenen}},
  \bibinfo{author}{\bibfnamefont{P.}~\bibnamefont{Nason}}, \bibnamefont{and}
  \bibinfo{author}{\bibfnamefont{J.}~\bibnamefont{Rojo}},
  \bibinfo{journal}{Nucl. Phys.} \textbf{\bibinfo{volume}{B834}},
  \bibinfo{pages}{116} (\bibinfo{year}{2010}).

\bibitem[{\citenamefont{Sanchez~Guillen
  et~al.}(1991)\citenamefont{Sanchez~Guillen, Miramontes, Miramontes, Parente,
  and Sampayo}}]{SanchezGuillen:1990iq}
\bibinfo{author}{\bibfnamefont{J.}~\bibnamefont{Sanchez~Guillen}},
  \bibinfo{author}{\bibfnamefont{J.}~\bibnamefont{Miramontes}},
  \bibinfo{author}{\bibfnamefont{M.}~\bibnamefont{Miramontes}},
  \bibinfo{author}{\bibfnamefont{G.}~\bibnamefont{Parente}}, \bibnamefont{and}
  \bibinfo{author}{\bibfnamefont{O.~A.} \bibnamefont{Sampayo}},
  \bibinfo{journal}{Nucl. Phys.} \textbf{\bibinfo{volume}{B353}},
  \bibinfo{pages}{337} (\bibinfo{year}{1991}).

\bibitem[{\citenamefont{van Neerven and Zijlstra}(1991)}]{vanNeerven:1991nn}
\bibinfo{author}{\bibfnamefont{W.~L.} \bibnamefont{van Neerven}}
  \bibnamefont{and} \bibinfo{author}{\bibfnamefont{E.~B.}
  \bibnamefont{Zijlstra}}, \bibinfo{journal}{Phys. Lett.}
  \textbf{\bibinfo{volume}{B272}}, \bibinfo{pages}{127} (\bibinfo{year}{1991}).

\bibitem[{\citenamefont{Zijlstra and van Neerven}(1991)}]{Zijlstra:1991qc}
\bibinfo{author}{\bibfnamefont{E.~B.} \bibnamefont{Zijlstra}} \bibnamefont{and}
  \bibinfo{author}{\bibfnamefont{W.~L.} \bibnamefont{van Neerven}},
  \bibinfo{journal}{Phys. Lett.} \textbf{\bibinfo{volume}{B273}},
  \bibinfo{pages}{476} (\bibinfo{year}{1991}).

\bibitem[{\citenamefont{Thorne}(2010)}]{Thorne:2010pa}
\bibinfo{author}{\bibfnamefont{R.}~\bibnamefont{Thorne}},
  \bibinfo{journal}{PoS} \textbf{\bibinfo{volume}{DIS2010}},
  \bibinfo{pages}{053} (\bibinfo{year}{2010}), \eprint{1006.5925}.

\bibitem[{\citenamefont{Alekhin et~al.}(2011)\citenamefont{Alekhin, Alioli,
  Ball, Bertone, \protect{Bl\"umlein} et~al.}}]{Alekhin:2011sk}
\bibinfo{author}{\bibfnamefont{S.}~\bibnamefont{Alekhin}},
  \bibinfo{author}{\bibfnamefont{S.}~\bibnamefont{Alioli}},
  \bibinfo{author}{\bibfnamefont{R.~D.} \bibnamefont{Ball}},
  \bibinfo{author}{\bibfnamefont{V.}~\bibnamefont{Bertone}},
  \bibinfo{author}{\bibfnamefont{J.}~\bibnamefont{\protect{Bl\"umlein}}},
  \bibnamefont{et~al.} (\bibinfo{year}{2011}), \eprint{arXiv:1101.0536}.

\bibitem[{\citenamefont{\protect{Pla\v{c}akyt\.{e}}}(2010)}]{Placakyte:2010}
\bibinfo{author}{\bibfnamefont{R.}~\bibnamefont{\protect{Pla\v{c}akyt\.{e}}}}
  (\bibinfo{year}{2010}), \bibinfo{note}{talk at the PDF4LHC meeting,
  http://indico.cern.ch/materialDisplay.py?contribId=6\&sessionId=2\&materialI%
d=slides\&confId=103872}.

\bibitem[{\citenamefont{Aaron et~al.}(2010)}]{:2009wt}
\bibinfo{author}{\bibfnamefont{F.}~\bibnamefont{Aaron}} \bibnamefont{et~al.}
  (\bibinfo{collaboration}{H1 and ZEUS Collaboration}), \bibinfo{journal}{JHEP}
  \textbf{\bibinfo{volume}{1001}}, \bibinfo{pages}{109} (\bibinfo{year}{2010}).

\bibitem[{\citenamefont{Laenen et~al.}(1993)\citenamefont{Laenen, Riemersma,
  Smith, and van Neerven}}]{Laenen:1992zk}
\bibinfo{author}{\bibfnamefont{E.}~\bibnamefont{Laenen}},
  \bibinfo{author}{\bibfnamefont{S.}~\bibnamefont{Riemersma}},
  \bibinfo{author}{\bibfnamefont{J.}~\bibnamefont{Smith}}, \bibnamefont{and}
  \bibinfo{author}{\bibfnamefont{W.~L.} \bibnamefont{van Neerven}},
  \bibinfo{journal}{Nucl. Phys.} \textbf{\bibinfo{volume}{B392}},
  \bibinfo{pages}{162} (\bibinfo{year}{1993}).

\bibitem[{\citenamefont{Riemersma et~al.}(1995)\citenamefont{Riemersma, Smith,
  and van Neerven}}]{Riemersma:1994hv}
\bibinfo{author}{\bibfnamefont{S.}~\bibnamefont{Riemersma}},
  \bibinfo{author}{\bibfnamefont{J.}~\bibnamefont{Smith}}, \bibnamefont{and}
  \bibinfo{author}{\bibfnamefont{W.~L.} \bibnamefont{van Neerven}},
  \bibinfo{journal}{Phys. Lett.} \textbf{\bibinfo{volume}{B347}},
  \bibinfo{pages}{143} (\bibinfo{year}{1995}).

\bibitem[{\citenamefont{Harris and Smith}(1995)}]{Harris:1995tu}
\bibinfo{author}{\bibfnamefont{B.}~\bibnamefont{Harris}} \bibnamefont{and}
  \bibinfo{author}{\bibfnamefont{J.}~\bibnamefont{Smith}},
  \bibinfo{journal}{Nucl.Phys.} \textbf{\bibinfo{volume}{B452}},
  \bibinfo{pages}{109} (\bibinfo{year}{1995}).

\bibitem[{\citenamefont{Buza et~al.}(1997)\citenamefont{Buza, Matiounine,
  Smith, and van Neerven}}]{Buza:1997nv}
\bibinfo{author}{\bibfnamefont{M.}~\bibnamefont{Buza}},
  \bibinfo{author}{\bibfnamefont{Y.}~\bibnamefont{Matiounine}},
  \bibinfo{author}{\bibfnamefont{J.}~\bibnamefont{Smith}}, \bibnamefont{and}
  \bibinfo{author}{\bibfnamefont{W.}~\bibnamefont{van Neerven}},
  \bibinfo{journal}{Phys.Lett.} \textbf{\bibinfo{volume}{B411}},
  \bibinfo{pages}{211} (\bibinfo{year}{1997}).

\bibitem[{\citenamefont{Bierenbaum et~al.}(2007)\citenamefont{Bierenbaum,
  \protect{Bl\"umlein}, and Klein}}]{Bierenbaum:2007qe}
\bibinfo{author}{\bibfnamefont{I.}~\bibnamefont{Bierenbaum}},
  \bibinfo{author}{\bibfnamefont{J.}~\bibnamefont{\protect{Bl\"umlein}}},
  \bibnamefont{and} \bibinfo{author}{\bibfnamefont{S.}~\bibnamefont{Klein}},
  \bibinfo{journal}{Nucl.Phys.} \textbf{\bibinfo{volume}{B780}},
  \bibinfo{pages}{40} (\bibinfo{year}{2007}).

\bibitem[{\citenamefont{Bierenbaum
  et~al.}(2009{\natexlab{a}})\citenamefont{Bierenbaum, \protect{Bl\"umlein},
  and Klein}}]{Bierenbaum:2009zt}
\bibinfo{author}{\bibfnamefont{I.}~\bibnamefont{Bierenbaum}},
  \bibinfo{author}{\bibfnamefont{J.}~\bibnamefont{\protect{Bl\"umlein}}},
  \bibnamefont{and} \bibinfo{author}{\bibfnamefont{S.}~\bibnamefont{Klein}},
  \bibinfo{journal}{Phys.Lett.} \textbf{\bibinfo{volume}{B672}},
  \bibinfo{pages}{401} (\bibinfo{year}{2009}{\natexlab{a}}).

\bibitem[{\citenamefont{\protect{Bl\"umlein}
  et~al.}(2006)\citenamefont{\protect{Bl\"umlein}, De~Freitas, van Neerven, and
  Klein}}]{Blumlein:2006mh}
\bibinfo{author}{\bibfnamefont{J.}~\bibnamefont{\protect{Bl\"umlein}}},
  \bibinfo{author}{\bibfnamefont{A.}~\bibnamefont{De~Freitas}},
  \bibinfo{author}{\bibfnamefont{W.}~\bibnamefont{van Neerven}},
  \bibnamefont{and} \bibinfo{author}{\bibfnamefont{S.}~\bibnamefont{Klein}},
  \bibinfo{journal}{Nucl.Phys.} \textbf{\bibinfo{volume}{B755}},
  \bibinfo{pages}{272} (\bibinfo{year}{2006}).

\bibitem[{\citenamefont{Bierenbaum
  et~al.}(2009{\natexlab{b}})\citenamefont{Bierenbaum, \protect{Bl\"umlein},
  and Klein}}]{Bierenbaum:2009mv}
\bibinfo{author}{\bibfnamefont{I.}~\bibnamefont{Bierenbaum}},
  \bibinfo{author}{\bibfnamefont{J.}~\bibnamefont{\protect{Bl\"umlein}}},
  \bibnamefont{and} \bibinfo{author}{\bibfnamefont{S.}~\bibnamefont{Klein}},
  \bibinfo{journal}{Nucl.Phys.} \textbf{\bibinfo{volume}{B820}},
  \bibinfo{pages}{417} (\bibinfo{year}{2009}{\natexlab{b}}).

\bibitem[{\citenamefont{Ablinger
  et~al.}(2011{\natexlab{a}})\citenamefont{Ablinger, \protect{Bl\"umlein},
  Klein, Schneider, and Wissbrock}}]{Ablinger:2010ty}
\bibinfo{author}{\bibfnamefont{J.}~\bibnamefont{Ablinger}},
  \bibinfo{author}{\bibfnamefont{J.}~\bibnamefont{\protect{Bl\"umlein}}},
  \bibinfo{author}{\bibfnamefont{S.}~\bibnamefont{Klein}},
  \bibinfo{author}{\bibfnamefont{C.}~\bibnamefont{Schneider}},
  \bibnamefont{and}
  \bibinfo{author}{\bibfnamefont{F.}~\bibnamefont{Wissbrock}},
  \bibinfo{journal}{Nucl.Phys.} \textbf{\bibinfo{volume}{B844}},
  \bibinfo{pages}{26} (\bibinfo{year}{2011}{\natexlab{a}}).

\bibitem[{\citenamefont{\protect{Bl\"umlein}
  et~al.}(2012)\citenamefont{\protect{Bl\"umlein}, Hasselhuhn, Klein, and
  Schneider}}]{Blumlein:2012vq}
\bibinfo{author}{\bibfnamefont{J.}~\bibnamefont{\protect{Bl\"umlein}}},
  \bibinfo{author}{\bibfnamefont{A.}~\bibnamefont{Hasselhuhn}},
  \bibinfo{author}{\bibfnamefont{S.}~\bibnamefont{Klein}}, \bibnamefont{and}
  \bibinfo{author}{\bibfnamefont{C.}~\bibnamefont{Schneider}}
  (\bibinfo{year}{2012}), \eprint{arXiv:1205.4184}.

\bibitem[{\citenamefont{Ablinger
  et~al.}(2011{\natexlab{b}})\citenamefont{Ablinger, \protect{Bl\"umlein},
  Klein, Schneider, and Wissbrock}}]{Ablinger:2011pb}
\bibinfo{author}{\bibfnamefont{J.}~\bibnamefont{Ablinger}},
  \bibinfo{author}{\bibfnamefont{J.}~\bibnamefont{\protect{Bl\"umlein}}},
  \bibinfo{author}{\bibfnamefont{S.}~\bibnamefont{Klein}},
  \bibinfo{author}{\bibfnamefont{C.}~\bibnamefont{Schneider}},
  \bibnamefont{and} \bibinfo{author}{\bibfnamefont{F.}~\bibnamefont{Wissbrock}}
  (\bibinfo{year}{2011}{\natexlab{b}}), \eprint{arXiv:1106.5937}.

\bibitem[{\citenamefont{Ablinger et~al.}(2012)\citenamefont{Ablinger,
  \protect{Bl\"umlein}, Hasselhuhn, Klein, Schneider et~al.}}]{Ablinger:2012qj}
\bibinfo{author}{\bibfnamefont{J.}~\bibnamefont{Ablinger}},
  \bibinfo{author}{\bibfnamefont{J.}~\bibnamefont{\protect{Bl\"umlein}}},
  \bibinfo{author}{\bibfnamefont{A.}~\bibnamefont{Hasselhuhn}},
  \bibinfo{author}{\bibfnamefont{S.}~\bibnamefont{Klein}},
  \bibinfo{author}{\bibfnamefont{C.}~\bibnamefont{Schneider}},
  \bibnamefont{et~al.} (\bibinfo{year}{2012}), \eprint{arXiv:1202.2700}.

\bibitem[{\citenamefont{Laenen and Moch}(1999)}]{Laenen:1998kp}
\bibinfo{author}{\bibfnamefont{E.}~\bibnamefont{Laenen}} \bibnamefont{and}
  \bibinfo{author}{\bibfnamefont{S.-O.} \bibnamefont{Moch}},
  \bibinfo{journal}{Phys.Rev.} \textbf{\bibinfo{volume}{D59}},
  \bibinfo{pages}{034027} (\bibinfo{year}{1999}).

\bibitem[{\citenamefont{Kawamura et~al.}(2012)\citenamefont{Kawamura, Presti,
  Moch, and Vogt}}]{Kawamura:2012cr}
\bibinfo{author}{\bibfnamefont{H.}~\bibnamefont{Kawamura}},
  \bibinfo{author}{\bibfnamefont{N.~L.} \bibnamefont{Presti}},
  \bibinfo{author}{\bibfnamefont{S.}~\bibnamefont{Moch}}, \bibnamefont{and}
  \bibinfo{author}{\bibfnamefont{A.}~\bibnamefont{Vogt}}
  (\bibinfo{year}{2012}), \eprint{arXiv:1205.5727}.

\bibitem[{\citenamefont{Aivazis
  et~al.}(1994{\natexlab{b}})\citenamefont{Aivazis, Olness, and
  Tung}}]{Aivazis:1993kh}
\bibinfo{author}{\bibfnamefont{M.~A.~G.} \bibnamefont{Aivazis}},
  \bibinfo{author}{\bibfnamefont{F.~I.} \bibnamefont{Olness}},
  \bibnamefont{and} \bibinfo{author}{\bibfnamefont{W.-K.} \bibnamefont{Tung}},
  \bibinfo{journal}{Phys. Rev.} \textbf{\bibinfo{volume}{D50}},
  \bibinfo{pages}{3085} (\bibinfo{year}{1994}{\natexlab{b}}).

\bibitem[{\citenamefont{Barnett}(1976)}]{Barnett:1976ak}
\bibinfo{author}{\bibfnamefont{R.~M.} \bibnamefont{Barnett}},
  \bibinfo{journal}{Phys.Rev.Lett.} \textbf{\bibinfo{volume}{36}},
  \bibinfo{pages}{1163} (\bibinfo{year}{1976}).

\bibitem[{\citenamefont{Nadolsky and Tung}(2009)}]{Nadolsky:2009ge}
\bibinfo{author}{\bibfnamefont{P.~M.} \bibnamefont{Nadolsky}} \bibnamefont{and}
  \bibinfo{author}{\bibfnamefont{W.-K.} \bibnamefont{Tung}},
  \bibinfo{journal}{Phys. Rev.} \textbf{\bibinfo{volume}{D79}},
  \bibinfo{pages}{113014} (\bibinfo{year}{2009}).

\bibitem[{\citenamefont{Witten}(1976)}]{Witten:1975bh}
\bibinfo{author}{\bibfnamefont{E.}~\bibnamefont{Witten}},
  \bibinfo{journal}{Nucl. Phys.} \textbf{\bibinfo{volume}{B104}},
  \bibinfo{pages}{445} (\bibinfo{year}{1976}).

\bibitem[{\citenamefont{Moch et~al.}(2005)\citenamefont{Moch, Vermaseren, and
  Vogt}}]{Moch:2004xu}
\bibinfo{author}{\bibfnamefont{S.}~\bibnamefont{Moch}},
  \bibinfo{author}{\bibfnamefont{J.~A.~M.} \bibnamefont{Vermaseren}},
  \bibnamefont{and} \bibinfo{author}{\bibfnamefont{A.}~\bibnamefont{Vogt}},
  \bibinfo{journal}{Phys. Lett.} \textbf{\bibinfo{volume}{B606}},
  \bibinfo{pages}{123} (\bibinfo{year}{2005}).

\bibitem[{\citenamefont{Vermaseren et~al.}(2005)\citenamefont{Vermaseren, Vogt,
  and Moch}}]{Vermaseren:2005qc}
\bibinfo{author}{\bibfnamefont{J.}~\bibnamefont{Vermaseren}},
  \bibinfo{author}{\bibfnamefont{A.}~\bibnamefont{Vogt}}, \bibnamefont{and}
  \bibinfo{author}{\bibfnamefont{S.}~\bibnamefont{Moch}},
  \bibinfo{journal}{Nucl.Phys.} \textbf{\bibinfo{volume}{B724}},
  \bibinfo{pages}{3} (\bibinfo{year}{2005}).

\bibitem[{\citenamefont{Bardeen et~al.}(1978)\citenamefont{Bardeen, Buras,
  Duke, and Muta}}]{Bardeen:1978yd}
\bibinfo{author}{\bibfnamefont{W.~A.} \bibnamefont{Bardeen}},
  \bibinfo{author}{\bibfnamefont{A.~J.} \bibnamefont{Buras}},
  \bibinfo{author}{\bibfnamefont{D.~W.} \bibnamefont{Duke}}, \bibnamefont{and}
  \bibinfo{author}{\bibfnamefont{T.}~\bibnamefont{Muta}},
  \bibinfo{journal}{Phys. Rev.} \textbf{\bibinfo{volume}{D18}},
  \bibinfo{pages}{3998} (\bibinfo{year}{1978}).

\bibitem[{\citenamefont{Altarelli et~al.}(1978)\citenamefont{Altarelli, Ellis,
  and Martinelli}}]{Altarelli:1978id}
\bibinfo{author}{\bibfnamefont{G.}~\bibnamefont{Altarelli}},
  \bibinfo{author}{\bibfnamefont{R.~K.} \bibnamefont{Ellis}}, \bibnamefont{and}
  \bibinfo{author}{\bibfnamefont{G.}~\bibnamefont{Martinelli}},
  \bibinfo{journal}{Nucl. Phys.} \textbf{\bibinfo{volume}{B143}},
  \bibinfo{pages}{521} (\bibinfo{year}{1978}).

\bibitem[{\citenamefont{Humpert and van Neerven}(1981)}]{Humpert:1980uv}
\bibinfo{author}{\bibfnamefont{B.}~\bibnamefont{Humpert}} \bibnamefont{and}
  \bibinfo{author}{\bibfnamefont{W.~L.} \bibnamefont{van Neerven}},
  \bibinfo{journal}{Nucl. Phys.} \textbf{\bibinfo{volume}{B184}},
  \bibinfo{pages}{225} (\bibinfo{year}{1981}).

\bibitem[{\citenamefont{Buza et~al.}(1996)\citenamefont{Buza, Matiounine,
  Smith, Migneron, and van Neerven}}]{Buza:1995ie}
\bibinfo{author}{\bibfnamefont{M.}~\bibnamefont{Buza}},
  \bibinfo{author}{\bibfnamefont{Y.}~\bibnamefont{Matiounine}},
  \bibinfo{author}{\bibfnamefont{J.}~\bibnamefont{Smith}},
  \bibinfo{author}{\bibfnamefont{R.}~\bibnamefont{Migneron}}, \bibnamefont{and}
  \bibinfo{author}{\bibfnamefont{W.}~\bibnamefont{van Neerven}},
  \bibinfo{journal}{Nucl.Phys.} \textbf{\bibinfo{volume}{B472}},
  \bibinfo{pages}{611} (\bibinfo{year}{1996}).

\bibitem[{\citenamefont{Adler}(1966)}]{Adler:1965ty}
\bibinfo{author}{\bibfnamefont{S.~L.} \bibnamefont{Adler}},
  \bibinfo{journal}{Phys. Rev.} \textbf{\bibinfo{volume}{143}},
  \bibinfo{pages}{1144} (\bibinfo{year}{1966}).

\bibitem[{\citenamefont{Altarelli}(1982)}]{Altarelli:1981ax}
\bibinfo{author}{\bibfnamefont{G.}~\bibnamefont{Altarelli}},
  \bibinfo{journal}{Phys. Rept.} \textbf{\bibinfo{volume}{81}},
  \bibinfo{pages}{1} (\bibinfo{year}{1982}).

\bibitem[{\citenamefont{Dokshitzer et~al.}(1996)\citenamefont{Dokshitzer,
  Marchesini, and Webber}}]{Dokshitzer:1995qm}
\bibinfo{author}{\bibfnamefont{Y.~L.} \bibnamefont{Dokshitzer}},
  \bibinfo{author}{\bibfnamefont{G.}~\bibnamefont{Marchesini}},
  \bibnamefont{and} \bibinfo{author}{\bibfnamefont{B.~R.}
  \bibnamefont{Webber}}, \bibinfo{journal}{Nucl. Phys.}
  \textbf{\bibinfo{volume}{B469}}, \bibinfo{pages}{93} (\bibinfo{year}{1996}).

\bibitem[{\citenamefont{Brock et~al.}(1995)}]{Brock:1993sz}
\bibinfo{author}{\bibfnamefont{R.}~\bibnamefont{Brock}} \bibnamefont{et~al.}
  (\bibinfo{collaboration}{CTEQ}), \bibinfo{journal}{Rev. Mod. Phys.}
  \textbf{\bibinfo{volume}{67}}, \bibinfo{pages}{157} (\bibinfo{year}{1995}).

\bibitem[{\citenamefont{Guzzi and Nadolsky}(2010)}]{SACOTNNLO2011}
\bibinfo{author}{\bibfnamefont{M.}~\bibnamefont{Guzzi}} \bibnamefont{and}
  \bibinfo{author}{\bibfnamefont{P.}~\bibnamefont{Nadolsky}}
  (\bibinfo{year}{2010}),
  \bibinfo{note}{http://hep.pa.msu.edu/cteq/public/SACOTNNLO2011/.}

\bibitem[{\citenamefont{Giele et~al.}(2002)}]{Giele:2002hx}
\bibinfo{author}{\bibfnamefont{W.}~\bibnamefont{Giele}} \bibnamefont{et~al.}
  (\bibinfo{year}{2002}), \eprint{hep-ph/0204316}.

\bibitem[{\citenamefont{Whalley et~al.}(2005)\citenamefont{Whalley, Bourilkov,
  and Group}}]{Whalley:2005nh}
\bibinfo{author}{\bibfnamefont{M.~R.} \bibnamefont{Whalley}},
  \bibinfo{author}{\bibfnamefont{D.}~\bibnamefont{Bourilkov}},
  \bibnamefont{and} \bibinfo{author}{\bibfnamefont{R.~C.} \bibnamefont{Group}}
  (\bibinfo{year}{2005}), \eprint{hep-ph/0508110;
  http://hepforge.cedar.ac.uk/lhapdf/}.

\bibitem[{\citenamefont{Salam and Rojo}(2009)}]{Salam:2008qg}
\bibinfo{author}{\bibfnamefont{G.~P.} \bibnamefont{Salam}} \bibnamefont{and}
  \bibinfo{author}{\bibfnamefont{J.}~\bibnamefont{Rojo}},
  \bibinfo{journal}{Comput. Phys. Commun.} \textbf{\bibinfo{volume}{180}},
  \bibinfo{pages}{120} (\bibinfo{year}{2009}).

\bibitem[{\citenamefont{Martin et~al.}(2009)\citenamefont{Martin, Stirling,
  Thorne, and Watt}}]{Martin:2009iq}
\bibinfo{author}{\bibfnamefont{A.~D.} \bibnamefont{Martin}},
  \bibinfo{author}{\bibfnamefont{W.~J.} \bibnamefont{Stirling}},
  \bibinfo{author}{\bibfnamefont{R.~S.} \bibnamefont{Thorne}},
  \bibnamefont{and} \bibinfo{author}{\bibfnamefont{G.}~\bibnamefont{Watt}},
  \bibinfo{journal}{Eur. Phys. J.} \textbf{\bibinfo{volume}{C63}},
  \bibinfo{pages}{189} (\bibinfo{year}{2009}).

\bibitem[{\citenamefont{Rojo et~al.}(2010)\citenamefont{Rojo, Forte, Huston,
  Nadolsky, Nason, Olness, Thorne, and Watt}}]{Binoth:2010ra}
\bibinfo{author}{\bibfnamefont{J.}~\bibnamefont{Rojo}},
  \bibinfo{author}{\bibfnamefont{S.}~\bibnamefont{Forte}},
  \bibinfo{author}{\bibfnamefont{J.}~\bibnamefont{Huston}},
  \bibinfo{author}{\bibfnamefont{P.}~\bibnamefont{Nadolsky}},
  \bibinfo{author}{\bibfnamefont{P.}~\bibnamefont{Nason}},
  \bibinfo{author}{\bibfnamefont{F.}~\bibnamefont{Olness}},
  \bibinfo{author}{\bibfnamefont{R.}~\bibnamefont{Thorne}}, \bibnamefont{and}
  \bibinfo{author}{\bibfnamefont{G.}~\bibnamefont{Watt}}
  (\bibinfo{collaboration}{SM and NLO Multileg Working Group})
  (\bibinfo{year}{2010}), p. \bibinfo{pages}{110}, \eprint{arXiv:1003.1241}.

\bibitem[{\citenamefont{Collins et~al.}(1978)\citenamefont{Collins, Wilczek,
  and Zee}}]{Collins:1978wz}
\bibinfo{author}{\bibfnamefont{J.~C.} \bibnamefont{Collins}},
  \bibinfo{author}{\bibfnamefont{F.}~\bibnamefont{Wilczek}}, \bibnamefont{and}
  \bibinfo{author}{\bibfnamefont{A.}~\bibnamefont{Zee}},
  \bibinfo{journal}{Phys.Rev.} \textbf{\bibinfo{volume}{D18}},
  \bibinfo{pages}{242} (\bibinfo{year}{1978}).

\bibitem[{\citenamefont{Alekhin et~al.}(2012)\citenamefont{Alekhin,
  \protect{Bl\"umlein}, and Moch}}]{Alekhin:2012ig}
\bibinfo{author}{\bibfnamefont{S.}~\bibnamefont{Alekhin}},
  \bibinfo{author}{\bibfnamefont{J.}~\bibnamefont{\protect{Bl\"umlein}}},
  \bibnamefont{and} \bibinfo{author}{\bibfnamefont{S.}~\bibnamefont{Moch}}
  (\bibinfo{year}{2012}), \eprint{arXiv:1202.2281}.

\bibitem[{\citenamefont{Stavreva et~al.}(2012)\citenamefont{Stavreva, Olness,
  Schienbein, Jezo, Kusina et~al.}}]{Stavreva:2012bs}
\bibinfo{author}{\bibfnamefont{T.}~\bibnamefont{Stavreva}},
  \bibinfo{author}{\bibfnamefont{F.}~\bibnamefont{Olness}},
  \bibinfo{author}{\bibfnamefont{I.}~\bibnamefont{Schienbein}},
  \bibinfo{author}{\bibfnamefont{T.}~\bibnamefont{Jezo}},
  \bibinfo{author}{\bibfnamefont{A.}~\bibnamefont{Kusina}},
  \bibnamefont{et~al.} (\bibinfo{year}{2012}), \eprint{arXiv:1203.0282}.

\end{thebibliography}

\end{document}